\pdfoutput=1
\documentclass[affil-it,auth-lg,compression,contents,custom-packages={enumitem},references={bigone}]{lib/preprint}

\tikzset{
    aux/.style={dashed},
    dottedline/.style={dotted},
    gluon/.style={
        decorate, 
        draw=black,
        decoration={
            snake,
            post=lineto,
            post length=0pt,
            segment length=4,
            amplitude=0.9
        }
    }
}

\newcommand{\dpar}{\partial}
\newcommand{\forw}{\uparrow}
\newcommand{\backw}{\downarrow}
\newcommand{\trans}{\perp}
\newcommand{\cb}{\tte}
\newcommand{\colfac}{\sfc}
\newcommand{\kinfac}{\sfn}

\begin{document}
    
    \date{\today}
    
    \email{l.borsten@hw.ac.uk,hk55@hw.ac.uk,branislav.jurco@gmail.com,tmacrelli@ phys.ethz.ch,c.saemann@hw.ac.uk,m.wolf@surrey.ac.uk}
    
    \preprint{EMPG--21--12,DMUS--MP--21/12}
    
    \title{Tree-Level Color--Kinematics Duality Implies\\[0.2cm] Loop-Level Color--Kinematics Duality up to Counterterms} 
    
    \author[a]{Leron~Borsten}
    \author[a]{Hyungrok~Kim}
    \author[b]{Branislav~Jur{\v c}o}
    \author[c,d]{Tommaso~Macrelli}
    \author[a]{Christian~Saemann}
    \author[c]{Martin~Wolf}
    
    \affil[a]{Maxwell Institute for Mathematical Sciences,\\ Department of Mathematics, Heriot--Watt University, Edinburgh EH14 4AS, UK}
    \affil[b]{Charles University Prague, Faculty of Mathematics and Physics,\\ Mathematical Institute, Prague 186 75, CZ}
    \affil[c]{Department of Mathematics, University of Surrey, Guildford GU2 7XH, UK}
    \affil[d]{Institute for Theoretical Physics, ETH Zurich\\ 8093 Z{\"u}rich, Switzerland}
    
    \abstract{Color--kinematics (CK) duality is a remarkable symmetry of gluon amplitudes that is the key to the double copy which links gauge theory and gravity  amplitudes. Here we show that the complete Yang--Mills action itself, including its gauge-fixing and ghost sectors required for quantization, can be recast to manifest CK duality using a series of field redefinitions and gauge choices. Crucially, the resulting loop-level integrands are automatically CK-dual, up to potential Jacobian counterterms required for unitarity. While these counterterms may break CK duality, they exist, are unique and, since the tree-level is unaffected, may be deduced from the action or the integrands. Consequently, CK duality is a symmetry of the action like any other symmetry, and it is anomalous in a controlled and mostly harmless sense. Our results apply to any theory with CK-dual tree-level amplitudes. We also show that two CK duality-manifesting parent actions may be factorized and fused into a consistent quantizable offspring, with the double copy as the prime example. This provides a direct proof of  the double copy to all loop orders.}
    
    \acknowledgements{We gratefully acknowledge stimulating conversations with Alexandros Anastasiou, Michael Duff, Henrik Johansson, Silvia Nagy, Alessandro Torrielli, and Michele Zoccali. We are also grateful to Clifford Cheung, James Mangan, and Julio Parra-Martinez for interesting questions on the first version of this paper. L.B., H.K., and C.S.~were supported by the Leverhulme Research Project Grant RPG--2018--329 \emph{The Mathematics of M5-Branes}. B.J.~was supported by the GA\v{C}R Grant EXPRO 19--28628X. T.M.~was partially supported by the EPSRC grant EP/N509772.}
    
    \datalicencemanagement{No additional research data beyond the data presented and cited in this work are needed to validate the research findings in this work. For the purpose of open access, the authors have applied a Creative Commons Attribution (CC BY) licence to any Author Accepted Manuscript version arising.} 
    
    \begin{body}
        
        \section{Introduction} 
        
        Two seemingly distinct principles govern the fundamental forces of Nature: the strong and electroweak forces on the one side and gravity on the other. While the former are described by Yang--Mills gauge theories that play out on space--time, gravity is a consequence of the curvature of space--time itself, promoted from stage to protagonist. In spite of this evident discrepancy, it has been shown in the seminal work~\cite{Bern:2008qj,Bern:2009kd,Bern:2014sna} that the tree-level scattering amplitudes of gravity can be derived as the `square' or `double copy' of  Yang--Mills  scattering amplitudes. This observation suggests a novel and fundamental unity between gravity and the other forces of Nature.
        
        The double copy does, however, rest crucially on the long-standing conjecture of Bern--Carrasco--Johansson (BCJ)~\cite{Bern:2008qj} that Yang--Mills scattering amplitudes satisfy what is known as  color--kinematics (CK) duality. Recall that the fundamental degrees of freedom in Yang--Mills theory are described by the gluon field $A^a_\mu(x)$. Beyond the space--time coordinate $x$, the field carries a color index $a$ and a Lorentz (co-)vector index $\mu$. The two sorts of indices serve rather different purposes: one manifests a \emph{color} internal symmetry, while the other instead reflects a \emph{kinematic} space--time symmetry. The CK duality conjecture, however, suggests  both indices should be put on an equal footing in a precise sense. Specifically, it posits that  gluon scattering amplitudes can be organized into sums of products of color and kinematic factors, such that the algebraic structure of the kinematic factors mirrors the Lie algebra relations of the color factors. The origins of this  unexpected property of the scattering amplitudes are nowhere to be found in the standard Yang--Mills action, making it all the more remarkable. The validity of CK duality at the tree level has been established long ago~\cite{BjerrumBohr:2010hn,BjerrumBohr:2009rd,Stieberger:2009hq,Mafra:2011kj,Broedel:2013tta,Du:2016tbc,Mizera:2019blq,Reiterer:2019dys}, but the extension to the loop level has remained an open issue for over a decade.
        
        In the present paper, we will extend the CK duality principle to the complete set of (off-shell) fields appearing in the Yang--Mills Becchi--Rouet--Stora--Tyutin (BRST) action, including the Faddeev--Popov ghosts capturing the gauge symmetry, as well as the Nakanishi--Lautrup fields and the anti-ghosts arising when gauge fixing. As a consequence, CK duality becomes a manifest infinite-dimensional symmetry of the BRST action. The latter serves as a powerful ordering principle for the required BCJ relations on the scattering amplitudes; in particular, the loop amplitude integrands derived directly from the Feynman diagrams of this action are automatically CK-dual. Importantly, they are the integrands of the theory, and they do not require to be validated by unitarity or other methods. At the loop level, these diagrams may have to be supplemented by Jacobian counterterms that ensure unitarity. While these will generically break CK duality, they exist, they are unique (in the appropriate sense), and they may be deduced from the action or the integrands themselves, since the tree level is unaffected.
        
        As observed in~\cite{Bern:2010yg}, on-shell CK duality can be manifested (up to six points) in terms of an action made cubic (or, in our terminology, strictified) through the introduction of auxiliary fields. The paper~\cite{Tolotti:2013caa} extended this to all orders, but without the auxiliary fields that render the action cubic; furthermore, their action was entirely geared towards the tree level, ignoring any and all loop-related desiderata (such as ghosts, BRST symmetry, etc.). In our previous papers~\cite{Borsten:2020zgj,Borsten:2021hua}, we extended the construction in~\cite{Tolotti:2013caa} to obtain an action whose Feynman diagrams manifest on-shell tree-level CK duality for both physical and unphysical fields. This suffices for the double copy of scattering amplitudes and actions to be well-defined~\cite{Borsten:2020zgj}, but not for full off-shell CK duality itself.\footnote{To be precise, the loop kinematic numerators derived directly from the Feynman rules in~\cite{Borsten:2020zgj,Borsten:2021hua} double-copy into genuine scattering amplitudes of $\caN=0$ supergravity. However, this neither requires nor implies that said loop kinematic numerators fulfill CK duality, cf.~also the discussion in~\cite{Bern:2015ooa}.} In the present paper, we go one step further by removing the on-shell restriction in the prior literature up to potentially arising counterterms. It then remains to show that these counterterms are absent or harmless in theories of interests.
        
        Realizing off-shell CK duality as a symmetry of the BRST action amounts to an advantageous shift in perspective. Like many other symmetries of an action, CK duality is generically anomalous at the loop level; in particular, it may be broken by the counterterms required to restore unitarity of the amplitudes. We will show, however, that these counterterms are well under control, and the quantum consistency of the theory is not affected. Moreover, our recent results~\cite{Borsten:2022vtg} strongly suggest that at least for some field theories, in particular for maximally supersymmetric Yang--Mills theory, these counterterms are absent.
        
        The new perspective allows us to replace the original proof of the loop-level double copy based on unitarity methods~\cite{Bern:2010yg} by the result that the CK-dual Yang--Mills BRST action itself manifestly double copies into a consistent, (perturbatively) quantizable gravitational BRST action. Given the tree-level double copy, which is implied by the arguments of~\cite{Bern:2010yg}, this  provides a direct proof that the double copy prescription for amplitudes holds to all loop orders. 

        Our action-focused interpretation opens up the path to a homotopy algebraic formulation of CK duality, parts of which were already visible in~\cite{Reiterer:2019dys,Borsten:2021hua}. From this vantage point, CK duality fails only up to homotopy for a generic representation of the amplitudes. It is rather evident that this homotopy algebraic interpretation will not only provide an explanation of the origins of CK duality and the double copy, linking both to string theory, but it is hoped that it will lead to improvements in the efficiency of the computation of scattering amplitudes based on homotopy algebraic techniques.
        
        This paper is organized as follows: 
        \begin{itemize}\itemsep-2pt
            \item[$\diamond$] \Cref{sec:outline} provides an outline of our constructions;
            \item[$\diamond$]  \Cref{sec:prelim} reviews some core concepts of quantum field theory used in this paper;
            \item[$\diamond$]  \Cref{sec:NLSM} works out the case of the non-linear sigma model for a semi-simple compact Lie group, where the complications of higher spin are absent;
            \item[$\diamond$]  \Cref{sec:loop_CK_duality_YM} details the algorithm to manifest CK duality for a gauge theory, using Yang--Mills theory as an example, and \Cref{sec:SYM} discusses the inclusion of adjoint-valued fermions, which requires supersymmetry;
            \item[$\diamond$]  \Cref{sec:double_copy} introduces the generic syngamy of actions;
            \item[$\diamond$]  The double copy of the non-linear sigma model action with manifest CK duality given in \cref{ssec:CK_dual_NLSM} yields the special galileon action, as described in \cref{ssec:double_copy_NLSM};
            \item[$\diamond$]  The double copy of the pure Yang--Mills action with manifest CK duality given in \cref{ssec:CK_manifest_YM} yields $\caN=0$ supergravity, as described in \cref{ssec:YM_double_copy}.
            \item[$\diamond$]  In \cref{ssec:DC_SUSY} the double copy of maximally supersymmetric Yang--Mills theory in $d=10$ space--time dimensions is given as an example, and it is shown in \cref{ssec:sen_mech} that the resulting double-copied action naturally yields a generalization of Sen's mechanism~\cite{Sen:2015nph} in the Ramond--Ramond sector;
            \item[$\diamond$] Finally, in \cref{sec:conclusion} we present our conclusions and give an outlook to future work.
        \end{itemize}
        We have intentionally kept our presentation self-contained; no further background in quantum field theory is required than that provided by a standard textbook, e.g.~\cite{Peskin:1995ev}.
        
        \section{Outline of our constructions}\label{sec:outline}
        
        Color--kinematics duality, as usually formulated, deals with factoring out color and kinematic components, i.e.~whole strings of structure constants, momenta, and polarizations, rather than dealing with just the color Lie bracket and the kinematic analogue thereof. This is natural if one is concerned purely with scattering amplitudes. If one rather works with actions, however, one can benefit from the ordering principle this provides for the amplitudes and the corresponding CK duality relations. In particular, the action of a fully CK-dual theory can be manipulated into a form resembling that of biadjoint scalar theory,
        \begin{equation}\label{eq:YM_prepared_action_intro}
            S^\text{YM}_\text{BRST,\,CK-dual}=\tfrac12\sfg_{\sfi\sfj}\bar{\sfg}_{\bar\sfa\bar\sfb}\caA^{\sfi\bar\sfa}\wave\caA^{\sfj\bar\sfb}+\tfrac1{3!}\bar\sff_{\bar\sfa\bar\sfb\bar\sfc}\sff_{\sfi\sfj\sfk}\caA^{\sfi\bar\sfa} \caA^{\sfj\bar\sfb}\caA^{\sfk\bar\sfc}~,
        \end{equation}
        where the only generalization is that the $\sff_{\sfi\sfj\sfk}$ are allowed to be differential operators which, just as the color structure constants $\bar\sff_{\bar \sfa\bar \sfb \bar \sfc}$, are totally anti-symmetric and satisfy the Jacobi identity. Put differently, the $\sff_{\sfi\sfj\sfk}$ are the structure constants of a \emph{kinematic Lie algebra}.
        
        Once one has cast the action into the form~\eqref{eq:YM_prepared_action_intro} for all fields, including potential ghosts and other Becchi--Rouet--Stora--Tyutin (BRST) fields, it automatically implies a very strong form of CK duality: the integrand corresponding to any correlator at any loop level is CK-dual. We stress already here, that the required manipulations of the action involve field redefinitions that may induce Jacobian contributions in the path integral measure which we do not include in the integrands, as we will discuss further below.
        
        From any two such parent theories, we can produce an offspring or syngamy\footnote{There is a clear analogy with the meiotic reproduction of cells with diploid chromosome sets.} by combining one type of structure constants of each parent theory into a new theory. The most prominent offspring of two parent BRST Yang--Mills actions cast in the form~\eqref{eq:YM_prepared_action_intro} is certainly the case in which the syngamy carries both kinematic structure constants from its parents, usually known as the \emph{double copy},
        \begin{equation}\label{eq:GR_prepared_action_intro}
            {S}^\text{DC}_\text{BRST,\,CK-dual}=\tfrac12 \sfg_{\sfi\sfk}\sfg_{\sfj\sfl}\caH^{\sfi\sfj}\wave\caH^{\sfk\sfl}+\tfrac1{3!} \sff_{\sfi\sfj\sfk}\sff_{\sfl\sfm\sfn}\caH^{\sfi\sfl}\caH^{\sfj\sfm}\caH^{\sfk\sfn}~.
        \end{equation}
        With some mild additional assumptions, a syngamy automatically comes with a consistent BRST operator, and in particular the theory~\eqref{eq:GR_prepared_action_intro} is indeed quantum equivalent to Einstein gravity coupled to a massless 2-form gauge field and a dilaton\footnote{To be precise, we claim that the action~\eqref{eq:YM_prepared_action_intro} can be supplemented by a suitable choice of regularization scheme and set of counterterms to produce an S-matrix equivalent to that of Yang--Mills theory and that, for any suitable choice of regularization scheme and set of counterterms for $\caN=0$ supergravity, one can construct a corresponding  set of counterterms for~\eqref{eq:GR_prepared_action_intro} such that the two theories produce equivalent S-matrices; cf.~also the more extensive discussion in \cref{ssec:field_redefinitions}.}. Although there is no supersymmetry involved, this theory is oxymoronically sometimes called $\caN=0$ supergravity,  due to the fact that it is the common bosonic sector of the type IIA and type IIB supergravity theories in ten dimensions.
        
        In this form, the double copy of the BRST symmetry and other symmetries (that do not explicitly depend on the space--time coordinates) is transparent. In particular,  the diffeomorphisms underpinning  general relativity are seen to be a consequence of the gauge symmetries  of the parent Yang-Mills theories, cf.~\cref{ssec:BRST_double_copy}.
        
        The very existence of the action~\eqref{eq:YM_prepared_action_intro} for Yang--Mills theory implies the usual scattering amplitude-theoretic CK duality  and double copy conjectures to all loop orders (up to counterterms, as discussed below): 
        \begin{enumerate}[label=(\roman*)]\itemsep-2pt
            \item construct all Feynman diagrams using~\eqref{eq:YM_prepared_action_intro} for a desired $n$-point, $L$-loop scattering amplitude, which are all cubic;
            \item compute the loop integrand corresponding to each Feynman diagram;
            \item group the diagrams according to their cubic graph topology.
        \end{enumerate}
        This yields kinematic numerators compliant with CK duality for (on-shell) tree and loop diagrams as well as (off-shell) tree and loop correlators and hence suitable for double copy. Our numerators will be formulated up to shifts in loop momenta, and the `labeling problem', cf.~e.g.~\cite{Casali:2020knc}, will turn out to be irrelevant for our considerations.
        
        The ordinary BV Yang--Mills action can  be manipulated into the semi-classically equivalent form~\eqref{eq:YM_prepared_action_intro} as follows. We assume that CK duality holds for physical on-shell tree-level scattering amplitudes and bootstrap our way to the desired form of the action using a very well worn set of tools: adding terms that sum to zero, tweaking choices of gauge, redefining fields, and introducing auxiliary fields. More explicitly: 
        \begin{enumerate}[label=(\roman*)]\itemsep-2pt
            \item Violations of tree-level CK duality for on-shell physical gluons are repaired by adding sets of terms that sum to zero due to the color Jacobi identity, but which have the effect of changing the partition of the scattering amplitude given by the Feynman diagrams. In this step, contributions to the scattering amplitude are shuffled around between different Feynman diagrams, while leaving the total scattering amplitude invariant.
            \item Violations of tree-level CK duality for scattering amplitudes with non-transverse modes of gauge fields on external legs are repaired by a choice of gauge that cancels them. By Ward identities, this also repairs tree-level CK duality violations for scattering amplitudes with Faddeev--Popov ghosts and anti-ghosts on external legs. In this step, the unphysical tree-level scattering amplitudes, being gauge-dependent, are changed by adding new kinds of Feynman vertices.
            \item\label{item:counterterms} Violations of tree-level CK duality due to off-shell momenta are repaired by a choice of possibly non-local field redefinitions (change of local coordinates in the space of fields). Violations of tree-level CK duality due to unphysically polarized fermions are also repaired by possibly non-local field redefinitions. In this step, tree-level scattering amplitudes do not change, but the loop integrands (and, thus, the corresponding counterterms) may change.
            \item Finally, the action can be tidied up to make the Feynman diagrams coincide literally with cubic graphs by suitably introducing auxiliary fields, a procedure that we call \emph{strictification}, borrowing a term from homotopical algebra. This step does not change any scattering amplitudes, loop integrands, or partition of the scattering amplitude, and  is merely cosmetic.
        \end{enumerate}

        \section{Quantum field theoretic preliminaries}\label{sec:prelim}
        
        In the following, we briefly review CK duality  as well as the double copy construction~\cite{Bern:2008qj,Bern:2010ue,Bern:2010yg}. We then continue with collecting a number of more general quantum field theoretic observations\footnote{which are ``theorems,'' satisfying the usual standard of rigor in theoretical physics} that we will use in our arguments in later sections.
        
        Note that we will always make the usual distinction between scattering amplitudes, corresponding to sums of amputated Feynman diagrams with physical states on external legs, and correlators, corresponding to sums of general Feynman diagrams with possibly off-shell momenta. Physical states are on shell, i.e.~they have light-like momenta with non-negative energies in the massless theories in which we are mostly interested, and, if they transform non-trivially under the Lorentz group, they are further restricted in their polarization or chirality.
        
        We will always have an action principle and a path integral measure in mind when speaking of correlators and scattering amplitudes. The homotopy algebraic point of view on quantum field theory unifies scattering amplitudes and actions, cf.~\cite{Doubek:2017naz,Jurco:2018sby,Borsten:2021hua}, but we refrain from using this language here bar some minor remarks.

        \subsection{Color--kinematics duality}\label{ssec:CKDuality}
        
        The perturbative tree-level scattering amplitudes of Yang--Mills theory allow for a parameterization in terms of Feynman diagrams with exclusively cubic (or trivalent) vertices, which each carry  a factor of the structure constants of the color or gauge Lie algebra as well as a kinematic factor. Furthermore, this parameterization can be chosen such that the symmetry properties of the Lie algebra structure constants match those of the kinematic factors~\cite{Bern:2008qj,Bern:2010ue}. This is known as \emph{color--kinematics (CK) duality} or \emph{Bern--Carrasco--Johansson (BCJ) duality}. 
        
        In more detail, we can parameterize the loop integrand of $n$-point $L$-loop scattering amplitudes of Yang--Mills theory as 
        \begin{equation}\label{eq:YM_amplitudes_parameterization_with_coupling}
                \scA_{n,L} =(-\rmi)^{n-3+3L}g^{n-2+2L}\sum_{i\in \Gamma_{n,L}}\int\left(\prod^L_{l=1}\frac{\rmd^dp_l}{(2\pi)^d}\right)\frac{\colfac_i\kinfac_i}{S_id_i}~,
        \end{equation}
        where $\rmi\coloneqq\sqrt{-1}$, $g$ is the Yang--Mills coupling constant, and $\Gamma_{n,L}$ is the set of cubic graphs\footnote{i.e.~graphs with vertices that all have degree three} with $n$ labeled external lines. The denominators $d_i$ are given by products of the Feynman--'t~Hooft propagators, i.e.~products of factors $\frac{1}{p^2_l}$ where $p_l$ is the momentum flowing through the internal line $l$. The $\colfac_i$ are the color numerators or color factors, consisting of contractions of the gauge Lie algebra structure constants and the Killing form according to the structure of the tree $i\in T$. The kinematic factors $\kinfac_i$ are sums of Lorentz-invariant contractions of external momenta, the Minkowski metric, and the polarization vectors labeling the external scattering states.
        
        Note that the color numerators $\colfac_i$ and propagator denominators $d_i$ are determined uniquely by the topology of the diagram $i\in \Gamma_{n,L}$ alone, while the kinematic numerators  $\kinfac_i$ are non-unique. It is this non-uniqueness that makes CK duality possible. 
        
        It will be convenient to absorb coupling constants and powers of $\rmi$ into the color and kinematic factors. Anticipating the double copy, we define\footnote{This redefinition of coupling constant is merely a bookkeeping trick. When renormalizing, the explicit coupling constants must be reintroduced by dimensional analysis.}
        \begin{equation}\label{eq:redfinitionColorKinematicFactors}
            \check\colfac_i=\left(g\sqrt{\tfrac{2}{\kappa\rmi}}\right)^{n-2+2L}\colfac_i
            \eand
            \check\kinfac_i=\left(\sqrt{\tfrac{\kappa}{2\rmi}}\right)^{n-2+2L}\kinfac_i~,
        \end{equation}
        where $\kappa\coloneqq4\sqrt{2\pi G}$ is Einstein's gravitational constant. After the rescaling, $\check\colfac_i$ and $\check\kinfac_i$ have the same mass dimensions, and we can write 
        \begin{equation}\label{eq:YM_amplitudes_parameterization}
            \scA_{n,L}=\rmi\sum_{i\in\Gamma_{n,L}}\int\left(\prod^L_{l=1}\frac{\rmd^dp_l}{(2\pi)^d\rmi}\right)\frac{\check\colfac_i\check\kinfac_i}{S_id_i}~.
        \end{equation}
        
        The parameterization~\eqref{eq:YM_amplitudes_parameterization} and the corresponding kinematic numerators are evidently not unique. CK duality is the existence of a choice\footnote{which is non-unique} such that~\cite{Bern:2008qj,Bern:2010ue}
        \begin{enumerate}[label=(\roman*)]\itemsep-2pt
            \item\label{item:antisymmetry} whenever anti-symmetry of the gauge Lie algebra structure constants or invariance of the Killing form imply that $\colfac_i+\colfac_j=0$ for two graphs $i,j\in\Gamma_{n,L}$, then $\kinfac_i+\kinfac_j=0$;
            \item\label{item:Jacobi} whenever the Jacobi identity of the gauge Lie algebra implies that $\colfac_i+\colfac_j+\colfac_k=0$ for graphs $i,j,k\in\Gamma_{n,L}$, then $\kinfac_i+\kinfac_j+\kinfac_k=0$.
        \end{enumerate}
        This has been shown to hold in the case of Yang--Mills theory tree amplitudes (i.e.\ $L=0$) from a number of different perspectives~\cite{BjerrumBohr:2010hn,BjerrumBohr:2009rd,Stieberger:2009hq,Mafra:2011kj,Broedel:2013tta,Du:2016tbc,Mizera:2019blq,Reiterer:2019dys}; see also~\cite{Bern:2019prr,Borsten:2020bgv} and references therein for other approaches and generalizations to a growing set of diverse gauge theories. There is, however, also evidence that CK duality must be generalized in order for it to extend to the loop level, see e.g.~\cite{Bern:2015ooa}.
        
        It is now an obvious conjecture that CK duality should extend to the full quantum or loop level, i.e.\ $L>0$, and much evidence supporting this conjecture has been collected, see e.g.~\cite{Bern:2009kd,Bern:2014sna,Carrasco:2011mn,Oxburgh:2012zr,Bern:2012uf,Du:2012mt, Yuan:2012rg,Boels:2013bi} as well as the reviews~\cite{Bern:2019prr,Borsten:2020bgv}.
        
        In the reparameterization~\eqref{eq:YM_amplitudes_parameterization_with_coupling}, we have ignored any considerations regarding regularization and renormalization and merely consider the loop integrands. We note that loop integrands themselves are certainly neither observable nor canonically extractable from the S-matrix; we will therefore discuss their definition in some detail later.
        
        By cutting open an $n$-point, $L$-loop Feynman diagram to a connected $(n+2L)$-point tree diagram with arbitrary operators on external legs\footnote{not to be confused with cutting apart a loop diagram into multiple tree diagrams}, we can simplify the CK duality conjecture.
        \begin{observation}\label{ob:tree_is_good_enough}
            The loop diagrams $\Gamma$ appearing in $\scA_{n,L}$ are glued together from (off-shell) tree diagrams, and if we can establish off-shell CK duality at the level of tree diagrams or correlators, then it will automatically hold at the loop level. 
        \end{observation}
        \noindent This is easily seen from the actual Feynman diagrams. If we have a triple of tree diagrams whose color numerators sum to zero due to the Jacobi identity, then their color numerators have to agree except for a factor quadratic in the structure constants of the color Lie algebra. Recall that the three color numerators fully determine  the three diagrams. It follows that the latter agree up to a common four-point subregion, where we have the three subdiagrams
        \begin{equation}\label{eq:stu_channels}
            \begin{tikzpicture}[
                scale=1,
                every node/.style={scale=1},
                baseline={([yshift=-.5ex]current bounding box.center)}
                ]
                \matrix (m) [
                matrix of nodes,
                ampersand replacement=\&,
                column sep=0.13cm,
                row sep=0.13cm
                ]{
                    {} \& {}\& {} \& {} \& {}
                    \\
                    {} \& {} \& {}\& {} \& {}
                    \\
                    {} \& {} \& {} \& {} \& {}
                    \\
                    {} \& {} \& {}\& {} \& {}
                    \\
                    {} \& {} \& {} \& {} \& {}
                    \\
                };
                \draw [gluon] (m-1-1) -- (m-3-2.center);
                \draw [gluon] (m-5-1) -- (m-3-2.center);
                \draw [gluon] (m-3-2.center) -| node[near start,below] {} (m-3-4.center);
                \draw [gluon] (m-1-5) -- (m-3-4.center);
                \draw [gluon] (m-5-5) -- (m-3-4.center);
                \node [circle,draw,dashed,minimum size=1.9cm] (c) at (0,0){};
                \foreach \x in {(m-3-2), (m-3-4)}{
                    \fill \x circle[radius=2pt];
                }
            \end{tikzpicture}
            ~~~~~~
            \begin{tikzpicture}[
                scale=1,
                every node/.style={scale=1},
                baseline={([yshift=-.5ex]current bounding box.center)}
                ]
                \matrix (m) [
                matrix of nodes,
                ampersand replacement=\&,
                column sep=0.13cm,
                row sep=0.13cm
                ]{
                    {} \& {} \& {} \& {} \& {}
                    \\
                    {} \& {} \& {} \& {} \& {}
                    \\
                    {} \& {} \& {} \& {} \& {}
                    \\
                    {} \& {} \& {} \& {} \& {}
                    \\
                    {} \& {} \& {} \& {} \& {}
                    \\
                };
                \draw [gluon] (m-1-1) -- (m-2-3.center);
                \draw [gluon] (m-5-1) -- (m-4-3.center);
                \draw [gluon] (m-2-3.center) -| node[near end,left] {} (m-4-3.center);
                \draw [gluon] (m-1-5) -- (m-2-3.center);
                \draw [gluon] (m-5-5) -- (m-4-3.center);
                \node [circle,draw,dashed,minimum size=1.9cm] (c) at (0,0){};
                \foreach \x in {(m-2-3), (m-4-3)}{
                    \fill \x circle[radius=2pt];
                }
            \end{tikzpicture}
            ~~~~~~
            \begin{tikzpicture}[
                scale=1,
                every node/.style={scale=1},
                baseline={([yshift=-.5ex]current bounding box.center)}
                ]
                \matrix (m) [
                matrix of nodes,
                ampersand replacement=\&,
                column sep=0.13cm,
                row sep=0.13cm
                ]{
                    {} \& {}\& {} \& {} \& {}
                    \\
                    {} \& {} \& {}\& {} \& {}
                    \\
                    {} \& {} \& {}\& {} \& {}
                    \\
                    {} \& {} \& {}\& {} \& {}
                    \\
                    {} \& {}\& {} \& {} \& {}
                    \\
                };
                \draw [gluon] (m-1-1) -- (m-3-4.center);
                \draw [gluon] (m-5-1) -- (m-3-2.center);
                \draw [gluon] (m-3-2.center) -| node[near start,below] {} (m-3-4.center);
                \draw [gluon] (m-1-5) -- (m-3-2.center);
                \draw [gluon] (m-5-5) -- (m-3-4.center);
                \node [circle,draw,dashed,minimum size=1.9cm] (c) at (0,0){};
                \foreach \x in {(m-3-2),(m-3-4)}{
                    \fill \x circle[radius=2pt];
                }
            \end{tikzpicture}
        \end{equation}
        If we now glue these together in the same way to form loop diagrams that differ again only in these three subdiagrams, then the corresponding sum of loop diagrams vanishes as well. Since all loop diagrams are glued together from tree diagrams, the observation is evident (up to the above mentioned labeling problem, which does not affect our discussion).
        
        \Cref{ob:tree_is_good_enough} allows us to lift the discussion to the level of Lagrangians, as we had done previously in the context of what we called the \emph{BRST--Lagrangian double copy}~\cite{Borsten:2020zgj,Borsten:2021hua}. More precisely, our formulation of CK duality will be the following: there is a renormalizable Lagrangian containing a Yang--Mills gauge potential as well as other fields, such that the scattering amplitudes between asymptotic states labeling gauge potentials are those of Yang--Mills theory and such that the implied Feynman diagram expansion manifests CK duality at the full quantum level. Note that this Feynman diagram expansion neglects possible counterterms. These counterterms may break  CK duality, which may be anomalous in this sense.
        
        CK duality has a number of generalizations, such as the flavor--kinematics duality exhibited by the non-linear sigma model discussed in \cref{sec:NLSM}; in the following, we will use the term \emph{color--kinematics duality} to capture all of these.
        
        \subsection{Master numerators at the tree level}\label{ssec:primaries}
        
        For later purposes, let us develop the decomposition of the tree-level scattering amplitudes in more detail, cf.~also the longer discussion in~\cite{Borsten:2021hua}. Given a general theory with CK-dual amplitudes, we can trivially write the $n$-point tree amplitude in the form 
        \begin{equation}
            \scA_{n,0}=\bmc^\sfT\bmD\bmn
            \ewith
            \bmD_{ij}=\frac{\delta_{ij}}{d_j}~,
        \end{equation}
        where we arranged the color and kinematic numerators $\sfc_i$ and $\sfn_i$ into column vectors $\bmc$ and $\bmn$. We note that there are $(2n-5)!!$ different cubic trees at $n$ points. Because of the color and kinematic Jacobi identities, however, only $(n-2)!$ of these are linearly independent. We can choose a basis $\bmc_{\rm m}$ of the color numerators, which we call \emph{master numerators}, and we denote the corresponding kinematic numerators by $\bmn_{\rm m}$. Thus,
        \begin{equation}
            \bmc=\bmJ\bmc_{\rm m}
            \eand
            \bmn=\bmJ\bmn_{\rm m}~,
        \end{equation}
        where $\bmJ$ is a $(2n-5)!!\times(n-2)!$-dimensional matrix capturing the linear dependence of the various numerators.
        
        At four points, for example, there are $(8-5)!!=3$ distinct cubic trees, corresponding to the $s$-, $t$-, and $u$-channels, as depicted in~\eqref{eq:stu_channels}. The color numerators of these are linearly dependent, as $\sfc_s=\sfc_t+\sfc_u$. We can choose a basis of $(4-2)!=2$ master numerators, for example $\sfc_t$ and $\sfc_u$. In this basis, the matrix $\bmJ$ reads as
        \begin{equation}\label{eq:nJn}
            \bmJ=
            \begin{pmatrix}
                1 & 1
                \\
                1 & 0
                \\ 
                0 & 1
            \end{pmatrix}~.
        \end{equation}
        
        In order to use~\cref{ob:tree_is_good_enough} in our later discussion, we need to continue the tree-level scattering amplitudes $\scA_{n,0}$ to tree-level \emph{amputated correlators}, which have external legs of arbitrary momentum, polarization (or chirality), and ghost number. The choice of relaxation to include off-shell and unphysical external states is essentially arbitrary, but it is helpful to think of the concrete realization  given by simply computing the amplitude with the amputated Feynman diagrams of the theory, where one does not impose any constraints on the external states and momenta. This yields the  amputated correlators $\hat\scA_{n,0}$ canonically associated to the underlying action. 
        
        Another example, which will be crucial for our discussion later, is the following. Consider a  generic   continuation of the  master numerators  {$\bmn_{\rm m}$} beyond the asymptotic states, allowing off-shell momenta, unphysical polarization/chirality, and Faddeev--Popov ghost states. We will denote the thus extended master numerators by $\hat\bmn_{\rm m}$, and   choose the complete set  of numerators to be given  by 
        \begin{equation}
            \hat\bmn=\bmJ\hat\bmn_{\rm m}~.
        \end{equation}
        If we contract these extended numerators with the color factors, we obtain the amputated correlators satisfying CK duality
        \begin{equation}
            \hat\scA'_{n,0}=\left[\bmc^{(n)\sfT}\bmD^{(n)}\hat\bmn^{(n)}\right]_\sigma~,
        \end{equation}
        where $\sigma$ denotes symmetrization over all external legs. 
        
        \subsection{Double copy of scattering amplitudes}
        
        Color--kinematics duality is the crucial ingredient in the amplitudes realization of the fact that $\text{gravity}=\text{gauge}\otimes\text{gauge}$, known as the double copy. Consider again the $L$-loop, $n$-gluons Yang--Mills scattering amplitude in the parameterization~\eqref{eq:YM_amplitudes_parameterization}. If CK duality holds, then we can replace the color factors $\colfac_i$ in~\eqref{eq:YM_amplitudes_parameterization} with a second copy of the kinematic factor $\kinfac_i$ to obtain the $\caN=0$ supergravity scattering amplitude~\cite{Bern:2008qj,Bern:2010ue,Bern:2010yg},
        \begin{equation}\label{eq:amp_double_copy} 
            \begin{aligned}
                \scH_{n,L}&=(-\rmi)^{n-3+3L}\left(\frac{\kappa}{2}\right)^{n-2+2L}\sum_{i\in\Gamma_{n,L}}\int\left(\prod^{L}_{l=1}\frac{\rmd^dp_l}{(2\pi)^d}\right)\frac{\kinfac_i\kinfac_i}{S_id_i}
                \\
                &=\rmi\sum_{i\in\Gamma_{n,L}}\int\left(\prod^{L}_{l=1}\frac{\rmd^dp_l}{(2\pi)^d\rmi}\right)\frac{\check\kinfac_i\check\kinfac_i}{S_id_i}~,
            \end{aligned}
        \end{equation}
        where $\check\kinfac_i$ is defined in~\eqref{eq:redfinitionColorKinematicFactors}. At the tree level, this is equivalent to the field theory limit of the famous Kawai--Lewellen--Tye (KLT) relations~\cite{Kawai:1985xq}, which relate the scattering amplitudes of open string theory (whose field theory limit contains Yang--Mills theory) to those of closed string theory (whose field theory limit contains gravity). It has recently been shown in~\cite{Chi:2021mio} that by bootstrapping the KLT relations~\cite{Kawai:1985xq} the tree-level double copy can be generalized, e.g.~to include higher derivative operators. However,~\eqref{eq:amp_double_copy} is an all-loop statement: one of the most startling consequences of full CK duality is the validity of the double copy prescription to all perturbative orders. This has been shown using the unitarity method in~\cite{Bern:2010yg}. Essentially, CK duality of loop diagrams implies CK duality of the sub-diagrams of the unitarity cuts, allowing one to reduce the validity of loop double copy  to the validity of the tree double copy.
        
        It is natural to think that the presented on-shell, scattering amplitudes-based picture of the paradigm $\text{gravity}=\text{gauge}\otimes\text{gauge}$ can be lifted to a field-theoretic or  action-based description~\cite{Bern:1999ji,Bern:2010yg,Hohm:2011dz,Borsten:2013bp,Anastasiou:2014qba,Cardoso:2016ngt,LopesCardoso:2018xes,Borsten:2015pla,Cheung:2016say,Cheung:2016prv,Luna:2016hge, Borsten:2017jpt, Anastasiou:2018rdx,Ferrero:2020vww,Borsten:2020xbt,Beneke:2021ilf}. It was argued in~\cite{Anastasiou:2014qba,Borsten:2017jpt,Anastasiou:2018rdx} that the complete BRST complex --- i.e.\ the gluon field $A_\mu$, the Nakanishi--Lautrup field $b$ for gauge-fixing, and the corresponding Faddeev--Popov ghosts $c, \bar c$ --- should be considered when formulating a field-theoretic double copy; using this, it was shown explicitly that the BRST Einstein--Hilbert action to cubic order follows~\cite{Borsten:2020xbt}. 
        
        In~\cite{Borsten:2020zgj}, we proposed the realization of a double-copied Yang--Mills action and BRST  operator valid to all orders in perturbation theory, a construction naturally interpreted in terms of factorizations of homotopy algebras~\cite{Borsten:2021hua}. We refer to this general construction as \emph{BRST--Lagrangian double copy}. Following this off-shell approach, and generalizing tree-level CK duality to the BRST-extended field space, we were able to show that the double-copied theory was a perturbative description of $\caN=0$ supergravity to all loop orders, bypassing the need of perfect CK duality. However, some technical difficulties of that proof came from the possibility of a violation of perfect off-shell CK duality. Having established CK duality for the BRST action  \cref{sec:loop_CK_duality_YM}, the validity of loop-level double copy becomes a manifest property of the BRST-Lagrangian  double copy, as discussed  in \cref{sec:double_copy}. In this regard, the shift in interpretation of CK duality to a property of the BRST action renders the consistency of the double copy theory plain to see.
        
        Just as in the case of CK duality, there is an evident generalization of the double copy to the case of general theories with some extended notion of CK duality, which we call syngamies; we will discuss these in \cref{ssec:BRST_double_copy}.
        
        \subsection{Equivalence of field theories}\label{ssec:equivalence}
        
        As is already clear from the discussion in \cref{ssec:CKDuality}, any lift of CK duality from tree-level diagrams to loop-level integrands will imply a reparameterization of the field theory scattering amplitudes, and we have to identify reparameterizations that link physically equivalent field theories. Before this, let us set up our discussion by introducing some precise nomenclature.
        
        First of all, we restrict our discussion to the perturbative situation. That is, we restrict our attention to perturbatively computed scattering amplitudes (or, at most, correlators that can be computed perturbatively by Feynman diagrams) at zero temperature, ignoring non-perturbative issues such as Wilson loops, black holes, confinement, phases, etc.; our fields take values in vector spaces, and all expressions, such as classical solutions, scattering amplitudes and correlators, are formal power series in the coupling constants. Homotopy algebraically, our classical field theories correspond to cyclic $L_\infty$-algebras, and we can use the homological perturbation lemma to compute the scattering amplitudes, cf.~e.g.~\cite{Borsten:2021hua,Jurco:2018sby}. This perspective suffices for our purposes, as we are merely interested in $n$-point, $L$-loop scattering amplitudes for finite $n$ and $L$ in this paper. It allows us to truncate accordingly any perturbative power series in the fields, and, e.g., to neglect higher-order terms in the action. We can thus ignore many questions regarding convergence.
        
        There are three notions of equivalence commonly found in the literature. First, if two actions have isomorphic solution spaces, then they are often called \emph{classically equivalent}. While this notion of equivalence can be useful for the generation or reparameterization of classical solutions, it is fairly coarse. For example, the solution spaces of both the free Klein--Gordon equation and the equation with polynomial potential over Minkowski space are isomorphic to boundary data on a Cauchy surface and therefore isomorphic.
        
        Second, there is the notion of \emph{semi-classical equivalence}, where we have a similarity transformation between the tree-level S-matrices of two theories. In other words, the tree-level scattering amplitudes of the theories agree up to a reparameterization of the external states. From the discussion in~\cite{Boulware:1968zz}, it follows that this is equivalent to an isomorphism between the classical solution spaces of both theories with arbitrary non-vanishing external sources. In the homotopy algebraic formulation of field theories, cf.~e.g.~\cite{Jurco:2018sby,Borsten:2021hua}, semi-classical equivalence implies that both theories have cyclic $L_\infty$-algebras that are related by a quasi-isomorphism. Mathematically, this is a very clear and natural notion of equivalence. Semi-classical equivalence implies classical equivalence, at least perturbatively. We can further distinguish between \emph{strict} semi-classical equivalence, where the field spaces of both theories are isomorphic, and \emph{weak} semi-classical equivalence, where the field spaces are not isomorphic, but auxiliary fields may have been introduced or integrated out.
        
        For our discussion, a suitable notion of \emph{quantum equivalence} will be important. By taking the limit $\hbar\rightarrow 0$, we can obtain a unique classical field theory from a perturbative quantum field theory; however, the reverse statement fails to hold. Quantization is not a unique process, but involves several choices such as the space of admissible fluctuations and its path integral measure. To render this measure well-defined, we usually have to choose additionally a regularization and renormalization procedure. Given a strict semi-classical equivalence between two theories, it is reasonably clear that a choice of quantization can be translated from one theory to the other, leading to S-matrices related by a similarity transform. For weakly semi-classically equivalent theories, one can always introduce trivial pairs, i.e.~physically irrelevant extra fields, such that the equivalence is enhanced to a strict semi-classical equivalence. This is essentially the decomposition theorem for $L_\infty$-algebras, cf.~e.g.~\cite{Jurco:2018sby,Borsten:2021hua}. Finally, we sometimes want to regard our S-matrix as the restriction of the S-matrix of a larger theory, in which we regard certain fields merely as auxiliary and not appearing on external legs of Feynman diagrams. In such a case, we must ensure that the restriction is compatible with unitarity and all the symmetries we want to preserve. This may be familiar from the BRST gauge-fixing procedure, where the unitarity of the full S-matrix together with the BRST symmetry guarantee that the restricted S-matrix is indeed unitary. 
        
        We thus conclude that for our purposes, the relevant notion of equivalence is always (perturbative) semi-classical equivalence, either in its strict or its weak form, because semi-classical theories admit quantizations leading to S-matrices related by a similarity transformation. 
        
        \subsection{Field redefinitions and the S-matrix}\label{ssec:field_redefinitions}
        
        Perturbatively, an S-matrix $\scS$ is a formal power series in $\hbar$,
        \begin{equation}\label{eq:S-matrix-expansion}cc
            \scS=\scS_0+\hbar\scS_1+\hbar^2\scS_2+\cdots~,
        \end{equation}
        and we recover a classical or tree-level S-matrix by taking the limit $\hbar\to0$ for the classes of theories that we consider.\footnote{In general, there are subtleties when comparing the \(\hbar\) expansion (whose leading order is the classical limit) with the loop counting (whose leading order is the tree level), cf.\ e.g.\ \cite{Donoghue:9310024, Brodsky:1009.2313, Holstein:0405239} and e.g.~\cite{Kosower:2018adc}. The class of theories that we consider, however, are massless theories where every \(n\)-ary vertex comes with \((n-2)\)th power of a coupling constant. Under these assumptions, in the perturbative regime (hence ignoring bound states, confinement, etc.)\ on a flat trivial classical background, these subtleties evaporate.
        } If two field theories are semi-classically equivalent, then their tree-level S-matrices are linked by a similarity transformation. Such a similarity transformation amounts to a coordinate change on the space of external states. If the theories are strictly semi-classically equivalent, and their full field spaces are isomorphic, then this coordinate change induces a (perturbatively invertible) field redefinition.
        
        At the quantum level, the infinities arising from loop integrals require regularization and renormalization, which make the situation more involved. As remarked above, a classical field theory corresponds to a family of quantum field theories, which differ in terms that are at least first order in $\hbar$. We will denote any such terms as \emph{counterterms}, irrespective of their roles in the quantization of the field theory. Clearly, some choices of counterterms will lead to the same S-matrix, while other choices will yield discrepancies proportional to positive powers of $\hbar$. 
        
        Generally, counterterms can restore or break unitarity (often due to a broken gauge symmetry) as well as desirable symmetries such as global symmetry or supersymmetry. If unitarity is broken, we usually regard the theory and the resulting S-matrix as pathological. We note that as usual in the context of CK duality and the double copy, we will merely be interested in the integrands of loop diagrams. Up to renormalizability and consistency of our field theory, we can therefore safely ignore all questions about regularization and renormalization. As we will explain in \cref{ssec:absence_unitarity}, ignoring the counterterms is harmless for most purposes.
        
        Despite these complications, the equivalence theorem of quantum field theory still allows us to consider very general field redefinitions. Consider an action functional $S[\phi]$ depending on some general field $\phi^I$ with $I$ a DeWitt index.\footnote{Recall that DeWitt indices bundle the space--time coordinate $x$ (or, after a Fourier transform, momentum) and all other required  indices for field species, Poincar\'e representations, global and local  symmetry representations, etc., as one might in a condensed-matter context.} The generating functional reads as 
        \begin{equation}\label{eq:gen_func_1}
            Z[J]=\int\caD\phi~\mu(\phi)~\rme^{\frac{\rmi}{\hbar}(S[\phi]+J_I\phi^I)}~,
        \end{equation}
        where $\mu(\phi)$ is the measure arising from the Hamiltonian form of the path integral, cf.~\cite[Section 9.3]{Weinberg:1995mt}. In the case of theories with canonical kinematic terms, the measure $\mu(\phi)$ is simply a constant that is absorbed in the normalization of $Z[0]$, and thus it can be dropped. Furthermore, after a field redefinition $\phi=F(\tilde \phi)$, we obtain the action $S[\phi]=S[F(\tilde \phi)]$ with a corresponding generating functional
        \begin{subequations}\label{eq:gen_func_2}
            \begin{equation}
                \tilde Z[J]=\int\caD\tilde\phi~\tilde\mu(\tilde\phi)~\rme^{\frac{\rmi}{\hbar}(S[F(\tilde \phi)]+J_I\tilde\phi^I)}
            \end{equation}
            with
            \begin{equation}
                \tilde\mu(\tilde\phi)=\mu(\phi)\det\left(\delder[F(\tilde\phi)]{\phi}\right)~.
            \end{equation}
        \end{subequations}
        We note that this is almost the generating functional $Z[J]$ after the coordinate change $\phi^I=F^I(\tilde\phi)$ except for a discrepancy in the source terms: we have $J_I\tilde\phi^I$ in~\eqref{eq:gen_func_2}, but the coordinate change on~\eqref{eq:gen_func_1} produces $J_IF^I(\tilde\phi)$. As far as the S-matrix is concerned, however, this discrepancy is irrelevant because we have to apply the Lehmann--Symanzik--Zimmermann reduction formula~\cite{Lehmann:1954rq}, and both $\tilde\phi^I$ and $F(\tilde\phi^I)$ are valid interpolating fields\footnote{i.e.~interpolating between the free and the interacting regimes}, cf.~e.g.~the discussion in~\cite[Section 5]{Srednicki:2007qs}. More explicitly, the S-matrix is obtained as a time-ordered exponential of derivatives of $Z[J]$ with respect to $J$ at $J=0$, cf.~e.g.~\cite[Section 9.2]{Itzykson:1980rh}. Since $Z[0]=\tilde Z[0]$, it follows that the S-matrix only changes up to a possible wave function renormalization. The correlators, on the other hand, are not protected by this argument.\footnote{We note that there is a more general approach~\cite{Vilkovisky:1984st} towards the coupling $J_I\phi^I$ that fixes also this difference.}
        
        Let us therefore consider the effect of field redefinitions on correlators more carefully. We will be interested in field redefinitions of the form
        \begin{equation}\label{eq:field_redefinition_form}
            F^I(\tilde\phi)=\tilde\phi^I+g\,G^I(\tilde\phi)~,
        \end{equation}
        where $G^I$ is analytic in the coupling constant $g$, such that all positive powers of $g$ are treated as interaction terms in perturbation theory. 
        
        For local $G^I$ that depend analytically on the field $\phi^I$ and its derivatives, the situation is uncontroversial. One can use the usual Faddeev--Popov trick and exponentiate the determinant arising from the field redefinition as a functional integral over ghost fields,
        \begin{equation}\label{eq:functional_determinant_to_ghosts}
            \det\left(\unit+g\frac{\delta G}{\delta\phi}\right)=\int\caD\bar c\,\caD c\,\rme^{\frac{\rmi}{\hbar}\left(\bar c_Ic^I+g\bar c_I\frac{\delta G^I}{\delta\phi^J}c^J\right)}~.
        \end{equation}
        The propagator between the ghosts is the identity, and the interaction vertex is a polynomial in the momenta by locality. The ghost loops then vanish in dimensional regularization~\cite{tHooft:1973wag}, see also the discussion in~\cite[Section 2]{Criado:2018sdb}.
        
        For our discussion, however, we need field redefinitions of the form~\eqref{eq:field_redefinition_form} with non-local functionals $G^I$. Specifically, $G^I$ may contain factors 
        \begin{subequations}
            \begin{equation}
                \frac{1}{\wave}\caO(x)=\int\rmd^dy\,\Delta(x-y)\caO(y)
            \end{equation}
            with
            \begin{equation}
                \wave\Delta(x-y)=\delta^{(d)}(x-y)~,
            \end{equation}
        \end{subequations}
        where $\caO(x)$ is an expression local in the fields. We can still re-exponentiate the determinant arising from the field redefinition as in~\eqref{eq:functional_determinant_to_ghosts}, but now we expect in general a non-vanishing contribution. This contribution will be proportional to a positive power of $\hbar$, which clearly identifies it as a non-local counterterm that will have to be added to the naively field-redefined action.
        
        The non-local nature of this counterterm is certainly unusual: it is incompatible with multiplicative renormalization, where we expect counterterms to be of the same forms as the terms in the bare Lagrangian. In the additive or Bogoliubov--Parasiuk--Hepp--Zimmermann renormalization scheme, however, this is not an issue. Moreover, such non-local terms are familiar from quantizing gauge theories in axial gauges such as the light-cone gauge. The problem with non-local terms is that they may destroy locality and unitarity (by introducing unphysical poles) of the S-matrix. In the case at hand, however, we already know that the quantum field theory is local, unitary, gauge invariant, and renormalizable in a different parameterization. These properties then translate in an evident way via the field redefinition, provided we add the non-local counterterm arising from the additional ghost loops.\footnote{An interesting example in this context for the opposite phenomenon is the curing of the gauge anomaly in axial quantum electrodynamics by a non-local counterterm~\cite{Adam:1997gj}. This counterterm then violates unitarity.} Comparing the renormalization before and after the field redefinition, we have the same set of counterterms related by a field redefinition up to the additional non-local counterterms arising from the Jacobian determinants.
        
        One may worry that the non-local field redefinition, while preserving renormalizability and unitary, somehow does not lead to a quantum equivalent theory. In~\cite[Section 10.4]{tHooft:1973wag}, it is shown diagrammatically\footnote{that is, purely combinatorially, without assuming the existence of a well-defined path integral} that the insertion of the ghost terms cancels the additional contributions arising from the field redefinition and that quantum equivalence indeed persists.
        \begin{observation}\label{ob:non-local-field_redefinitions}
            Non-local field redefinitions lead to physically equivalent field theories, which will require the addition of non-local counterterms to the renormalized action. These counterterms, however, neither lead to anomalies nor affect the renormalizability of the theory.
        \end{observation}
        \noindent 
        This observation is essentially self-evident from our discussion in \cref{ssec:equivalence}: field theories that are related by general field redefinitions are certainly semi-classically equivalent, and therefore they admit a choice of quantization such that their S-matrices agree.
        
        Besides field redefinitions, we could also introduce or integrate out auxiliary fields in order to produce equivalent actions. A specialization of this transformation will be the subject of the following section.
        
        \subsection{Strictification}\label{ssec:strictification}
        
        A standard technique in the theory of scattering amplitudes arises from the following observation.
        \begin{observation}\label{ob:strictification}
            Any field theory can be reformulated such that all interaction vertices are cubic.
        \end{observation}
        \noindent
        Abstractly, this is a corollary, cf.~\cite{Borsten:2021hua} for details, of the strictification theorem for homotopy algebras~\cite{igor1995,Berger:0512576}. More concretely, we can blow up interaction vertices by inserting auxiliary fields, as we shall show in the following. See also~\cite{Bern:2010yg} for an example in the context of the double copy.
        
        In many situations, fields carry internal labels that are connected at interaction vertices by sets of cubic structure constants. This is certainly the case for Yang--Mills theory coupled to matter, where the structure constants are those of a Lie algebra or those of a Lie algebra action on a particular representation. This internal structure then automatically induces a preferred way of blowing up a vertex. For example,
        \begin{subequations}
            \begin{equation}
                \sff_{ab}{}^f\sff_{fcg}\sff_{de}{}^gA^aB^bC^cD^dE^e
            \end{equation}
            for some structure constants $\sff_{ab}{}^c$ and fields $A,\ldots,E$ is pictorially represented as
            \begin{equation}
                \begin{tikzpicture}[
                    scale=1,
                    every node/.style={scale=1},
                    baseline={([yshift=-.5ex]current bounding box.center)}
                    ]
                    \matrix (m) [
                    matrix of nodes,
                    ampersand replacement=\&,
                    column sep=0.13cm,
                    row sep=0.13cm
                    ]{
                        $A$ \& {} \& {} \& {} \& {} \& {} \& {} \& {} \& $E$
                        \\
                        {} \& {} \& {} \& {} \& {} \& {} \& {} \& {} \& {}
                        \\
                        {} \& {} \& {} \& {} \& {} \& {} \& {} \& {} \& {}
                        \\
                        {} \& {} \& {} \& {} \& {} \& {} \& {} \& {} \& {}
                        \\
                        $B$ \& {} \& {} \& {} \& $C$ \& {} \& {} \& {} \& $D$
                        \\
                    };
                    \draw (m-1-1) -- (m-3-3.center);
                    \draw (m-5-1) -- (m-3-3.center);
                    \draw (m-3-3.center) -- (m-3-7.center);
                    \draw (m-3-5.center) -- (m-5-5);
                    \draw (m-3-7.center) -- (m-1-9);
                    \draw (m-3-7.center) -- (m-5-9);
                    \foreach \x in {(m-3-3), (m-3-5), (m-3-7)}{
                        \fill \x circle[radius=2pt];
                    }
                \end{tikzpicture}
            \end{equation}
        \end{subequations}
        For each internal line $i$ in each blown up interaction vertex, we then introduce a pair $(G_i,\bar G_i)$ of auxiliary fields, resulting in a Lagrangian with exclusively cubic interaction vertices. In the above case, for example, we would use
        \begin{equation}
                \bar G_{1a} G^a_1+\bar G_{2a}G_2^a+\sff_{ab}{}^fA^aB^b\bar G_{1f}
                +\sff_{fcg}G_1^fC^cG_2^g+\sff_{de}{}^g\bar G_{2g}D^dE^e~.
        \end{equation}
        Because such a strictification is undone by integrating out the auxiliary fields that appear purely algebraically and at most quadratically, strictification produces a quantum equivalent field theory in which all scattering amplitudes agree.
        
        For our purposes, we will also need to strictify non-local actions of the form
        \begin{equation}\label{eq:typically_non-local_term}
            E_1^M\frac{1}{\wave}E^2_M~,
        \end{equation}
        where $E_1$ and $E_2$ are polynomials in fields and their derivatives and $M$ is some multi-index. This can be done by introducing auxiliary fields as follows:
        \begin{equation}
            -G^M\wave\bar G_M+G^ME^2_M+E_1^M\bar G_M~.
        \end{equation}
        Because the strictified theory is weakly semi-classically equivalent to the original theory, quantum equivalence of both theories is immediate. Explicitly, it is clear from the perturbative expansion in terms of Feynman diagrams that there are no new loops formed by auxiliary fields; this is discussed in some more detail in~\cite{Borsten:2021hua}.
        
        When strictifying a full Batalin--Vilkovisky (BV) action, one must take into account the gauge transformations of the auxiliary fields. These are deduced from their on-shell gauge transformations, and the results are expressions that are cubic and higher in the fields. As a consequence, the BV action cannot be fully strictified for the auxiliary fields by simply blowing up vertices. Fortunately, this is also not needed for our argument. We note, however, that one can use the full machinery of homotopy algebras, leading to an enlargement of the field space to the loops on this field space. For further details, see~\cite{Borsten:2021hua}.
        
        We can thus rewrite any field theory in the form 
        \begin{equation}\label{eq:simple_cubic_action}
            S[\Phi]=\tfrac12\sfG_{IJ}\Phi^I \Phi^J+\tfrac{1}{3!}\sfF_{IJK}\Phi^I\Phi^J\Phi^K~,
        \end{equation}
        where $I,J$ are DeWitt indices encoding all field labels, including particle species and position. The structure constants $\sfG_{IJ}$ and $\sfF_{IJK}$ encode the free action and the interaction terms, respectively. As usual, we perform all summations and integrations over repeated indices.
        
        \subsection{Actions with manifest color--kinematics duality}
        
        First, let us make the following observation:
        \begin{observation}\label{ob:TW-terms}
            Given a field theory with CK-dual tree-level scattering amplitudes, we can always find a corresponding action whose Feynman diagrams yield a CK-dual parameterization of the tree-level scattering amplitudes.
        \end{observation}
        \noindent 
        For Yang--Mills theory, this observation was already made in~\cite{Bern:2010ue}, and in~\cite{Tolotti:2013caa} a general algorithm for the construction of the relevant action (with an implicit reparameterization of $n$-ary interaction vertices in terms of cubic vertices) was given. In the case of a generic CK-dual theory, we can construct the desired action using the following straightforward algorithm:
        \begin{enumerate}[label=(\roman*)]\itemsep-2pt
            \item Consider a perturbative quantum field theory given by an action principle $S$ and tree-level scattering amplitudes that permit a CK-dual parameterization; choose one such parameterization. The fields will have some internal labels (e.g.~color or flavor labels), and there will be a set of structure constants in interaction terms. We explicitly allow terms in $S$ that vanish once algebraic relations for these structure constants (e.g.~anti-symmetry and the Jacobi identity of the Lie algebra structure constants as well as symmetry and invariance of the metric) are taken into account. Set $S_{3}=S$, where $S_{n}$ denotes the action with CK duality  manifest up to $n$ points. Now proceed with the algorithm starting at   $n=4$. 
            \item\label{item:compareFeynmanDiagrams} The $n$-point tree Feynman diagrams produced by $S_{n-1}$ naturally partition the $n$-point tree scattering amplitude into pieces corresponding to different pole structures produced by propagators corresponding to internal edges. Compare these partitions to the CK-dual parameterization of the $n$-point scattering amplitudes. The sum of the differences must vanish, as the tree-level scattering amplitudes must agree.
            \item Add the (vanishing) sum of the differences to the action $S_{n-1}$ as an $n$-point vertex, producing the action $S_n$. The $m$-point tree Feynman diagrams produced by $S_n$ then agree with those in the CK-dual parameterization of the $m$-point scattering amplitudes for all $m\leq n$.
            \item If we have reached the maximal order that is of relevance for the scattering amplitudes we are interested in, halt. Otherwise, increment $n$, and go back to~\ref{item:compareFeynmanDiagrams}. 
        \end{enumerate}
        We note that the form of the action $S_n$ automatically comes with a preferred choice of strictification, which we can readily perform. The Feynman diagrams of the strictified action are then literally the cubic trees of the CK-dual parameterization of the scattering amplitudes with each cubic vertex and interaction vertex. 
        
        We will only ever be interested in $n$-point, $L$-loop scattering amplitudes for finite $n$ and $L$, allowing us to ignore interaction vertices beyond order $n+2L$. For our purposes, the above algorithm thus completes after finitely many steps. 
        
        Using the above algorithm, we can specialize our \cref{ob:TW-terms} a bit further:
        \begin{observation}\label{ob:iterative_TW_terms}
            Given a field theory with a CK-dual parameterization of all its tree-level scattering amplitudes and an action whose Feynman diagrams produce this parameterization up to $n$ points, we can add vanishing terms to this action such that the Feynman diagrams produce this parameterization up to $n+1$ points.
        \end{observation}
        We will present the lowest order terms arising from our algorithm for the case of Yang--Mills theory in \cref{ssec:CK_manifest_YM}.
        
        \subsection{Loop integrands}
        
        As mentioned in the discussion of CK duality in \cref{ssec:CKDuality}, the notion of loop integrand is unphysical and not uniquely defined. Let us therefore briefly consider how they arise and what the implications of quantum equivalence are. 
        
        Given a particular action, we can evidently read off the corresponding Feynman rules and produce the expressions corresponding to Feynman diagrams involving loops. Working in momentum space, we will encounter a momentum integral for each individual loop, and the loop integrands will be rational functions with poles at the locations of propagators.
        
        Performing a field redefinition, however, the loop integrands will change. The question
        \begin{question}
            Do the loop integrands of a theory satisfy CK duality?
        \end{question}
        \noindent
        should therefore be replaced by the refinement
        \begin{question}\label{quest:CKDualityLoopIntegrands}
            Is there a formulation of the theory such that the resulting loop integrands satisfy CK duality?
        \end{question}
        \noindent
        The need for this refinement should not be surprising and is similar to the fact that there are many ways of strictifying, i.e.~rewriting the action in terms of exclusively cubic interaction vertices, even though most of these will not produce manifestly CK-dual parameterizations of the tree-level scattering amplitudes.
        
        Concretely, we note again that the fundamental physical object of a perturbative field theory is the S-matrix $\scS$. We assume that we are given the expansion of the S-matrix in powers of $\hbar$ as in~\eqref{eq:S-matrix-expansion} so that we can extract a classical limit, implying a family of actions belonging to semi-classically equivalent field theories. From our discussion in \cref{ssec:equivalence} it is then clear that, for these actions, there are choices of quantization that recover the full, original S-matrix $\scS$. We can thus reword \cref{quest:CKDualityLoopIntegrands} more precisely:
        \begin{question}
            Given the S-matrix $\scS$ of a theory, is there a renormalizable classical action that reproduces the classical limit $\scS_0$, possibly as a restriction of its full tree-level S-matrix, and that allows for a quantization such that the resulting loop-level scattering amplitude integrands satisfy CK duality?
        \end{question}
        \noindent
        Starting from a concrete action that leads to the S-matrix $\scS$, this question merely amounts to a parameterization of the field space in the vicinity of the vacuum which we use for quantization. This parameterization or coordinatization is evidently not unique --- different parameterizations are linked by field redefinitions or coordinate changes --- and we end up with the following further specialization:
        \begin{question}
            Given an action functional $S$, is there a weakly semi-classically equivalent action\footnote{i.e.~an action arising from $S$ by introducing auxiliary fields and performing field redefinitions} such that the resulting Feynman diagram expansion produces CK-dual loop integrands?
        \end{question}
        \noindent
        It is this last question that we will answer affirmatively for a number of field theories.
        
        \subsection{Potential absence of unitarity in loop integrands}\label{ssec:absence_unitarity}
        
        As stated in \cref{sec:outline}, we will use a sequence of potentially non-local field redefinitions in order to rephrase our BRST action in a manifestly CK-dual form. These field redefinitions may introduce Jacobians into the path integral, which we will drop. This may break BRST invariance of the path integral measure when dealing with gauge theories. More fundamentally, unitarity may also be broken. The question is now whether this is a problem, and the answer depends largely on whether one wants to study CK duality in its own right or as a means for the double copy.
        
        \paragraph{Relevance of unitarity.}
        Let us briefly recall why unitarity of loop integrands is so ubiquitous in the usual discussion of CK duality. 
        \begin{enumerate}[label=(\roman*)]\itemsep-2pt
            \item\label{item:unitarity:1} It is clear that the S-matrix of a physically meaningful theory must be unitary as to be compatible with the usual probabilistic interpretation.
            \item\label{item:unitarity:2} Many works on amplitudes start from an ansatz for the amplitudes or integrands of a field theory. In order to verify that the ansatz is valid and in particular that it belongs indeed to the field theory under consideration, agreement on unitarity cuts is verified. 
            \item\label{item:unitarity:3} There is the above-mentioned proof~\cite{Bern:2010ue} that CK duality of the integrands implies the validity of the double copy prescription, which is based on unitarity methods. The same holds for establishing CK duality at the tree level.
            \item\label{item:unitarity:4} It is more convenient to work with unitary integrands, both because of the tools available and because of the computational simplicity.
            \item\label{item:unitarity:5} One of the key practical goals for considering CK duality and double copy at the loop level is to study the UV behavior of supergravity theories, cf.~\cite{Bern:2019prr}. That is, the focus lies on finding out whether potential counterterms are absent or not. One may expect that the important cancellations originate from unitarity.
        \end{enumerate}

        We can now show that these points are circumvented from our perspective on CK duality.
        \begin{enumerate}[label=(\roman*)]\itemsep-2pt
            \item It is clear that unitarity of the S-matrix is indispensable. As stressed in the previous section, however, we are concerned with unphysical objects (the BRST action and loop integrands) which require renormalization to be turned into meaningful quantities. For that reason, the existence of a unitarity restoring counterterms is sufficient, which will be guaranteed in our formalism.
            \item We will never work with an ansatz for an amplitude or an integrand. Our amplitudes and integrands arise from the Feynman rules of a field theory action principle, which is manifestly classically equivalent to the field theory we are interested in. There are no validity checks to be performed.
            \item We are interested in establishing the double copy at the loop level, and for this we have an independent proof that merely relies on the validity of the double copy at the tree level. Our proof is independent on how the tree-level validity is demonstrated.
            \item We currently study the computational simplicity of our framework from a homotopy algebraic perspective, and our results so far are very encouraging~\cite{Borsten:2022aa}. They seem to imply that, particularly at higher loop, the non-unitary action perspective may be simpler.
            \item There is evidence that a CK-dual set of loop level integrands for ordinary Yang--Mills theory with all ideally desired properties does not exists and one has to compromise. Consider e.g.~\cite{Bern:2015ooa}, where full CK duality was relaxed to CK duality on a spanning set of generalized unitarity cuts of the amplitude. The question is now which property is more relevant for identifying cancellations between loop integrands and thus for the absence of counterterms. One may argue~\cite{Borsten:2022aa} that gravity is properly regularized within string theory and string theory exhibits our kind of loop level CK duality. Therefore, it may be the latter and not unitarity which ought to be preserved when dropping desired properties from field theory integrands. 
            \item Our recent results~\cite{Borsten:2022vtg} strongly suggest that for particular field theories, such as $\caN=4$ super Yang--Mills theory, there is an implementation of our algorithm that neither breaks BRST symmetry nor unitarity.
        \end{enumerate}
        
        \paragraph{Recovering the counterterms after the double copy.}
        Points (iii) and (v) are the crucial ones, and the key question is whether one can make statements about the UV structure of the double copied supergravity theory working with our non-unitary formulation. Let us therefore sketch a way of discriminating between unitarity-restoring counterterms and `true' counterterms that need to be inserted in order to make supergravity finite also in a unitary formulation. 
        
        First, we note that, from the raw double-copied BRST action, the unitarity-restoring counterterms are trivially recovered. Having integrated out the auxiliary fields, suppose that the double-copied action contains a non-local term of the form $\wave h\scO$ where $\scO$ is some non-local operator. This non-local term can be removed by a field redefinition of the form $h\mapsto h+\alpha\scO$ for some coefficient $\alpha$. (This will produce more non-local terms at higher order which are dealt with iteratively.) Such redefinitions then correspond to the Jacobian counterterms that are required to restore unitarity of the raw double-copied action, modulo any Jacobians corresponding to local field redefinitions (which are irrelevant, as they are automatically removed in dimensional regularization, for example). The other type of non-local terms that can appear in the double-copied action are those proportional to e.g.~$\partial_\mu h^{\mu}{}_\nu$, originating from non-local terms in the gauge-fixing sector of Yang--Mills theory; such terms correspond to non-local terms in the gauge-fixing sector of gravity and, as such, do not necessitate Jacobians.
        
        Thus, we do not have to refer to the action explicitly to identify which apparent UV divergences are actually due to Jacobian counterterms and which are due to `real' divergences requiring `true' counterterms. Any divergences arising from a term in the numerator proportional to $p^2$, where $p$ is the momentum through an internal edge, is due to a possibly non-local term in the action that can be canceled by a field redefinition. In this way, one can determine whether a given gravity loop diagram `really' diverges and requires `true' counterterms in a way that accords with traditional methods.

        Altogether, we believe that the advantages of having a manifestly CK-dual BRST action as an organizing principle at our disposal is well worth the slight disadvantages of working with integrands that do not manifest the unitarity of the scattering amplitudes.
        
        \section{Example: the non-linear sigma model}\label{sec:NLSM}
        
        As in~\cite{Borsten:2021hua}, let us start with a discussion of the non-linear sigma model, also known as the principal chiral model, to illustrate our construction lifting on-shell tree-level CK duality to the loop level. This model, despite its simplicity, provides a good description of mesons in the regime where their masses can be neglected.
        
        \subsection{Field theoretic setup}
        
        Consider $d$-dimensional Minkowski space $\IM^d\coloneqq\IR^{1,d-1}$ equipped with the mostly-plus metric $(\eta_{\mu\nu})=\diag(-1,1,\ldots,1)$ with $\mu,\nu,\ldots=0,1,\ldots,d-1$ and coordinates $x^\mu$ together with a semi-simple compact matrix Lie group $\sfG$. The kinematical data of the non-linear sigma model on $\sfG$ are maps $g\colon\IM^d\rightarrow\sfG$, which yield the flat current $g^{-1}\partial_\mu g$, that takes values in the Lie algebra $\frg$ of $\sfG$. With respect to the anti-Hermitian basis $\cb_a$ of $\frg$ with $a,b,\ldots=1,2,\ldots,\dim(\frg)$, we introduce structure constants by $[\cb_a,\cb_b]\eqqcolon{f_{ab}}^c\cb_c$ with $[-,-]$ the Lie bracket on $\frg$, and $\inner{\cb_a}{\cb_b}\coloneqq-\tr(\cb_a\cb_b)=g_{ab}$ with $\tr$ the matrix trace. The corresponding action reads as
        \begin{equation}\label{eq:PCMAction}
            S^\text{NLSM}=-\tfrac{1}{2\lambda^2}\int\rmd^dx\,(g^{-1}\partial_\mu g)_a(g^{-1}\partial^\mu g)^a~,
        \end{equation}
        where $\lambda$ is a coupling constant of mass dimension $1-\frac d2$. This action is invariant under the global left--right $\sfG$-action $g\mapsto g_\text{L}^{-1}gg_\text{R}$ for $g_\text{L,\,R}\in\sfG$.
        
        Using the exponential parameterization\footnote{Another often used parameterization in this context is the Cayley parameterization, cf.~e.g.~\cite{Kostant:0109066}.}  $g\coloneqq\exp(\lambda\phi)$ for $\phi\colon\IM^d\rightarrow\frg$ and setting $\ad_{\lambda\phi}(-)\coloneqq\lambda[\phi,-]$, the formula
        \begin{equation}
            g^{-1}\partial_\mu g=\frac{1-\rme^{-\ad_{\lambda\phi}}}{\ad_{\lambda\phi}}(\lambda\partial_\mu\phi)=\sum_{n=0}^\infty\frac{(-1)^n}{(n+1)!}\,\lambda^{n+1}\ad_\phi^n(\partial_\mu\phi)
        \end{equation}
        allows us to rewrite~\eqref{eq:PCMAction} as
        \begin{equation}\label{eq:PCMActionLieAlgebra}
            S^\text{NLSM}=\int\rmd^dx\,\sum_{n=0}^\infty\frac{\lambda^{2n}}{(2n+2)!}\,\tr\big\{\partial_\mu\phi\,\ad_\phi^{2n}(\partial^\mu\phi)\big\}\,.
        \end{equation}
        Because of the evident symmetry $\phi\mapsto-\phi$ which, in turn, corresponds to the symmetry $g\mapsto g^{-1}$ of~\eqref{eq:PCMAction}, there are only interaction terms (and, correspondingly, Feynman vertices) of even degrees in $\phi$, each of which contains exactly two derivatives.
        
        \subsection{Example: the four-point scattering amplitude}
        
        Before discussing the general case, let us consider the four-point tree-level scattering amplitude as it nicely illustrates the key idea. Explicitly, the first two terms in the action~\eqref{eq:PCMActionLieAlgebra} are
        \begin{equation}\label{eq:PCMActionLieAlgebra4pointA}
            S^\text{NLSM}=-\tfrac12\int\rmd^dx\,\partial_\mu\phi_a\partial^\mu\phi^a-\tfrac{\lambda^2}{4!}\int\rmd^dx\,f_{ab}{}^ef_{cde}\phi^a\partial_\mu\phi^b\phi^c\partial^\mu\phi^d+\cdots~,
        \end{equation}
        where we used the shorthand $f_{abc}\coloneqq f_{ab}{}^dg_{dc}$, which is totally anti-symmetric. Hence, upon taking all the momenta incoming, the four-point tree-level correlator is
        \begin{subequations}\label{eq:NLSM4pointAmplitudeA}
            \begin{equation}
                \scC_{4,0}=\!\! 
                \begin{tikzpicture}[
                    scale=1,
                    every node/.style={scale=1},
                    baseline={([yshift=-.5ex]current bounding box.center)}
                    ]
                    \matrix (m) [
                    matrix of nodes,
                    ampersand replacement=\&, 
                    column sep=0.2cm, 
                    row sep=0.2cm
                    ]{
                        1 \& {}\& {} \& {} \& 4
                        \\
                        {} \& {} \& {}\& {} \& {}
                        \\
                        {} \& {} \& {}\& {} \& {}
                        \\
                        {} \& {} \& {}\& {} \& {}
                        \\
                        2 \& {} \& {} \& {} \& 3
                        \\
                    };
                    \draw (m-1-1) -- (m-3-3.center);
                    \draw (m-5-1) -- (m-3-3.center);
                    \draw (m-1-5) -- (m-3-3.center);
                    \draw (m-5-5) -- (m-3-3.center);
                    \foreach \x in {(m-3-3)}{
                        \fill \x circle[radius=2pt];
                    }
                \end{tikzpicture}
                =\begin{aligned}
                    &\frac{\rmi\lambda^2}{2\cdot3!}\big[f^{a_1a_2b}f_b{}^{a_3a_4}(p_1\cdot p_3+p_2\cdot p_4-p_1\cdot p_4-p_2\cdot p_3)
                    \\
                    &\kern1cm+f^{a_1ba_4}f_b{}^{a_2a_3}(p_1\cdot p_3+p_2\cdot p_4-p_1\cdot p_2-p_3\cdot p_4)
                    \\
                    &\kern1cm+f^{a_1ba_3}f_b{}^{a_4a_2}(p_1\cdot p_2+p_3\cdot p_4-p_1\cdot p_4-p_2\cdot p_3)\big]~,
                \end{aligned}
            \end{equation}
            and analogously to scattering amplitudes, we can introduce (off-shell) kinematic and color numerators
            \begin{equation}\label{eq:NLSM4pointAmplitudeAb}
                \begin{aligned}
                    \scC_{4,0}&=\frac{\rmi\lambda^2}{4!}\big[\underbrace{f^{a_1a_2b}f_b{}^{a_3a_4}}_{\eqqcolon\,\colfac_{s_{12}}}\underbrace{(s_{13}+s_{24}-s_{14}-s_{23})}_{\eqqcolon\,\frac{2\kinfac_{s_{12}}}{s_{12}}}
                    \\
                    &\kern2cm+\underbrace{f^{a_1ba_4}f_b{}^{a_2a_3}}_{\eqqcolon\,\colfac_{s_{14}}}\underbrace{(s_{13}+s_{24}-s_{12}-s_{34})}_{\eqqcolon\,\frac{2\kinfac_{s_{14}}}{s_{14}}}
                    \\
                    &\kern2cm+\underbrace{f^{a_1ba_3}f_b{}^{a_4a_2}}_{\eqqcolon\,\colfac_{s_{13}}}\underbrace{(s_{12}+s_{34}-s_{14}-s_{23})}_{\eqqcolon\,\frac{2\kinfac_{s_{13}}}{s_{13}}}\big]
                    \\
                    &=\frac{\rmi\lambda^2}{2\cdot3!}\left(\frac{\colfac_{s_{12}}\kinfac_{s_{12}}}{s_{12}}+\frac{\colfac_{s_{14}}\kinfac_{s_{14}}}{s_{14}}+\frac{\colfac_{s_{13}}\kinfac_{s_{13}}}{s_{13}}\right)~,
                \end{aligned}
            \end{equation}
        \end{subequations}
        where $s_{ij}\coloneqq(p_i+p_j)^2$ for $i=1,\ldots,4$ are the usual Mandelstam variables. Using momentum conservation $\sum_{i=1}^4p_i=0$, we obtain
        \begin{equation}\label{eq:NLSMKinFac4point}
            \kinfac_{s_{12}}=s_{12}(s_{13}-s_{14})~,
            \quad
            \kinfac_{s_{14}}=s_{14}(s_{13}-s_{12})~,
            \quad
            \kinfac_{s_{13}}=s_{13}(s_{12}-s_{14})~.
        \end{equation}
        Evidently, the kinematic numerators~\eqref{eq:NLSMKinFac4point} satisfy 
        \begin{equation}\label{eq:KinJacbi4point}
            \kinfac_{s_{12}}-\kinfac_{s_{14}}-\kinfac_{s_{13}}=0~,
        \end{equation}
        and likewise, by the Jacobi identity, we also have 
        \begin{equation}
            \colfac_{s_{12}}-\colfac_{s_{14}}-\colfac_{s_{13}}=f^{ba_1a_2}f_b{}^{a_3a_4}+f^{ba_1a_3}f_b{}^{a_4a_2}+f^{ba_1a_4}f_b{}^{a_2a_3}=0~.
        \end{equation}
        The tree-level scattering amplitude $\scA_{4,0}$ is obtained by putting the external momenta $p_1,\ldots,p_4$ on shell. Even without explicitly introducing strictification auxiliary fields, we  see that $\scA_{4,0}$ can be understood as a sum over cubic graphs, and both $\scA_{4,0}$ and $\scC_{4,0}$ enjoy CK duality. CK duality at four points is already manifest in the standard action.
        
        However, we can further improve our action to obtain the final line in~\eqref{eq:NLSM4pointAmplitudeAb} directly from the Feynman diagrams. When expressing the scattering amplitude in Mandelstam variables, we have effectively used the fact that terms containing squared momenta, $p_i^2$ for $i=1,\ldots,4$, mutually cancel from the expressions. This is similar to what happens in the case of Yang--Mills theory, where the CK-compliant form\footnote{off shell at four points and on shell beyond four points at the tree level} is a result of suitably adding terms to the Lagrangian which vanish due to the Jacobi identity~\cite{Bern:2010yg,Tolotti:2013caa}. In order to see what terms we need in the present case, we rewrite~\eqref{eq:PCMActionLieAlgebra4pointA} as
        \begin{equation}\label{eq:PCMActionLieAlgebra4pointB}
            S^\text{NLSM}_4=\tfrac12\int\rmd^dx\,\phi_a\wave\phi^a-\tfrac{\lambda^2}{2\cdot 4!}\int\rmd^dx\,f_{ab}{}^ef_{cde}\phi^a\phi^c\big[\wave(\phi^b\phi^d)-\phi^d\wave\phi^b-\phi^b\wave\phi^d\big]+\cdots~,
        \end{equation}
        which, using the anti-symmetry $f_{ab}{}^c=-f_{ba}{}^c$ of the structure constants, we can simplify to
        \begin{equation}\label{eq:PCMActionLieAlgebra4pointC}
            S^\text{NLSM}_4=\tfrac12\int\rmd^dx\,\phi_a\wave\phi^a-\tfrac{\lambda^2}{2\cdot 4!}\int\rmd^dx\,f_{ab}{}^ef_{cde}\phi^a\phi^c\wave(\phi^b\phi^d)+\cdots~.
        \end{equation}
        The four-point tree-level correlator is now manifestly of the form
        \begin{equation}
            \scC_{4,0}=\tfrac{\rmi\lambda^2}{4!}\big[\colfac_{s_{12}}(s_{13}+s_{24}-s_{14}-s_{23})+\colfac_{s_{14}}(s_{13}+s_{24}-s_{12}-s_{34})+\colfac_{s_{13}}(s_{12}+s_{34}-s_{14}-s_{23})\big]~,
        \end{equation}
        where the color numerators $\colfac_{s_{ij}}$ are defined as in~\eqref{eq:NLSM4pointAmplitudeA} and the $s_{ij}$ are again the Mandelstam variables. Hence, contrary to the case of Yang--Mills theory, where the vanishing of the extra terms in the action is guaranteed by the Jacobi identity, the vanishing in~\eqref{eq:PCMActionLieAlgebra4pointB} is due to the anti-symmetry of the structure constants $f_{ab}{}^c$.
        
        \subsection{Lifting on-shell to off-shell color--kinematics duality}
        
        To illustrate our procedure that lifts on-shell CK duality off shell by field redefinitions, let us assume that we started not from $S^\text{NLSM}$, but from the following action, which is semi-classically equivalent to $S^\text{NLSM}$ and satisfies on-shell CK-duality:
        \begin{equation}
            {S^\text{NLSM}_4}'=S^\text{NLSM}_4+\lambda^2\int\rmd^dx\,\Big(\underbrace{f_{abc}f_{de}{}^a\phi^b(\wave\phi^c)\partial_{\mu}\phi^d\wave\partial^\mu\phi^e}_{\eqqcolon\,\scL'_4}+\underbrace{\tfrac12f_{abc}f_{de}{}^a\phi^b(\wave\phi^c)\phi^d\wave{}^2\phi^e}_{\eqqcolon\,\scL''_4}\Big)~.
		\end{equation}
        The two additional contributions to the original correlator~\eqref{eq:NLSM4pointAmplitudeA} are
        \begin{subequations}
            \begin{equation}\label{eq:NLSM4pointAmplitudeA_1}
                \begin{aligned}
                    &-\rmi\lambda^2\big[\underbrace{f^{a_1a_2b}f_b{}^{a_3a_4}}_{\eqqcolon\,\colfac_{s_{12}}}\underbrace{((p_1^2-p_2^2)(p_3^2-p_4^2)(p_1\cdot p_2+p_3\cdot p_4))}_{\eqqcolon\,-\frac{\kinfac'_{s_{12}}}{2\cdot 3!}s_{12}}
                    \\
                    &\kern2cm+\underbrace{f^{a_1ba_4}f_b{}^{a_2a_3}}_{\eqqcolon\,\colfac_{s_{14}}}\underbrace{((p_1^2-p_4^2)(p_3^2-p_2^2)(p_1\cdot p_4+p_3\cdot p_2))}_{\eqqcolon\,-\frac{\kinfac'_{s_{14}}}{2\cdot 3!s_{14}}}
                    \\
                    &\kern2cm+\underbrace{f^{a_1ba_3}f_b{}^{a_4a_2}}_{\eqqcolon\,\colfac_{s_{13}}}\underbrace{((p_1^2-p_3^2)(p_2^2-p_4^2)(p_1\cdot p_3+p_2\cdot p_4))}_{\eqqcolon\,-\frac{\kinfac'_{s_{13}}}{2\cdot 3!s_{13}}}\big]
                    \\
                    &\hspace{1cm}=\frac{\rmi\lambda^2}{2\cdot 3!}\left(\frac{\colfac_{s_{12}}\kinfac'_{s_{12}}}{s_{12}}+\frac{\colfac_{s_{14}}\kinfac'_{s_{14}}}{s_{14}}+\frac{\colfac_{s_{13}}\kinfac'_{s_{13}}}{s_{13}}\right)
                \end{aligned}
            \end{equation}
            and
            \begin{equation}\label{eq:NLSM4pointAmplitudeA_2}
                \begin{aligned}
                    &-\frac{\rmi\lambda^2}{2}\big[\underbrace{f^{a_1a_2b}f_b{}^{a_3a_4}}_{\eqqcolon\,\colfac_{s_{12}}}\underbrace{(p_1^2 p_3^2(p_1^2+p_3^2)-p_1^2 p_4^2(p_1^2+p_4^2)-p_2^2 p_3^2(p_2^2+p_3^2)+p_2^2 p_4^2(p_2^2+p_4^2))}_{\eqqcolon\,-\frac{\kinfac''_{s_{12}}}{3!s_{12}}}
                    \\
                    &\kern1.6cm+\underbrace{f^{a_1ba_4}f_b{}^{a_2a_3}}_{\eqqcolon\,\colfac_{s_{14}}}\underbrace{(p_1^2 p_3^2(p_1^2+p_3^2)-p_1^2 p_2^2(p_1^2+p_2^2)-p_3^2 p_4^2(p_3^2+p_4^2)+p_2^2 p_4^2(p_2^2+p_4^2))}_{\eqqcolon\,-\frac{\kinfac''_{s_{14}}}{3!s_{14}}}
                    \\
                    &\kern1.6cm+\underbrace{f^{a_1ba_3}f_b{}^{a_4a_2}}_{\eqqcolon\,\colfac_{s_{13}}}\underbrace{(p_1^2 p_2^2(p_1^2+p_2^2)-p_1^2 p_4^2(p_1^2+p_4^2)-p_2^2 p_3^2(p_2^2+p_3^2)+p_3^2 p_4^2(p_3^2+p_4^2))}_{\eqqcolon\,-\frac{\kinfac''_{s_{13}}}{3!s_{13}}}\big]
                    \\
                    &\hspace{0.6cm}=\frac{\rmi\lambda^2}{2\cdot3!}\left(\frac{\colfac_{s_{12}}\kinfac''_{s_{12}}}{s_{12}}+\frac{\colfac_{s_{14}}\kinfac''_{s_{14}}}{s_{14}}+\frac{\colfac_{s_{13}}\kinfac''_{s_{13}}}{s_{13}}\right)~,
                \end{aligned}
            \end{equation}
        \end{subequations}
        coming from $\scL'_4$ and $\scL''_4$, respectively, where again the $s_{ij}$ are the Mandelstam variables.
        
        On shell, the contributions $\kinfac'_{s_{ij}}$ and $\kinfac''_{s_{ij}}$ vanish, so that the kinematic Jacobi identity~\eqref{eq:KinJacbi4point} is not modified. Off shell, we have
        \begin{subequations}\label{eq:KinJacobiModification}
            \begin{equation}
                \begin{aligned}
                    -(\kinfac'_{s_{12}}-\kinfac'_{s_{14}}-\kinfac'_{s_{13}})&=12(p_1+p_2)^2\big[(p_1^2-p_2^2)(p_3^2-p_4^2)(p_1\cdot p_2+p_3\cdot p_4)\big]
                    \\
                    &\kern1cm-12(p_1+p_4)^2\big[(p_1^2-p_4^2)(p_3^2-p_2^2)(p_1\cdot p_4+p_3\cdot p_2)\big]
                    \\
                    &\kern1cm-12(p_1+p_3)^2\big[(p_1^2-p_3^2)(p_2^2-p_4^2)(p_1\cdot p_3+p_2\cdot p_4)\big]
                \end{aligned}
            \end{equation}
            and
            \begin{equation}
                \begin{aligned}
                    &-(\kinfac''_{s_{12}}-\kinfac''_{s_{14}}-\kinfac''_{s_{13}})
                    \\
                    &~~=6(p_1+p_2)^2\big[p_1^2p_3^2(p_1^2+p_3^2)-p_1^2 p_4^2(p_1^2+p_4^2)-p_2^2 p_3^2(p_2^2+p_3^2)+p_2^2 p_4^2(p_2^2+p_4^2)\big]
                    \\
                    &\kern1cm-6(p_1+p_4)^2\big[p_1^2 p_3^2(p_1^2+p_3^2)-p_1^2 p_2^2(p_1^2+p_2^2)-p_3^2 p_4^2(p_3^2+p_4^2)+p_2^2 p_4^2(p_2^2+p_4^2)\big]
                    \\
                    &\kern1cm-6(p_1+p_3)^2\big[p_1^2 p_2^2(p_1^2+p_2^2)-p_1^2 p_4^2(p_1^2+p_4^2)-p_2^2 p_3^2(p_2^2+p_3^2)+p_3^2 p_4^2(p_3^2+p_4^2)\big]~.
                \end{aligned}
            \end{equation}
        \end{subequations}
        Summing the terms in~\eqref{eq:KinJacobiModification} and using momentum conservation $p_4=-p_1-p_2-p_3$, we obtain 
        \begin{equation}\label{eq:KinJacobi4}
            \begin{aligned}
                &-(\kinfac'_{s_{12}}-\kinfac'_{s_{14}}-\kinfac'_{s_{13}})-(\kinfac''_{s_{12}}-\kinfac''_{s_{14}}-\kinfac''_{s_{13}})
                \\ 
                &\kern4cm=12s_{12}^2(p_1^2-p_2^2)(s_{12}+s_{23}+s_{13}-p_1^2-p_2^2-2p_3^2)
                \\
                &\kern5cm-12s_{23}^2(p_3^2-p_2^2)(s_{12}+s_{23}+s_{13}-2p_1^2-p_2^2-p_3^2)
                \\
                &\kern5cm-12s_{13}^2(p_1^2-p_3^2)(s_{12}+s_{23}+s_{13}-p_1^2-2p_2^2-p_3^2)~,
            \end{aligned}
        \end{equation}
        which is generically non-vanishing, rendering the kinematic Jacobi identity invalid.
        
        In our prescription, we now perform the appropriate field redefinition to fix the failure of the kinematic Jacobi identity, which is here the inverse of the field redefinition 
        \begin{equation}\label{eq:NLSM_fieldredefinition}
            \phi^a\mapsto\phi^a+\lambda f_{bc}{}^a\phi^b\wave\phi^c~,
        \end{equation}
        linking $S^\text{NLSM}_4$ to ${S^\text{NLSM}_4}'$. 
        
        Altogether, we saw an example of a situation in which the kinematic Jacobi identity is valid on shell but violated off shell by terms containing $\wave$. These terms could be absorbed in a field redefinition. This neatly captures the basic intuition that off-shell failures of CK duality are due to physically irrelevant terms that may be encoded in the action by taking advantage of its redundancies. More precisely, for an analytic action any failure of CK duality due to any off-shell extension of the S-matrix can be repaired  by  introducing new vertices in   the action that  are invisible for on-shell field configurations, i.e.\ those fields $\phi$ satisfying $\wave \phi=0$. These vertices are necessarily proportional to  $\wave \phi$ and, thus, may be realized through the kinetic term $\phi\wave \phi$ by a field redefinition.
        
        \subsection{Lifting on-shell color--kinematics duality to the loop level}\label{ssec:offshell_lift_NLSM}
        
        With this intuition in mind, let us now present the general algorithm that lifts on-shell tree-level CK duality to the loop level, up to potential counterterms. This becomes mostly straightforward with the observations we made in \cref{sec:prelim}. Recall that on-shell CK duality of tree-level scattering amplitudes of the non-linear sigma model has been established in~\cite{Chen:2013fya}.
        \begin{enumerate}[label=(\roman*)]\itemsep-2pt
            \item Choose a CK-dual parameterization of the tree-level scattering amplitudes, and pick a set of master color numerators $\bmc^{(n)}_{\rm m}$ for all numbers $n$ of external legs we are interested in, cf.~\cref{ssec:primaries}.
            \item Choose one of the equivalent actions for the non-linear sigma model, for example~\eqref{eq:PCMActionLieAlgebra}; the specific chosen form is inconsequential as long as the kinetic term is of the canonical form. Denote this action by $S_3$\footnote{For the choice of action~\eqref{eq:PCMActionLieAlgebra} we already have manifest CK duality at four points, but let us us ignore this as we would like to make the argument for a generic choice of weakly semi-classically equivalent action.}, where the subscript indicates the number of points up to which CK duality is manifest, and start with $n=4$.
            \item\label{item:tolotti} Use~\cref{ob:iterative_TW_terms} to add a term that is identically zero to the action $S_{n-1}$ such that its Feynman rules produce the chosen form of the CK-dual tree-level scattering amplitudes up to $n$~points. Note that the thus produced action $S^\text{on-shell}_{n}$ generically contains non-local terms of the type $E_1^M\frac{1}{\wave}E^2_M$, cf.~\eqref{eq:typically_non-local_term}, as is well-known from the case of Yang--Mills theory, cf.~\cite{Bern:2010ue,Tolotti:2013caa}.
            \item\label{item:redefinition} Using the Feynman diagrams\footnote{As previously mentioned, this yields  a canonical choice of off-shell continuation to the amputated correlator $\hat\scA'_{n,0}$, but we could have used any other  reasonably well-behaved choice.} of $S^\text{on-shell}_{n}$,   we can now compute the off-shell master numerators $\hat\bmn^{(n)}_{\rm m}$ corresponding to the color master numerators $\bmc^{(n)}_{\rm m}$ and complete them to a full set of CK duality manifesting kinematic numerators $\hat \bmn^{(n)}=\bmJ^{(n)}\hat\bmn^{(n)}_{\rm m}$. These numerators give rise to a manifestly CK-dual amputated correlator $\hat\scA'_{n,0}$ that is obtained from $\bmc^{(n)\sfT}\bmD^{(n)}\hat\bmn^{(n)}$ by symmetrizing over the external legs\footnote{Note that symmetrization preserves manifest CK duality.}:
            \begin{equation}\label{eq:amputated_correlator_NLSM}
                \hat\scA'_{n,0}=\left[\bmc^{(n)\sfT}\bmD^{(n)}\hat\bmn^{(n)}\right]_\sigma~.
            \end{equation}
            This expression necessarily agrees with the actual amputated correlator $\hat\scA_{n,0}$ of the theory on shell. Off shell, however, the two expressions may differ by sums of terms $p^2F_i$, where $p$ is the momentum of an external line and $F_i$ is the sum of products of rational functions of momenta and Lie algebra structure constants. These terms can be canceled by terms in the Lagrangian of the form $\tilde F_i\wave\phi$, where $\tilde F_i$ is a polynomial of $(n-1)$-st order in the fields containing derivatives as well as the operator $\frac{1}{\wave}$, acting on certain parts of the contained monomials. Such terms can be constructed from the kinematic terms by field redefinitions of the form
            \begin{equation}
                \phi\mapsto\phi+\sum_i\tilde F_i~,
            \end{equation}
            where we suppressed all flavor indices. We perform this field redefinition to obtain an action $S^\text{off-shell,\,non-local}_n$. The Jacobian determinants for the required field redefinitions lead to additional counterterms that must be included in the renormalization. They do not affect the consistency of our theory, cf.~\cref{ssec:field_redefinitions}.
            \item If we have not reached the maximum order that is of relevance for the scattering amplitudes in which we are interested, then increment $n$, and go back to~\ref{item:tolotti}. Otherwise, continue to~\ref{item:strictify}.
            \item\label{item:strictify} Use \cref{ob:strictification} to strictify the order~$k$ interaction vertices in the action $S_n$ for $k\le n$. Non-local terms are strictified to local ones. For details on this strictification, see \cref{ssec:strictification} or~\cite[Section 3]{Borsten:2021hua}.
        \end{enumerate}
        For arbitrarily large but finite $n$, the action $S_n$ obtained in this manner is local up to order $n$, and its Feynman rules produce fully CK-dual tree-level scattering amplitudes up to $n$ points. By \cref{ob:tree_is_good_enough}, this suffices to ensure loop-level CK duality of the scattering amplitude integrands, up to potential counterterms.
        
        As an aside, we note that the field redefinitions performed in~\ref{item:redefinition} in the above algorithm can also be obtained by a combination of strictifications and purely local field redefinitions. Let us assume that the field redefinition we need to implement is
        \begin{equation}\label{eq:field_redef}
            \phi\mapsto\phi+E_1\frac{1}{\wave}E_2
        \end{equation}
        for some expressions $E_1$ and $E_2$ which are local in the fields, and such that $E_1$ is at least of linear order in the fields. We can simply multiply the integrand in the partition function by the expression
        \begin{equation}
            \int\caD G\,\caD\bar G~\rme^{\int\rmd^d x\left(\bar G\wave G-\bar G E_2\right)}~,
        \end{equation}
        where $G$ and $\bar G$ are auxiliary fields. This merely adds a volume factor, which is evidently irrelevant in perturbative quantum computations. We can then perform the local field redefinition
        \begin{equation}\label{eq:field-redef-aux-argument}
            \phi\mapsto\phi+E_1G~,
        \end{equation}
        which, after imposing the equation of motion for the auxiliary field $G$,
        \begin{equation}
            \wave G=E_2~,
        \end{equation}
        is equivalent to the desired field redefinition~\eqref{eq:field_redef}. The fact that this field redefinition is local then makes it uncontroversial that the tree-level scattering amplitudes have not been affected.\footnote{If $E_1$ were to have a non-zero constant term, then the field redefinition~\eqref{eq:field-redef-aux-argument} would be non-trivial at the linear order, thus mixing the scattering amplitudes of $\phi$ with those of $G$; then the scattering amplitudes of the redefined $\phi$ would differ from those of the original $\phi$.} The new loop corrections arising off shell then reproduce the additional counterterms that we would have expected from the non-local field redefinition.
        
        The above algorithm proves that up to potential counterterms, on-shell CK duality can be lifted off shell and, thus, to loop-level CK duality for all diagrams having the sigma model scalar field $\phi$ on their external legs. We now argue that this trivially extends to arbitrary diagrams, including those with auxiliary fields on external legs.
        Evidently, any diagram with one or more auxiliary fields on external legs, called \emph{auxiliary diagrams} in the following, is necessarily a subdiagram of a diagram with fields from exclusively the BRST-extended Hilbert space on external legs, called \emph{standard diagrams} in the following. Correspondingly, any pair or triple of auxiliary diagrams whose sum vanishes due to the algebraic properties of the structure constants and the Killing form, for example 
        \begin{equation}
            \begin{tikzpicture}[
                scale=1,
                every node/.style={scale=1},
                baseline={([yshift=-.5ex]current bounding box.center)}
                ]
                \matrix (m) [
                matrix of nodes,
                ampersand replacement=\&,
                column sep=0.13cm,
                row sep=0.13cm
                ]{
                    {} \& {}\& {} \& {} \& {}
                    \\
                    {} \& {} \& {}\& {} \& {}
                    \\
                    {} \& {} \& {} \& {} \& {}
                    \\
                    {} \& {} \& {}\& {} \& {}
                    \\
                    {} \& {} \& {} \& {} \& {}
                    \\
                };
                \draw [gluon] (m-1-1) -- (m-3-2.center);
                \draw [gluon] (m-5-1) -- (m-3-2.center);
                \draw [gluon] (m-3-2.center) -| node[near start,below] {} (m-3-4.center);
                \draw [gluon] (m-1-5) -- (m-3-4.center);
                \draw [dashed] (m-5-5) -- (m-3-4.center);
                \foreach \x in {(m-3-2), (m-3-4)}{
                    \fill \x circle[radius=2pt];
                }
            \end{tikzpicture}
            ~~~~~~
            \begin{tikzpicture}[
                scale=1,
                every node/.style={scale=1},
                baseline={([yshift=-.5ex]current bounding box.center)}
                ]
                \matrix (m) [
                matrix of nodes,
                ampersand replacement=\&,
                column sep=0.13cm,
                row sep=0.13cm
                ]{
                    {} \& {} \& {} \& {} \& {}
                    \\
                    {} \& {} \& {} \& {} \& {}
                    \\
                    {} \& {} \& {} \& {} \& {}
                    \\
                    {} \& {} \& {} \& {} \& {}
                    \\
                    {} \& {} \& {} \& {} \& {}
                    \\
                };
                \draw [gluon] (m-1-1) -- (m-2-3.center);
                \draw [gluon] (m-5-1) -- (m-4-3.center);
                \draw [gluon] (m-2-3.center) -| node[near end,left] {} (m-4-3.center);
                \draw [gluon] (m-1-5) -- (m-2-3.center);
                \draw [dashed] (m-5-5) -- (m-4-3.center);
                \foreach \x in {(m-2-3), (m-4-3)}{
                    \fill \x circle[radius=2pt];
                }
            \end{tikzpicture}
            ~~~~~~
            \begin{tikzpicture}[
                scale=1,
                every node/.style={scale=1},
                baseline={([yshift=-.5ex]current bounding box.center)}
                ]
                \matrix (m) [
                matrix of nodes,
                ampersand replacement=\&,
                column sep=0.13cm,
                row sep=0.13cm
                ]{
                    {} \& {}\& {} \& {} \& {}
                    \\
                    {} \& {} \& {}\& {} \& {}
                    \\
                    {} \& {} \& {}\& {} \& {}
                    \\
                    {} \& {} \& {}\& {} \& {}
                    \\
                    {} \& {}\& {} \& {} \& {}
                    \\
                };
                \draw [gluon] (m-1-1) -- (m-3-4.center);
                \draw [gluon] (m-5-1) -- (m-3-2.center);
                \draw [gluon] (m-3-2.center) -| node[near start,below] {} (m-3-4.center);
                \draw [gluon] (m-1-5) -- (m-3-2.center);
                \draw [dashed] (m-5-5) -- (m-3-4.center);               
                \foreach \x in {(m-3-2), (m-3-4)}{
                    \fill \x circle[radius=2pt];
                }
            \end{tikzpicture}
        \end{equation}
        where a dashed line denotes an external auxiliary field,
        is contained in a pair or triple of standard diagrams, for example
        \begin{equation}
            \begin{tikzpicture}[
                scale=1,
                every node/.style={scale=1},
                baseline={([yshift=-.5ex]current bounding box.center)}
                ]
                \matrix (m) [
                matrix of nodes,
                ampersand replacement=\&,
                column sep=0.13cm,
                row sep=0.13cm
                ]{
                    {} \& {}\& {} \& {} \& {}
                    \\
                    {} \& {} \& {}\& {} \& {}
                    \\
                    {} \& {} \& {} \& {} \& {}\& {}\& {}
                    \\
                    {} \& {} \& {}\& {} \& {} \& {}
                    \\
                    {} \& {} \& {} \& {} \& {} \& {}
                    \\
                    {} \& {} \& {} \& {} \& {} \& {}
                    \\
                    {} \& {} \& {} \& {} \& {} \& {}
                    \\
                };
                \draw [gluon] (m-1-1) -- (m-3-2.center);
                \draw [gluon] (m-5-1) -- (m-3-2.center);
                \draw [gluon] (m-3-2.center) -| node[near start,below] {} (m-3-4.center);
                \draw [gluon] (m-1-5) -- (m-3-4.center);
                \draw [dashed] (m-5-5.center) -- (m-3-4.center);
                \draw [gluon] (m-5-5.center) -- (m-4-8);
                \draw [gluon] (m-5-5.center) -- (m-7-4);
                \foreach \x in {(m-3-2), (m-3-4),(m-5-5)}{
                    \fill \x circle[radius=2pt];
                }
            \end{tikzpicture}
            ~~~
            \begin{tikzpicture}[
                scale=1,
                every node/.style={scale=1},
                baseline={([yshift=-.5ex]current bounding box.center)}
                ]
                \matrix (m) [
                matrix of nodes,
                ampersand replacement=\&,
                column sep=0.13cm,
                row sep=0.13cm
                ]{
                    {} \& {}\& {} \& {} \& {}
                    \\
                    {} \& {} \& {}\& {} \& {}
                    \\
                    {} \& {} \& {} \& {} \& {}\& {}\& {}
                    \\
                    {} \& {} \& {}\& {} \& {} \& {}
                    \\
                    {} \& {} \& {} \& {} \& {} \& {}
                    \\
                    {} \& {} \& {} \& {} \& {} \& {}
                    \\
                    {} \& {} \& {} \& {} \& {} \& {}
                    \\
                };
                \draw [gluon] (m-1-1) -- (m-2-3.center);
                \draw [gluon] (m-5-1) -- (m-4-3.center);
                \draw [gluon] (m-2-3.center) -| node[near end,left] {} (m-4-3.center);
                \draw [gluon] (m-1-5) -- (m-2-3.center);
                \draw [dashed] (m-5-5.center) -- (m-4-3.center);
                \draw [gluon] (m-5-5.center) -- (m-4-8);
                \draw [gluon] (m-5-5.center) -- (m-7-4);
                \foreach \x in {(m-2-3), (m-4-3),(m-5-5)}{
                    \fill \x circle[radius=2pt];
                }
            \end{tikzpicture}
            ~~~~
            \begin{tikzpicture}[
                scale=1,
                every node/.style={scale=1},
                baseline={([yshift=-.5ex]current bounding box.center)}
                ]
                \matrix (m) [
                matrix of nodes,
                ampersand replacement=\&,
                column sep=0.13cm,
                row sep=0.13cm
                ]{
                    {} \& {}\& {} \& {} \& {}
                    \\
                    {} \& {} \& {}\& {} \& {}
                    \\
                    {} \& {} \& {} \& {} \& {}\& {}\& {}
                    \\
                    {} \& {} \& {}\& {} \& {} \& {}
                    \\
                    {} \& {} \& {} \& {} \& {} \& {}
                    \\
                    {} \& {} \& {} \& {} \& {} \& {}
                    \\
                    {} \& {} \& {} \& {} \& {} \& {}
                    \\
                };
                \draw [gluon] (m-1-1) -- (m-3-4.center);
                \draw [gluon] (m-5-1) -- (m-3-2.center);
                \draw [gluon] (m-3-2.center) -| node[near start,below] {} (m-3-4.center);
                \draw [gluon] (m-1-5) -- (m-3-2.center);
                \draw [dashed] (m-5-5.center) -- (m-3-4.center);
                \draw [gluon] (m-5-5.center) -- (m-4-8);
                \draw [gluon] (m-5-5.center) -- (m-7-4);
                \foreach \x in {(m-3-2), (m-3-4),(m-5-5)}{
                    \fill \x circle[radius=2pt];
                }
            \end{tikzpicture}
        \end{equation}
        where a dashed line denotes the propagator for a conjugate pair of auxiliary fields. Note that the completion of an auxiliary diagram to a standard diagram is unique. This completion is moreover a non-degenerate operation by definition: if the completion were to have a non-trivial kernel, then we could remove these auxiliary field modes from our theory because their whole purpose is to reproduce a standard diagram. As no modes or terms can be projected out, the duality for standard diagrams necessarily implies CK duality for auxiliary diagrams.\footnote{The same evidently also holds in any strictified CK-dual theory, e.g.~Yang--Mills theory discussed in \cref{ssec:offshell_lift_YM}.} This completes the proof of the validity of off-shell CK duality up to potential counterterms for the non-linear sigma model.
        
        \subsection{Color--kinematics duality manifesting action}\label{ssec:CK_dual_NLSM}
        
        Using the preceding algorithm we can construct an action whose Feynman rules lead to manifestly CK dual parameterizations of the scattering amplitudes for the non-linear sigma model. In the following, we comment on its explicit form. For a rather distinct approach to a CK duality manifesting action see~\cite{Cheung:2016prv}.
        
        As explained in detail in~\cite{Borsten:2021hua}, a CK-dual parameterization of scattering amplitudes involves a factorization of fields, propagators, and interactions of the schematic form
        \begin{equation}\label{eq:typical_factorization}
            \mbox{color/flavor}~\otimes~\mbox{kinematics}~\otimes~\mbox{scalar field}~.
        \end{equation}
        CK duality relates properties of the color or flavor part and the kinematics part; the scalar field part does not participate. The action of the kinematic algebra on fields involves differential operators, and therefore the fields' components transforming under the kinematic algebra are most conveniently labeled by DeWitt indices $\sfi,\sfj,\ldots$, i.e. indices indicating field species as well as space–time position (or, after a Fourier transform, momentum). This, however, breaks the symmetry between the color/flavor and kinematics parts; it is therefore convenient to promote also the color indices to DeWitt indices $\bar\sfa,\bar\sfb,\ldots$ encoding both the color or flavor index and the space--time position, with the understanding that the space--time positions in both types of DeWitt indices are taken to be the same. For a very detailed discussion of this issue in the case of Yang--Mills theory, see~\cite[Section 9.3]{Borsten:2021hua}.
        
        Using the DeWitt indices over a common space--time point, we can manifest full CK duality in the action of the non-linear sigma model using the concise and suggestive action
        \begin{equation}\label{eq:CK_dual_NLSM}
            S^\text{NLSM}_\text{CK-dual}=\tfrac12\sfg_{\sfi\sfj}\bar\sfg_{\bar\sfa\bar\sfb}\Phi^{\sfi\bar\sfa}\wave\Phi^{\sfj\bar\sfb}+\tfrac1{3!}\sff_{\sfi\sfj\sfk}\bar\sff_{\bar\sfa\bar\sfb\bar\sfc}\Phi^{\sfi\bar\sfa}\Phi^{\sfj\bar\sfb}\Phi^{\sfk\bar\sfc}~.
        \end{equation}
        CK duality then amounts to the fact that the flavor and kinematic algebraic data
        \begin{equation}
            (\,\bar\sfg_{\bar\sfa\bar\sfb}\,,\,\bar\sff_{\bar\sfb\bar\sfc}{}^{\bar\sfa}\,)
            \eand
            (\,\sfg_{\sfi\sfj}\,,\,\sff_{\sfj\sfk}{}^\sfi\,)
        \end{equation}
        with
        \begin{equation}\label{eq:raise_lower_NLSM}
            \sff_{\bar\sfa\bar\sfb\bar \sfc}=\bar\sfg_{\bar \sfc\bar \sfd}\bar\sff_{\bar\sfa\bar\sfb}{}^{\bar\sfd}
            \eand
            \sff_{\sfi\sfj\sfk}=\sfg_{\sfk\sfl}\sff_{\sfi\sfj}{}^\sfk
        \end{equation}
        have the same algebraic properties, namely anti-symmetry and the Jacobi identity of the Lie algebra structure constants as well as symmetry and invariance of the metric:
        \begin{equation}\label{eq:alg_rel_NLSM}
            \begin{aligned}
                \bar\sfg_{[\bar\sfa\bar\sfb]}&=0~,~
                &\bar\sff_{(\bar\sfb\bar\sfc)}{}^{\bar\sfa} &=0~,~
                &\bar\sff_{\bar\sfe[\bar\sfa}{}^{\bar \sfd}\sff_{\bar\sfb\bar\sfc]}{}^{\bar\sfe}&=0~,~
                &\bar\sfg_{\bar\sfa(\bar\sfb}\sff_{\bar\sfc)\bar\sfe}{}^{\bar\sfa}=0~,
                \\
                \sfg_{[\sfi\sfj]}&=0~,~
                &\sff_{(\sfj\sfk)}{}^\sfi &=0~,~
                &\sff_{\sfn[\sfi}{}^\sfm\sff_{\sfj\sfk]}{}^\sfn&=0~,~
                &\sfg_{\sfi(\sfj}\sff_{\sfk)\sfm}{}^\sfi=0~.
            \end{aligned}
        \end{equation}
        Here $(-)$ and $[-]$ denote total symmetrization and antisymmetrization of enclosed indices, as usual.
        
        The suggestive but condensed notation used in~\eqref{eq:CK_dual_NLSM} certainly requires some more detailed explanation. We have the fields 
        \begin{equation}
            \begin{aligned}
                (\Phi^{\sfi\bar\sfa})&=\big(\phi^a(x),Y^{ma}(x)\big)~,
                \\
                \big(Y^{ma}(x)\big)&=\big(\phi^a(x),C^a_\mu(x),\bar C^a_\mu(x),D^a(x),\bar D^a(x),\ldots\big)~,
            \end{aligned}
        \end{equation}
        with $m$ labeling the various auxiliary fields $C_\mu^a(x)$, $\bar C^a_\mu(x)$, $\ldots$ introduced to strictify the theory. Note that $\Phi^{\sfi\bar\sfa}$ carries DeWitt indices, while the indices of the individual fields are standard flavor indices, running over the adjoint of $\frg$, or Lorentz indices, and the space--time dependence is made explicit. 
        
        The \emph{flavor metric} $\bar\sfg_{\bar\sfa\bar\sfb}$ is simply the Cartan--Killing form $g_{ab}$ on $\frg$, decorated by a space--time delta-function as not to affect the space--time index. We also have the \emph{kinematic metric} $\sfg_{\sfi\sfj}$, which can be described by the block-diagonal symmetric matrix
        \begin{equation}
            (\sfg_{\sfi\sfj})=\diag(\unit_2,\sigma_+\otimes\eta^{\otimes n_1},\sigma_+\otimes\eta^{\otimes n_2},\ldots)\delta^{(d)}(x-y)~,
        \end{equation}
        where
        \begin{equation}\label{eq:conj_field_pairing_matrix}
            \sigma_\pm=
            \begin{pmatrix}
                0 & 1
                \\
                \pm1 & 0
            \end{pmatrix}
        \end{equation}
        (with $\sigma_-$ defined for later use),
        $n_i$ is the tensor rank of the $i$th pair of auxiliary fields, $\eta$ is the Minkowski metric, and $x$ and $y$ are the space--time coordinates of the first and second arguments, respectively.
        
        The interactions are encoded in the \emph{flavor algebra structure constants} $\bar\sff_{\bar\sfa\bar\sfb}{}^{\bar\sfc}$ and the \emph{kinematic structure constants} $\sff_{\sfi\sfj}{}^{\sfk}$. As indicated in~\eqref{eq:raise_lower_NLSM}, the indices of these structure constants are raised and lowered with the respective metrics.\footnote{The position-space inner product $\int\rmd^dx\int\rmd^dy\,\delta^{(d)}(x-y)$ Fourier-transforms to momentum space as $(2\pi)^{-d}\int\rmd^dp\int\rmd^dq\,\delta^{(d)}(p+q)$. The difference is important: in momentum space, the structure constants $ \sff_{\sfi\sfj}{}^\sfk$ only depend on $p_2$ and $p_3$, where $p_1,p_2,p_3$ are the momenta corresponding to the indices $\sfi,\sfj,\sfk$ respectively. When the index $\sfi$ is lowered with the kinematic metric $\sfg_{\sfi\sfj}$, we get an overall $\delta^{(d)}(p_1+p_2+p_3)$, which allows the kinematic structure constants $\sff_{\sfi\sfj\sfk}$ to be totally anti-symmetric.}
        
        For our discussion, it is convenient to absorb the coupling constant $\lambda$ into the flavor algebra structure constants. Moreover, in order to have both the flavor and the kinematic algebras on an equal footing (which will be particularly useful in the discussion of the double copy to the special theory of galileons in \cref{ssec:double_copy_NLSM}), we want to ensure that their respective structure constants have matching mass dimensions. We thus identify the flavor algebra structure constants $\bar\sff_{\bar\sfb\bar\sfc}{}^{\bar\sfa}$ with the product of two space--time delta-functions and the rescaled Lie algebra structure constants $\frac{\lambda}{\sqrt{\mu}}f_{bc}{}^a$, where $\mu$ is the coupling of the special galileon. This additional factor is then compensated by regarding the $\sff_{\sfi\sfj}{}^{\sfk}$ as the structure constants for a CK-dual kinematic algebra for the non-linear sigma model, rescaled by a factor of $\sqrt{\mu}$. Note that the $\sff_{\sfi\sfj\sfk}$ are differential operators (with constant coefficients), acting on each of the three fields $\Phi^{\sfi\bar\sfa}$, $\Phi^{\sfj\bar\sfb}$, and $\Phi^{\sfk\bar\sfc}$ individually before taking the product of the results. 
        
        Finally, we recall that an overall space--time integration is implicit in the contraction of DeWitt indices (with the caveat that the color and kinematic DeWitt indices are to be taken over the same point, so that one integration drops out).
        
        The action~\eqref{eq:CK_dual_NLSM} now manifests full CK duality in the sense that, by construction, the kinematic structure constants obey the same algebraic identities as the flavor structure constants, cf.~\eqref{eq:alg_rel_NLSM}. Consequently, the amplitude integrands given directly by the Feynman diagrams of~\eqref{eq:CK_dual_NLSM} satisfy CK duality to all orders in perturbation theory.
        
        \section{Example: Yang--Mills theory}\label{sec:loop_CK_duality_YM}
        
        We now turn our attention to Yang--Mills theory. Our field theoretic setup is the Batalin--Vilkovisky (BV) formulation of Yang--Mills theory, which allows us to  work with general choices of gauge~\cite{Batalin:1981jr}. Usually, the BV formalism is introduced in situations where the gauge algebra is \emph{open}, i.e.~the gauge transformations only close on shell or, equivalently, $Q_\text{BRST}^2=0$ only on shell. In the BV formalism, we introduce a dual, second copy of each field appearing in the BRST complex; such a second copy is called an \emph{anti-field}. The BRST differential $Q_\text{BRST}$ can then be lifted to the BV differential $Q_\text{BV}$ that captures both gauge symmetries and equations of motion, allowing for $Q_\text{BV}^2=0$ also off shell.  This lift also uniquely defines the BV action of a field theory. The dual anti-fields enhance ordinary field space to a symplectic manifold similar to classical phase space, and gauge fixing amounts to a canonical transformation that selects a Lagrangian submanifold of this BV field space. For a suitably good choice of this Lagrangian submanifold, the Hessian of the kinetic term in the path integral becomes non-degenerate, and the goal of gauge fixing has been attained. For a general description of BV quantization, see e.g.~\cite{Gomis:1994he}.\footnote{We note at this point that the BV formalism also provides a direct link between classical field theories and homotopy algebras~\cite{Jurco:2018sby}.}
        
        Here, we will only use the BV formalism to impose a very general choice of gauge fixing that we can fine tune according to our needs later. This choice of gauge fixing is conveniently encoded in the \emph{gauge fixing fermion} $\Psi$, which is a functional on ordinary field space of degree $-1$. The canonical transformation then reads as
        \begin{equation}
            (\Phi^I,\Phi^+_I)\mapsto(\tilde \Phi^I,\tilde \Phi^+_I)=\left(\Phi^I,\Phi_I^++\delder[\Psi]{\Phi^I}\right)~,
        \end{equation}
        where $\Phi^I$ and $\Phi^+_I$ denote the collection of (ordinary) fields and their corresponding anti-fields. The gauge fixed action  is then the BRST action $S_\text{BRST}$ that is given by the BV action for the transformed field, restricted to vanishing anti-fields:
        \begin{equation}
            S_\text{BRST}=\left.S_\text{BV}[\tilde \Phi^I,\tilde \Phi^+_I]\right|_{\Phi^+_I=0}~.
        \end{equation}
        For a detailed exposition of our formalism, see also~\cite{Borsten:2021hua}.
        
        \subsection{Gauge-fixed Batalin--Vilkovisky action}
        
        We consider Yang--Mills theory over Minkowski space $\IM^d\coloneqq\IR^{1,d-1}$ with a semi-simple compact matrix Lie algebra $\frg$ as the gauge Lie algebra, and we use the same notation as in \cref{sec:NLSM}.
        
        \begin{table}[ht]
            \begin{center}
                \begin{tabular}{|c|l|c|c|c|c|c|c|c|}
                    \hline
                    \multicolumn{4}{|c|}{fields} & \multicolumn{3}{c|}{anti-fields}
                    \\
                    \hline
                    & role & $|-|_\text{gh}$ & dim & & $|-|_\text{gh}$ &  dim
                    \\
                    \hline
                    $c^a$ & ghost field & 1 &  $\tfrac{d}{2}-2$ & $c^{+a}$ & $-2$ & $\tfrac{d}{2}+2$ 
                    \\
                    $A_\mu^a$ & gluon field & 0 &  $\tfrac{d}{2}-1$ & $A^{+a}_\mu$ & $-1$ &  $\tfrac{d}{2}+1$
                    \\
                    $b^a$ & Nakanishi--Lautrup field & 0 & $\tfrac{d}{2}$ & $b^{+a}$ & $-1$ &  $\tfrac{d}{2}$ 
                    \\
                    $\bar c^a$ & anti-ghost field & $-1$  & $\tfrac{d}{2}$ & $\bar c^{+a}$ & 0 &  $\tfrac{d}{2}$
                    \\
                    \hline
                \end{tabular}
            \end{center}
            \caption{The full set of BV fields for Yang--Mills theory on $\IM^d$ with gauge Lie algebra $\frg$, including their ghost numbers and their mass dimensions.}\label{tab:fields:YM}
        \end{table}
        
        The BV action of Yang--Mills theory contains the fields listed in \cref{tab:fields:YM} and reads as 
        \begin{equation}\label{eq:BVActionYM}
            S_\text{BV}^\text{YM}\coloneqq\int\rmd^dx\,\Big\{\!-\tfrac14F_{a\mu\nu}F^{a\mu\nu}+A^+_{a\mu}\nabla^\mu c^a+\tfrac g2{f_{bc}}^ac^+_ac^bc^c-b^a\bar c^+_a\Big\}
        \end{equation}
        with $g$ the Yang--Mills coupling constant. All the fields are rescaled in such a way that the Yang--Mills coupling constant $g$ appears in all interaction vertices; the mass dimension of the coupling constant $g$ is $2-\frac{d}{2}$. The curvatures and covariant derivatives are defined as 
        \begin{equation}
            F^a_{\mu\nu}\coloneqq\partial_\mu A_\nu^a-\partial_\nu A_\mu^a+g{f_{bc}}^aA^b_\mu A^c_\nu
            \eand
            \nabla_\mu c^a\coloneqq\partial_\mu c^a+g{f_{bc}}^aA^b_\mu c^c~.
        \end{equation}
        
        As usual, the canonical symplectic form on the field--anti-field space induces a Poisson bracket, which in turn leads to the BV operator $Q_\text{BV}=\{S_\text{BV},-\}$ with $Q_\text{BV}^2=0$. This operator acts on individual fields as
        \begin{equation}\label{eq:BVOperatorYM}
            \begin{aligned}
                Q_\text{BV}c^a&=-\tfrac g2{f_{bc}}^ac^bc^c,~~~
                &Q_\text{BV}c^{+a}&=-\nabla^\mu A^{+a}_\mu-g{f_{bc}}^ac^bc^{+c}~,
                \\
                Q_\text{BV}A_\mu^a&=\nabla_\mu c^a~,~~~
                &Q_\text{BV}A^{+a}_\mu&=\nabla^\nu F_{\nu\mu}^a-g{f_{bc}}^aA^{+b}_\mu c^c~,
                \\
                Q_\text{BV}b^a&=0~,~~~
                &Q_\text{BV}b^{+a}&=-\bar c^{+a}~,
                \\
                Q_\text{BV}\bar c^a&=b^a~,~~~
                &Q_\text{BV}\bar c^{+a}&=0~.
            \end{aligned}
        \end{equation}
        
        As explained above, we implement gauge fixing as a canonical transformation encoded by a gauge fixing fermion, and we start from the choice
        \begin{equation}\label{eq:gaugeFixingFermionYM}
            \Psi=-\int\rmd^dx\,\bar c_a\big(\partial^\mu A^a_\mu+\tfrac{1}{2} b^a\big)
        \end{equation}
        for Feynman gauge. After gauge fixing, we perform a further field redefinition, given by the following symplectomorphism or canonical transformation on the field--anti-field space:
        \begin{equation}\label{eq:canonicalFieldRedefinitionYM}
            \begin{aligned}
                \tilde c^a&\coloneqq c^a~,~~~
                &\tilde c^{+a}&\coloneqq c^{+a}~,
                \\
                \tilde A^a_\mu&\coloneqq A^a_\mu~,~~~
                &\tilde A^{+a}_\mu&\coloneqq A_\mu^{+a}+\partial_\mu b^{+a}~,
                \\
                \tilde b^a&\coloneqq b^a+\partial^\mu A^a_\mu~,~~~
                &\tilde b^{+a}&\coloneqq b^{+a}~,
                \\
                \tilde{\bar c}^a&\coloneqq\bar c^a~,&\tilde{\bar c}^{+a}&\coloneqq\bar c^{+a}~.
            \end{aligned}
        \end{equation}
        Restricting to the Lagrangian submanifold $\Phi_I^+=0$, we thus arrive at the action  
        \begin{equation}\label{eq:YM_field_redefined_action}
            \tilde S^\text{YM}_\text{BRST}\coloneqq\int\rmd^dx\,\Big\{\tfrac12\tilde A_{a\mu}\wave\tilde A^{a\mu}-\tilde{\bar c}_a\wave\tilde c^a+\tfrac12\tilde b_a\tilde b^a\Big\}+\tilde S^\text{YM,\,int}_\text{BRST}~.
        \end{equation}
        
        \subsection{BRST-extended Hilbert space} 
        
        The tree-level scattering amplitudes of Yang--Mills theory can be expressed in terms of the incoming and outgoing gluons' momenta $(p_\mu)=(p_0,\vec p)$ and helicities. Instead of using discrete labels parameterizing the helicities, it is convenient to work with a linearly independent set of polarization vectors $\eps_\mu$ satisfying
        \begin{equation}
            (\eps_\mu)=\colvec{0,\vec\eps}~,~~~
            \vec p\cdot\vec\eps=0~,
            \eand
            |\vec\eps\,|=1~.
        \end{equation}
        
        As we will see, it will prove useful to consider states from a larger Hilbert space\footnote{or, more properly, a Krein space, as it comes with an indefinite metric}, extending the conventional Hilbert space $\frH^\text{YM}_\text{phys}$ of perturbative physical states\footnote{i.e.\ the Fock space of on-shell transverse gluons; this differs greatly from the true physical Hilbert space due to confinement} to the full BRST field space $\frH^\text{YM}_\text{BRST}$, cf.~\cite{Kugo:1977yx} or~\cite[Section 16.4]{Peskin:1995ev}. This extension, among others, includes the two additional unphysical polarizations of the gluon, the forward and backward ones, denoted by $A^{\forw\,a}_\mu$ and $A^{\backw\,a}_\mu$, respectively. For general gluons with non-zero light-like momenta, the forward polarization vector $\eps^\forw_\mu$ is the normalized momentum, and the backward polarization vector $\eps^\backw_\mu$ is obtained by reversing the spatial part,
        \begin{subequations}\label{eq:on-shell_polarisation_vectors}
            \begin{equation}
                (\eps_\mu^\forw)=\frac{1}{\sqrt{2}|\vec p\,|}\colvec{p_0,\vec{p}}
                \eand
                (\eps_\mu^\backw)=\frac{1}{\sqrt{2}|\vec p\,|}\colvec{p_0,-\vec{p}}~,
            \end{equation}
            so that
            \begin{equation}
                \eps^\forw\cdot\eps^\forw=0~,~~~
                \eps^\backw\cdot\eps^\backw=0~,
                \eand
                \eps^\forw\cdot\eps^\backw=-1~.
            \end{equation}
        \end{subequations}
        A further addition are states containing ghosts or anti-ghosts. We will be interested in scattering amplitudes for the states belonging to the extended Hilbert space $\frH^\text{YM}_\text{BRST}$. The perturbative physical S-matrix for the states in the physical Hilbert space $\frH^\text{YM}_\text{phys}$ is a restriction of the S-matrix for the states in the BRST-extended Hilbert space $\frH^\text{YM}_\text{BRST}$. Note that the latter contains transitions of physical gluons in $\frH^\text{YM}_\text{phys}$ to unphysical particles in $\frH^\text{YM}_\text{BRST}$. The restriction of the S-matrix, however, is consistent, and unitarity of the restricted S-matrix is a consequence of the BRST symmetry and of the S-matrix on $\frH^\text{YM}_\text{BRST}$ being unitary (with respect to the canonical indefinite metric on $\frH^\text{YM}_\text{BRST}$), cf.~\cite[Section~16.4]{Peskin:1995ev}.
        
        After gauge fixing, the BV transformations~\eqref{eq:BVOperatorYM} restrict to the full BRST transformations because the gauge-fixing fermion is independent of the anti-fields. We have
        \begin{equation}\label{eq:BRSTOperatorYM}
            \begin{aligned}
                Q^\text{YM}_\text{BRST}c^a&=-\tfrac12g{f_{bc}}^ac^bc^c~,~~~
                &Q^\text{YM}_\text{BRST}\bar c^a&=b^a~,
                \\
                Q^\text{YM}_\text{BRST}A^a_\mu&=\nabla_\mu c^a~,~~~
                &Q^\text{YM}_\text{BRST}b^a&=0~,
            \end{aligned}
        \end{equation}
        and $(Q^\text{YM}_\text{BRST})^2=0$ off shell. 
        
        Importantly, the linearized BRST operator $Q^\text{lin}_\text{BRST}$ still acts on $\frH^\text{YM}_\text{BRST}$. The physical (i.e.\ transversely polarized) gluon states $A^{\trans\,a}_\mu$ are singlets under the action of the linearized BRST operator, $Q^\text{YM,\,lin}_\text{BRST}A^{\trans\,a}_\mu=0$. The remaining four states form two doublets,
        \begin{equation}
            Q^\text{YM,\,lin}_\text{BRST}A^{\forw\,a}_\mu=\partial_\mu c^a
            \eand
            Q^\text{YM,\,lin}_\text{BRST}\bar c^a=b^a=\tfrac{1}{2}\partial^\mu A^{\backw\,a}_\mu~.
        \end{equation}
        This is most easily seen in the momentum representation.
        
        \subsection{On-shell color--kinematics duality on the BRST-extended Hilbert space}
        
        As mentioned in \cref{ssec:CKDuality}, CK duality of the Yang--Mills tree-level scattering amplitudes is well-established. In order to lift CK duality to the loop level using \cref{ob:tree_is_good_enough}, we first must extend CK duality to the extended BRST Hilbert space introduced above\footnote{See~\cite{Mafra:2014gja} for a distinct  approach to establishing CK duality respecting loop-level amplitudes using the  pure spinor superstring BRST cohomology.}. The heuristic rationale is clear. We would like to glue  the tree-level CK-dual diagrams into loop-level CK-dual  diagrams. In the latter, however, unphysical modes run inside the loops, so we must allow them to also live on the external lines of the tree-level diagrams prior to gluing. As shown in~\cite{Borsten:2020zgj,Borsten:2021hua}, this can be achieved via field redefinitions and the on-shell BRST Ward identities, and we will review these two ingredients in some detail in the following.  
        
        The first step is to ensure CK duality for scattering amplitudes with arbitrarily polarized gluons on external legs. The key insight here is that the amputated amplitude with non-transverse  gluons is \textit{not} gauge-invariant and so can be adjusted through the choice of gauge fixing. 
        
        First, forward-polarized gluons can be absorbed by residual gauge transformations, so they cannot affect CK duality. For backward-polarized gluons, CK duality violating terms can be corrected by the introduction of terms $S_\text{corr}$ in the action that are proportional to $\partial^\mu A_\mu^a$ and do not contain the Nakanishi--Lautrup field $b^a$. Clearly, we can produce such terms by performing a shift of the Nakanishi--Lautrup field
        \begin{equation}
            b^a\mapsto b^a+Z^a
        \end{equation}
        with $Z^a$ at least of cubic order in the fields. This shift does not introduce a Jacobian functional determinant, and in order to avoid the introduction of new interaction terms linear in $b^a$, we adjust our gauge by shifting the gauge-fixing fermion by
        \begin{equation}
            \Psi\mapsto\Psi+\Xi\ewith Z_a-\delder[\Xi]{\bar c^a}=0~,
        \end{equation}
        cf.~\cite{Borsten:2021hua} for more details. Both operations clearly leave the scattering amplitudes of our theory invariant, and we can conclude the following.
        \begin{observation}\label{ob:longitudinal_states}
            Terms in an action that are proportional to $\partial^\mu A_\mu^a$ can be absorbed by a field redefinition of the Nakanishi--Lautrup field as well as choice of gauge.
        \end{observation}
        \noindent
        More specifically:
        \begin{observation}\label{ob:lift_to_BRST_extended_ghost_no_0}
            Let $S$ be a BRST action of Yang--Mills theory whose Feynman rules produce on-shell CK-dual scattering amplitudes for $0<k<n$ external physically polarized gluons and $n-k$ external gluons of arbitrary polarization. Then there is a field redefinition and a change of gauge such that the Feynman rules of the resulting action $\tilde S$ produce on-shell CK-dual scattering amplitudes for $n-k+1$ arbitrarily polarized gluons.
        \end{observation}
        
        The change of gauge as well as the shift of the Nakanishi--Lautrup field modify the BRST operator $Q_\text{BRST}$. Its action on the states in the BRST-extended Hilbert space, however, is determined by its linearization $Q_\text{BRST}|_{g=0}$, which remains unmodified. We thus have the on-shell BRST Ward identities at our disposal, which follow from the BRST invariance of the vacuum:
        \begin{equation}\label{eq:on_shell_Ward}
            0=\langle 0|[Q^\text{YM,\,lin}_\text{BRST},{\scO_1\cdots\scO_n}]|0\rangle~.
        \end{equation}
        Considering the special case 
        \begin{equation}
            \scO_1\cdots\scO_n=A^\forw\bar c(c\bar c)^k A^\trans_1\cdots A^\trans_{n-2k-2}
        \end{equation}
        with $A^{\forw\,a}_\mu$ forward-polarized and all other gluons of physical polarization, we obtain
        \begin{equation}
            \langle 0|(c\bar c)^{k+1}A_1^\trans\cdots A_{n-2k-2}^\trans|0\rangle\ \sim\ \langle 0|A^\forw(c\bar c)^kbA_1^\trans\cdots A_{n-2k-2}^\trans|0\rangle~,
        \end{equation}
        and we conclude:
        \begin{observation}\label{ob:onshell_Ward_identities}
            Any scattering amplitude in the BRST-extended Hilbert space with $k+1$ ghost--anti-ghost pairs and all gluons transversely polarized is given by a sum of scattering amplitudes with $k$ ghost pairs. 
        \end{observation}
        As a result of this observation, we have an iteratively induced CK-dual parameterization of the scattering amplitudes with $k+1$ ghost--anti-ghost pairs from the CK-dual form of the scattering amplitudes with $k$ ghost--anti-ghost pairs, starting from $k=0$. This parameterization is obtained by literally copying the relevant cubic Feynman diagrams for $k$ and replacing the two unphysically polarized gluon modes by a ghost--anti-ghost pair. We can then use \cref{ob:TW-terms} to obtain an action whose Feynman rules produce these parameterizations.
        \begin{observation}\label{ob:ghost_pair_reduction}
            If the Feynman rules for a Yang--Mills theory action produces CK-dual $n$-point tree-level scattering amplitudes for $k$ ghost--anti-ghost pairs, then its $n$-point tree-level scattering amplitudes are also CK-dual for $k+1$ ghost--anti-ghost pairs.
        \end{observation}
        Starting from a Yang--Mills action whose Feynman rules yield a CK-dual parameterization of tree-level scattering amplitudes for physical gluons, we can use iteratively \cref{ob:lift_to_BRST_extended_ghost_no_0,ob:ghost_pair_reduction} to create an action whose Feynman rules produce CK-dual parameterizations of scattering amplitudes for the whole BRST-extended Hilbert space. CK duality for non-transverse gluons amounts to a gauge choice, which then implies CK duality for ghosts by the BRST symmetry.
        
        \subsection{Lifting on-shell color--kinematics duality to the loop level}\label{ssec:offshell_lift_YM}
        
        We are now in the same situation as in the case of the non-linear sigma model in \cref{ssec:offshell_lift_NLSM}: CK duality holds on shell, and any violation can be compensated by terms that can be produced by field redefinitions. The field redefinitions necessary for $n$-point scattering amplitudes, however, will affect the on-shell scattering amplitudes at higher points. Therefore, we will refine the algorithm presented in \cref{ssec:offshell_lift_NLSM} as follows:
        \begin{enumerate}[label=(\roman*)]\itemsep-2pt
            \item\label{item:choice} Choose a CK-dual parameterization of the tree-level scattering amplitudes for physical (transversely polarized) gluons, and pick a set of master color numerators $\bmc^{(n)}_{\rm m}$ for all numbers $n$ of external legs we are interested in, cf.~\cref{ssec:primaries}.
            \item Set $S^\text{off-shell}_{3}=\tilde S^\text{YM}_\text{BRST}$ as given in~\eqref{eq:YM_field_redefined_action}. Then proceed with the algorithm starting at $n=4$.
            \item\label{item:tolotti_YM} Use~\cref{ob:iterative_TW_terms} to add a term that is identically zero to the action $S_{n-1}$ such that its Feynman rules produce the chosen form of the CK-dual tree-level scattering amplitudes up to $n$~points. Note that the thus produced action $S^\text{on-shell}_{n}$ generically contains non-local terms of the type $E_1^M\frac{1}{\wave}E^2_M$, cf.~\cite{Bern:2010ue,Tolotti:2013caa}.
            \item In the following step, we extend CK duality for amplitudes to amputated correlators on the BRST-extended Hilbert space. We do this by iterating over the number $k$ of ghost--anti-ghost pairs, starting at $k=0$.
            \begin{enumerate}[label=(\alph*),ref=(\roman{enumi}.\alph*)]\itemsep-2pt
                \item\label{item:inner_iteration} Prepare the action $S^\text{on-shell}_{n,k}$. For $k=0$, this action is simply $S^\text{on-shell}_{n,k}=S^\text{on-shell}_{n}$. For $k>0$, we use \cref{ob:ghost_pair_reduction} to rewrite the ghost sector in such a way that the tree-level amplitudes are given in CK-dual form. 
                \item Compute the off-shell master numerators $\hat\bmn^{(n)}_{\rm m}$ for arbitrary external legs that are unrestricted in their polarization, momentum and species, up to the fact that precisely $k$ ghosts and $k$ anti-ghosts are present. Extend these to a full set of numerators $\hat \bmn^{(n)}=\bmJ^{(n)}\hat\bmn^{(n)}_{\rm m}$ and further to the amputated correlator 
                \begin{equation}
                    \hat\scA'_{n,0}=\left[\bmc^{(n)\sfT}\bmD^{(n)}\hat\bmn^{(n)}\right]_\sigma~,
                \end{equation}
                cf.~\eqref{eq:amputated_correlator_NLSM}. The latter agrees with the actual amputated correlator $\hat\scA_{n,0}$ of Yang--Mills theory for external states that are physical (on-shell) external states in the BRST-extended Hilbert space. Any difference between $\hat\scA_{n,0}$ and $\hat\scA'_{n,0}$ must therefore be due to the unphysical polarization of a gluon or to the fact that a gluon is off shell. Contributions of forward-polarized gluons can be absorbed by residual gauge transformations, and therefore the difference must be given by terms proportional to $\eps_i\cdot p_i$ or $p_i^2$, where $\eps_i$ and $p_i$ are polarization and momentum vectors associated with the external leg~$i$. We note that potential differences due to unphysical polarizations can only appear for $k=0$, as the rewriting of the ghost sector in~\ref{item:inner_iteration} yields on-shell CK duality for the full BRST-extended Hilbert space. Any difference between $\hat\scA_{n,0}$ and $\hat\scA'_{n,0}$ for $k>0$ is thus proportional to $p_i^2$.
                \item Any potential difference proportional to $\eps_i\cdot p_i$ is removed by introducing appropriate terms into the action that are proportional to $\dpar^\mu A_\mu^a$. This can be done using~\cref{ob:lift_to_BRST_extended_ghost_no_0}, i.e.~by a field redefinition of the Nakanishi--Lautrup field $b^a$ and a subsequent change of gauge. 
                \item Any difference in the amputated correlators proportional to $p^2_i$ can now be absorbed by a field redefinition, just as in the case of the non-linear sigma model. Explicitly, such expressions can be canceled by adding interaction terms of the form $\tilde F_i\wave\Phi_i$, where $\Phi_i\in\{A_\mu^a,c^a,\bar c^a\}$ and $\tilde F_i$ is a polynomial of $(n-1)$-st order in the fields containing derivatives as well as the operator $\frac{1}{\wave}$. The color indices are fully contracted by the structure constants of the gauge Lie algebra. Such interaction terms can be produced by a field redefinition of the form
                \begin{equation}
                    \Phi_i\mapsto\Phi_i+\sum_i \tilde F_i~.
                \end{equation}            
                \item If $2k\leq n-2$, increase $k$ by one and return to~\ref{item:inner_iteration}. 
            \end{enumerate}
            \item If we have not reached the maximum order that is of relevance for the scattering amplitudes we are interested in, then increment $n$, and go back to~\ref{item:tolotti_YM}. Otherwise, continue to~\ref{item:strictify2}.
            \item\label{item:strictify2} Use \cref{ob:strictification} to strictify the order~$k$ interaction vertices (including those containing ghosts and gauge-fixing terms) in the action $S^\text{off-shell,\,non-local}_n$ for $k\le n$.
        \end{enumerate}
        The above algorithm proves that on-shell CK duality can be lifted off shell and, a~fortiori, loop-level CK duality for all diagrams, as always up to potential counterterms. Moreover, just as in the case of the non-linear sigma model, off-shell CK duality extends to Feynman diagrams having auxiliary fields on external legs, cf.~the discussion in \cref{ssec:offshell_lift_NLSM}. This completes the proof of off-shell CK duality for Yang--Mills theory up to potential counterterms.

        Importantly, the algorithm requires tree-level computations only, avoiding all need to tackle the loop-level CK duality relations directly. The most direct approach to the latter involves super-exponentially growing ans\"atze, which quickly become computationally expensive. This tree-level-only route, therefore, potentially offers computational advantages, particularly in the context of the various applications of the double copy and especially when combined with the homotopy-algebraic framework that we are currently developing~\cite{Borsten:2022aa}.
        
        \subsection{Color--kinematics duality manifesting action}\label{ssec:CK_manifest_YM}
        
        The preceding discussion implies the existence of a  strict, local Yang--Mills BRST action with manifest exact CK duality. Analogously to \cref{ssec:CK_dual_NLSM}, the factorization of the fields and kinematic and interaction terms into the form
        \begin{equation}
            \mbox{color}~\otimes~\mbox{kinematics}~\otimes~\mbox{scalar field}
        \end{equation}
        is most conveniently described by pairs of DeWitt indices over a common space--time point. We use again $\sfi,\sfj,\ldots$ and $\bar\sfa,\bar\sfb,\ldots$ for the kinematic and color DeWitt indices, respectively. A detailed discussion of the factorization is found in~\cite[Section 9.3]{Borsten:2021hua}.
        The action obtained from our algorithm is then of the form
        \begin{equation}\label{eq:leftYM}
            S^\text{YM}_\text{BRST,\,CK-dual}=\tfrac12\sfg_{\sfi\sfj}\bar{\sfg}_{\bar\sfa\bar\sfb}\caA^{\sfi\bar\sfa}\wave\caA^{\sfj\bar\sfb}+\tfrac1{3!}\sff_{\sfi\sfj\sfk}\bar\sff_{\bar\sfa\bar\sfb\bar\sfc}\caA^{\sfi\bar\sfa}\caA^{\sfj\bar\sfb}\caA^{\sfk\bar\sfc}~,
        \end{equation}
        and CK duality amounts again to the fact that the color and kinematical data
        \begin{equation}
            (\,\bar\sfg_{\bar\sfa\bar\sfb}\,,\,\bar\sff_{\bar\sfb\bar\sfc}{}^{\bar\sfa}\,)
            \eand
            (\,\sfg_{\sfi\sfj}\,,\,\sff_{\sfj\sfk}{}^\sfi\,)
        \end{equation}
        with
        \begin{equation}\label{eq:raise_lower_SYM}
            \sff_{\bar\sfa\bar\sfb\bar \sfc}=\bar\sfg_{\bar \sfc\bar \sfd}\bar\sff_{\bar\sfa\bar\sfb}{}^{\bar\sfd}
            \eand
            \sff_{\sfi\sfj\sfk}=\sfg_{\sfk\sfl}\sff_{\sfi\sfj}{}^\sfk
        \end{equation}
        have the same algebraic properties, namely anti-symmetry and the Jacobi identity of the Lie algebra structure constants as well as symmetry and invariance of the metric:
        \begin{equation}\label{eq:alg_rel_YM}
            \begin{aligned}
                \bar\sfg_{[\bar\sfa\bar\sfb]}&=0~,~~~
                \bar\sff_{(\bar\sfb\bar\sfc)}{}^{\bar\sfa} &=0~,~~~
                &\bar\sff_{\bar\sfe[\bar\sfa}{}^{\bar \sfd}\sff_{\bar\sfb\bar\sfc]}{}^{\bar\sfe}&=0~,~~~
                &\bar\sfg_{\bar\sfa(\bar\sfb}\sff_{\bar\sfc)\bar\sfe}{}^{\bar\sfa}=0~,
                \\
                \sfg_{[\sfi\sfj]}&=0~,~~~
                \sff_{(\sfj\sfk)}{}^\sfi &=0~,~~~
                &\sff_{\sfn[\sfi}{}^\sfm\sff_{\sfj\sfk]}{}^\sfn&=0~,~~~
                &\sfg_{\sfi(\sfj}\sff_{\sfk)\sfm}{}^\sfi=0~.
            \end{aligned}
        \end{equation}
        As before, $(-)$ and $[-]$ denote total symmetrization and antisymmetrization of enclosed indices, as usual. 
        
        To summarize, the $\sff_{\sfj\sfk}{}^\sfi$ are the structure constants of an infinite-dimensional \emph{kinematic  Lie algebra},  generalizing the kinematic algebras~\cite{Monteiro:2011pc, Bjerrum-Bohr:2012kaa, Monteiro:2013rya, Chen:2019ywi,Chen:2021chy} to the complete BRST field space.\footnote{See also~\cite{Frost:2020eoa} for a Lie algebra structure on the data of the tree-level amplitudes directly related to the BCJ amplitude relations.} This provides a natural field theoretic notion of CK duality, as opposed to one based on on-shell amplitudes. CK duality is now a manifest property of the BRST action. Consequently, the loop integrands computed directly from the corresponding Feynman rules will satisfy CK duality automatically. CK duality of the loop integrands may be anomalous since the Jacobian determinants generated by the field redefinitions induce counterterms that may break it. However, CK duality of the BRST action itself   is not only a very natural interpretation of the concept; it also suffices for the double copy.
        
        Let us unpack the above. As the Nakanishi--Lautrup field does not participate in the interactions, we can integrate it out, which will also prove to be useful in the discussion of the double copy later. The field content is then
        \begin{subequations}
            \begin{equation}\label{eq:YM_field}
                (\caA^{\sfi\bar\sfa})=\big(A^{a}_\mu(x),\bar c^{a}(x),c^{a}(x),Y^{m{a}}(x)\big)~,
            \end{equation}
            containing the infinite list of auxiliary fields
            \begin{equation}
                \big(Y^{ma}(x)\big)=\big(Y_{\mu\nu\rho}^{a}(x),Y^{a}_{\mu\nu}(x),\bar Y^{a}_{\mu\nu}(x),\ldots\big)~.
            \end{equation} 
        \end{subequations}
        The index $m$ ranges over the various auxiliary fields, and all other indices on the component fields are color or Lorentz indices with space--time dependence made manifest.
        
        All structure constants are relatively evident analogues of those encountered in \cref{ssec:CK_dual_NLSM}, and more detailed explanations are found in~\cite[Section 9.3]{Borsten:2021hua}. The \emph{color metric} $\bar\sfg_{\bar\sfa\bar\sfb}$ is the invariant quadratic form on the gauge Lie algebra $\frg$ with accompanying space--time delta functions. The \emph{kinematic metric} $\sfg_{\sfi\sfj}$ can be described explicitly by the block-diagonal graded-symmetric matrix
        \begin{equation}\label{eq:kin_metric_YM}
            (\sfg_{\sfi\sfj})=\diag(\eta,\sigma_-,\sigma_{s_1}\otimes\eta^{\otimes n_1},\sigma_{s_2}\otimes\eta^{\otimes n_2},\ldots)\,\delta^{(d)}(x-y)~,
        \end{equation}
        where $\sigma_\pm$ was defined in~\eqref{eq:conj_field_pairing_matrix}, $n_i$ is the tensor rank of the $i$th pair of auxiliary fields, $s_i\in\{\pm\}$ is the ghost parity of the $i$th pair of auxiliary fields, $\eta$ is the Minkowski metric, and $x$ and $y$ are the space--time coordinates of the first and second arguments, respectively. Upon restricting to the field subspace parameterized by $(A^a_\mu,\bar c^a, c^a)$, the kinematic metric $\sfg_{\sfi\sfj}$ restricts to the matrix
        \begin{equation}\label{eq:kin_metric}
            \begin{pmatrix}
                \eta^{\mu\nu} & 0 & 0
                \\
                0 & 0 & 1
                \\
                0 & -1 & 0
            \end{pmatrix}\delta^{(d)}(x-y)~.
        \end{equation}
        The rest of $\sfg_{\sfi\sfj}$ is simply a constant, graded-symmetric matrix pairing the conjugate strictification auxiliary fields, i.e.
        \begin{equation}
            \tfrac12\sfg_{\sfi\sfj}\bar\sfg_{\bar\sfa\bar\sfb}\caA^{\sfi\bar\sfa}\wave\caA^{\sfj\bar\sfb}=\int\rmd^dx\,\left(\cdots+Y_{\mu\nu\rho}\wave\bar Y^{\mu\nu\rho}+Y_{\mu\nu}\wave\bar Y^{\mu\nu}+\cdots\right)~,
        \end{equation}    
        where the grading accounts for the ghost degrees of the auxiliary fields.
        
        We can use the color and kinematic metrics and their respective inverses to raise and lower the indices on the \emph{color algebra structure constants} $\bar\sff_{\bar\sfa\bar\sfb}{}^{\bar\sfc}$ and the \emph{kinematic algebra structure constants} $\sff_{\sfi\sfj}{}^{\sfk}$. 
        
        We identify the $\bar\sff_{\bar\sfa\bar\sfb}{}^{\bar\sfc}$ with products of two space--time delta functions with rescaled\footnote{Just as in \cref{ssec:CK_dual_NLSM}, this rescaling ensures that color and kinematic Lie algebras are truly on an equal footing with structure constants of matching mass dimensions, making the discussion of the double copy in \cref{ssec:YM_double_copy} more convenient.} color Lie algebra structure constants $g\sqrt{\frac{2}{\kappa}}f_{ab}{}^c$, where $\kappa=4\sqrt{2\pi G}$ is Einstein's gravitational constant. This rescaling is compensated by a corresponding rescaling by $\sqrt{\frac{\kappa}{2}}$ of the kinematic algebra structure constants $\sff_{\sfi\sfj}{}^{\sfk}$. Note that in the action, the $\sff_{\sfi\sfj\sfk}$ are differential operators acting on each field $\Phi^{\sfi\bar\sfa}$, $\Phi^{\sfj\bar\sfb}$, and $\Phi^{\sfk\bar\sfc}$ individually before taking the product of the results.        
        
        As usual, repeated DeWitt indices involve a space--time integration. Note, however, that since we are considering DeWitt indices over the same point, there is only one integration for each pair of color and kinematic DeWitt indices.
        
        The action~\eqref{eq:leftYM} now manifests full CK duality in the sense that, by construction, the kinematic structure constants obey the same algebraic identities as the color structure constants, cf.~\eqref{eq:alg_rel_YM}. Consequently, the amplitude integrands given directly by the Feynman diagrams of~\eqref{eq:leftYM} satisfy CK duality to all orders in perturbation theory.         
        
        Already here, we note that the action of the BRST operator can similarly be written as
        \begin{equation}\label{eq:leftQ}
            Q_\text{BRST}\caA^{\sfi\bar\sfa}=\sfq^\sfi_\sfj\delta^{\bar\sfa}_{\bar\sfb}\caA^{\sfj\sfb}+\tfrac12\sfq_{\sfj\sfk}^\sfi\bar\sff_{\bar\sfb\bar\sfc}{}^{\bar\sfa}\caA^{\sfj\bar\sfb}\caA^{\sfk\bar\sfc}~.
        \end{equation}
        This observation will be helpful in discussing the BRST--Lagrangian double copy in \cref{ssec:YM_double_copy}. In~\eqref{eq:leftQ}, $\sfq^\sfi_\sfj$ is a linear differential operator, and $\sfq_{\sfj\sfk}^\sfi$ is a bidifferential operator, both of ghost degree~$1$. For example, the  restriction of $\sfq^\sfi_\sfj$ to the $(A^a_\mu,\bar c^a,c^a)$ subspace is given by the matrix
        \begin{equation}
            \begin{pmatrix}
                0 & 0 & \partial_\mu
                \\
                -\partial^\nu & 0& 0
                \\
                0 & 0 & 0
            \end{pmatrix}~.
        \end{equation}
        Again, a more detailed discussion of the structure of the BRST operator is found in~\cite[Section 9.3]{Borsten:2021hua}. 
        
        \subsection{Supersymmetric Yang--Mills theories}\label{sec:SYM}
        
        To underline the generality of our lift of CK duality from on-shell tree to off-shell loop level, let us briefly consider the extension to supersymmetric Yang--Mills theories. This is particularly relevant as our recent result~\cite{Borsten:2022vtg} strongly suggests that the potentially problematic counterterms can be avoided in our procedure for $\caN=4$ super Yang--Mills theory.
        
        Again, it is well-established that the scattering amplitudes of supersymmetric Yang--Mills theory can be parameterized in a CK-dual manner, cf.~\cite{BjerrumBohr:2009rd,Stieberger:2009hq} from string theoretic arguments as well as~\cite{Jia:2010nz} for a field theoretic argument based on recursion relations. Instead of relying on these arguments, we could have simply employed the supersymmetric on-shell Ward identities, similar to \cref{ob:ghost_pair_reduction}, in order to proof that a CK-dual parameterization of the physically polarized Yang--Mills scattering amplitudes at the tree level implies the existence of a CK-dual parameterization of all tree-level scattering amplitudes in supersymmetric Yang--Mills theory. Evidently, this argument applies only in the case of irreducible supersymmetric Yang--Mills multiplets. Even in this case, if one supersymmetrically adds higher-dimensional operators then this argument can fail. For example, $\caN=2$ supersymmetric Yang--Mills theory with a $\sim(\phi F^2+\text{fermions})$ deformation  has non-trivial superamplitudes where the all gluon components are identically zero.\footnote{We are grateful to Henrik Johansson for bringing this  possibility to our attention.}
        
        Once the existence of a parameterization of the scattering amplitudes in terms of Feynman diagrams that exhibit CK duality has been established, we can follow again our algorithm from \cref{ssec:offshell_lift_YM} to extend it first to the BRST-extended Hilbert space and then further to the general tree-level Feynman diagrams. We note that the kinematic terms of fermionic fields $\psi$ are of the form
        \begin{equation}
            S^\text{Dirac}=\int\rmd^dx\,\bar\psi\slashed{\partial}\psi~,
        \end{equation}
        where $\slashed{\partial}\coloneqq\gamma^\mu\partial_\mu$ with $\gamma^\mu$ the gamma matrices. Using field redefinitions, we can thus produce both the unphysical polarizations $\varepsilon$ that violate $\slashed p\varepsilon=0$ as well as off-shell violations proportional to $\wave \psi$. All other arguments remain the same,\footnote{An alternative, slightly more involved argument can be constructed using (fractional) off-shell supersymmetry together with off-shell supersymmetric Ward identities.} and we will present some more details in \cref{ssec:DC_SUSY}.
        
        Because the loop integrands in Feynman diagrams behave nicely under dimensional reduction, we can start with $\caN=1$ supersymmetric Yang--Mills theory in any space--time dimension $d=3,4,6,10$, establish full off-shell CK duality, and then infer the same statement for theories with extended supersymmetry.
        
        \subsection{No-go theorems about the kinematic algebra}\label{ssec:coleman-mandula}
        
        After the Yang--Mills action has been recast into the cubic form~\eqref{eq:leftYM}, color and kinematics are on an equal footing. In particular, the kinematic algebra, like that for color, is a Lie algebra equipped with an invariant inner product. It is infinite-dimensional (because of the DeWitt indices), but otherwise it is just another symmetry of the action, in apparent contradiction with well-known no-go theorems about exotic symmetries. In particular, the Coleman--Mandula theorem~\cite{Coleman:1967ad} states that, under certain weak assumptions\footnote{and leaving out supersymmetry}, the most general symmetry algebra of a quantum field theory that leaves its S-matrix invariant is the direct sum of the space--time symmetry algebra and a finite-dimensional Lie algebra. In other terms, space--time symmetry cannot combine with internal symmetry in any non-trivial way. Also, the Weinberg--Witten theorem~\cite{Weinberg:1980kq} states that, under certain weak assumptions, a theory with conserved vector currents cannot contain massless higher-spin particles. The reader may now wonder whether the Yang--Mills action manifesting the kinematic algebra violates these theorems. After all, the kinematic algebra is a continuous symmetry whose generators carry non-trivial Lorentz representations; and the reformulated Yang--Mills action contains auxiliary fields with multiple Lorentz indices, indicative of high helicity.
        
        Potential contradictions can only arise if we include the strictification auxiliaries in the scattering states on which the S-matrix acts. If we do not include the auxiliaries on external legs, that is, if we regard them as mere composites of the gluons, ghosts, and anti-ghosts, then the kinematic algebra is, strictly speaking, not a symmetry of the S-matrix insofar as the auxiliary fields transform under the kinematic algebra in a way inconsistent with their decomposition into gluons etc. Hence, the Coleman--Mandula and Weinberg--Witten theorems simply do not apply. We can, however, include the strictification auxiliaries on external legs and formally regard them as independent fields. In this case, the Coleman--Mandula and Weinberg--Witten theorems break down because the kinematic metric for the auxiliaries is not positive definite in the bosonic sector, cf.~\eqref{eq:kin_metric_YM}.\footnote{Another obvious loophole is that the cubic form of the action contains an infinite number of fields, in contradiction with the assumptions in the Coleman--Mandula theorem. But this proves a red herring: one can truncate the theory to only those auxiliaries that strictify (for example) the quartic interaction --- this is consistent if the theory does not contain gauge symmetry, like the non-linear sigma model --- and then one obtains a fully CK-dual action with finitely many fields.} We note that both no-go theorems are readily violated in theories without positive-definiteness of the Hilbert space norm. For example, it is quite straightforward to couple a gauge symmetry to a vector field $V^{a\mu}$ whose kinetic term is $V^{a\mu}\wave V_{a\mu}$.
        
        \section{Double copy and generalizations}\label{sec:double_copy}
        
        As we show in the following, we can double-copy a perfect-CK duality-manifesting parent action, yielding an action ${S}^\text{DC}_\text{BRST}$ that is, under some mild assumptions, invariant with respect to the double-copied BRST operator ${Q}^\text{DC}_\text{BRST}$. Working with  a parent action  in  Feynman gauge to quadratic order, the double copy action ${S}^\text{DC}_\text{BRST}$   is clearly consistently  gauge-fixed.  The existence of  a nilquadratic ${Q}^\text{DC}_\text{BRST}$ annihilating   ${S}^\text{DC}_\text{BRST}$ then implies that we have consistent BRST action ready for quantization. In the case of a parent Yang--Mills theory, the off-spring  is weakly semi-classically equivalent to $\caN=0$ supergravity, essentially by construction. Then, by the existence of ${Q}^\text{DC}_\text{BRST}$, there is a choice of quantization such that this theory is fully quantum equivalent to  $\caN=0$ supergravity to all orders in perturbation theory. From this point of view, the field theory notion of CK duality is particularly natural; it is a symmetry that makes the validity of the double copy  transparent. In particular,  diffeomorphisms, as well as any other required  local symmetries, arise as the double copy of gauge symmetries.
        
        This streamlines the arguments for a loop-level double copy procedure which we gave in~\cite{Borsten:2020zgj,Borsten:2021hua}, where the starting theory satisfies CK duality only on shell and, correspondingly, ${Q}^\text{DC}_\text{BRST}{S}^\text{DC}_\text{BRST}$ will be proportional to terms that vanish when imposing the equations of motion. Moreover, as our discussion showed that tree-level CK duality implies full CK duality up to counterterms, our observation directly leads to a growing family of consistent, double-copy constructible theories.
        
        The standard argument to extend the validity of the double copy prescription to the loop level relies on generalized unitarity~\cite{Bern:2010yg}. Anticipating the discussion in \cref{sec:conclusion}, we remark here that in our approach counterterms are needed to restore unitarity for CK-dual loop integrands. The violation of unitarity before the introduction of counterterms prevents us from following the loop-level proof of~\cite{Bern:2010yg}. As summarized above, however, our Lagrangian approach affords an alternative and independent demonstration of the double copy.
        
        \subsection{BRST--Lagrangian double copy and syngamies}\label{ssec:BRST_double_copy}
        
        In order to explain the general procedure, we start from a general cubic theory as obtained by the strictification procedure explained in \cref{ssec:strictification}. We directly specialize the form of the cubic action~\eqref{eq:simple_cubic_action} as follows:
        \begin{equation}\label{eq:cubic_action_standard_form}
            S=\tfrac12\sfG_{IJ}\Phi^I\wave \Phi^J+\tfrac{1}{3!}\sfF_{IJK}\Phi^I\Phi^J\Phi^K~,
        \end{equation}
        where, as before, $I,J$ are DeWitt indices and $\sfG_{IJ}$ and $\sfF_{IJK}$ are some structure constants. Recall that by \cref{ob:strictification}, any Poincar\'e-invariant field theory can be cast in this form (after gauge fixing if necessary), even with the introduction of the wave operator in the kinematic term.
        
        If the theory is invariant under a gauge symmetry, this will be encoded in the action of a BRST operator $Q_\text{BRST}$. As mentioned in \cref{ssec:strictification}, it requires some serious modifications of the theory to cast the BRST operator in a cubic ansatz; we therefore allow for a generic ansatz 
        \begin{equation}\label{eq:cubic_BRST_standard_form}
            Q_\text{BRST}\Phi^I=\sfQ_J^I\Phi^J+\tfrac12\sfQ_{JK}^I\Phi^J\Phi^K+\tfrac1{3!}\sfQ_{JKL}^I\Phi^J\Phi^K\Phi^L+\cdots~,
        \end{equation}
        in which $\sfQ^I_{JK\dotsm}$ is a multilinear differential operator acting on the subsequent fields individually, before the product of the results is taken. Poincar\'e invariance restricts the field theory such that all the structure constants $\sfG_{IJ}$, $\sfF^I_{JK}$, $\sfQ^I_J$, etc.~introduced are constructed from constant coefficient differential operators and have no further space--time dependence.\footnote{In particular, since they have constant coefficients, they commute on the flat Minkowski space--time that we consider.} 
        
        We further specialize to theories whose fields split into left and right components over common space--time points similar to the factorization~\eqref{eq:typical_factorization} exhibiting CK~duality. This allow us to express the DeWitt indices as $I=(\alpha, \bar\alpha)$, $J=(\beta, \bar\beta)$ etc., which are, just as in the special cases discussed in \cref{ssec:CK_dual_NLSM} and \cref{ssec:CK_manifest_YM} to be regarded over a common space--time point. We can thus write
        \begin{equation}
            \begin{gathered}
                G_{IJ}=\sfg_{(\alpha, \bar\alpha);(\beta,\bar\beta)} \eqqcolon \sfg_{\alpha\beta} \,\bar\sfg_{\bar\alpha\bar\beta}~,
                \\
                F_{IJK}=\sff_{(\alpha, \bar\alpha);(\beta,\bar\beta);(\gamma,\bar\gamma)} \eqqcolon \sff_{\alpha\beta\gamma}\,\bar\sff_{\bar\alpha\bar\beta\bar\gamma}~,
                \\
                \sff_{\alpha\beta\gamma}=\sfg_{\gamma\delta}\sff_{\alpha\beta}{}^{\delta}~,\eand
                \bar\sff_{\bar\alpha\bar\beta\bar\gamma}=\bar\sfg_{\bar\gamma\bar\delta}\bar\sff_{\bar\alpha\bar\beta}{}^{\bar\delta}~.
            \end{gathered}
        \end{equation}
        We attach no specific meaning to the indices $\alpha$ and $\bar\alpha$ here. In the examples of the non-linear sigma model considered in~\cref{ssec:CK_dual_NLSM}, $\bar\alpha$ is identified with the flavor DeWitt index $\bar\sfa$, and $(\bar\sfg_{\bar\alpha\bar\beta},\bar\sff_{\bar\alpha\bar\beta}{}^{\bar\gamma})$ specialize to the flavor metric and structure constants $(\bar\sfg_{\bar\sfa\bar\sfb},\bar\sff_{\bar\sfa\bar\sfb}{}^{\bar\sfc})$. In the case of Yang--Mills theory, considered in~\cref{ssec:CK_manifest_YM}, the index $\bar\alpha$ is identified with the color DeWitt index $\bar\sfa$, and the metric and structure constants are those of the color algebra. In both cases, $\alpha$ is identified with the kinematic DeWitt index $\sfi$,  and $(\sfg_{\alpha\beta},\sff_{\alpha\beta}{}^{\gamma})$ are the metric and structure constants of the kinematic algebra.
        
        The action~\eqref{eq:cubic_action_standard_form} becomes 
        \begin{subequations}\label{eq:left_theory}
            \begin{equation}\label{eq:cubic_action_decomposition}
                S=\tfrac12\sfg_{\alpha\beta}\,\bar\sfg_{\bar\alpha\bar\beta}\Phi^{\alpha\bar\alpha}\wave\Phi^{\beta\bar\beta}+\tfrac1{3!}\sff_{\alpha\beta\gamma}\,\bar\sff_{\bar\alpha\bar\beta\bar\gamma}\Phi^{\alpha\bar\alpha}\Phi^{\beta\bar\beta}\Phi^{\gamma\bar\gamma}~,
            \end{equation}
            and (full) CK duality is tantamount to the same algebraic relations\footnote{i.e.~anti-symmetry and the Jacobi identity of the Lie algebra structure constants as well as symmetry and invariance of the metric} being satisfied by $(\sfg_{\alpha\beta},\sff_{\alpha\beta}{}^{\gamma})$ and $(\bar\sfg_{\bar\alpha\bar\beta},\bar\sff_{\bar\alpha\bar\beta}{}^{\bar\gamma})$.
            
            Given the factorization of the indices and its linearity, it is natural to assume that the BRST operator factorizes similarly,
            \begin{equation}\label{eq:left_BRST}
                \sfQ^I_J=\sfq^\alpha_\beta\delta^{\bar\alpha}_{\bar\beta}+\delta^\alpha_\beta\bar\sfq^{\bar\alpha}_{\bar\beta}~,
                \quad
                \sfQ^{I}_{JK}=\sff_{\beta\gamma}{}^\alpha\,\bar\sfq^{\bar\alpha}_{\bar\beta\bar\gamma}+\sfq^\alpha_{\beta\gamma}\,\bar\sff_{\bar\beta\bar\gamma}{}^{\bar\alpha}~,
                \quad
                \ldots,
            \end{equation}
            so that
            \begin{equation}\label{eq:left_BRST_action}
                Q_{\text{BRST}}\Phi^{\alpha\bar\alpha}=\sfq^{\alpha}_{\beta}\delta^{\bar\alpha}_{\bar\beta}\Phi^{\beta\bar\beta}+\delta^{\alpha}_{\beta} \bar\sfq^{\bar\alpha}_{\bar\beta}\Phi^{\beta\bar\beta}+\tfrac12\sff_{\beta\gamma}{}^{\alpha}\,\bar\sfq^{\bar\alpha}_{\bar\beta\bar\gamma}\Phi^{\beta\bar\beta}\Phi^{\gamma\bar\gamma}+\tfrac12\sfq^{\alpha}_{\beta\gamma}\,\bar\sff_{\bar\beta\bar\gamma}{}^{\bar\alpha}\Phi^{\beta\bar\beta}\Phi^{\gamma\bar\gamma}+\cdots~.
            \end{equation}    
        \end{subequations}
        
        Let us now introduce a second such theory, with action
        \begin{subequations}\label{eq:right_theory}
            \begin{equation}\label{eq:right_cubic_action_decomposition}
                S=\tfrac12\sfg_{\sfa\sfb}\,\bar\sfg_{\bar\sfa\bar\sfa}\Phi^{\sfa\bar\sfa}\wave\Phi^{\sfb\bar\sfb}+\tfrac1{3!}\sff_{\sfa\sfb\sfc}\,\bar\sff_{\bar\sfa\bar\sfb\bar\sfc}\Phi^{\sfa\bar\sfa}\Phi^{\sfb\bar\sfb}\Phi^{\sfc\bar\sfc}
            \end{equation}
            and BRST operator 
            \begin{equation}\label{eq:right_BRST}
                Q_{\text{BRST}}\Phi^{\sfa\bar\sfa}=\sfq^{\sfa}_{\sfb}\delta^{\bar\sfa}_{\bar\sfb}\Phi^{\sfb\bar\sfb}+\delta^{\sfa}_{\sfb}\bar\sfq^{\bar\sfa}_{\bar\sfb}\Phi^{\sfb\bar\sfb}+\tfrac12\sff_{\sfb\sfc}{}^{\sfa}\,\bar\sfq^{\bar\sfa}_{\bar\sfb\bar\sfc}\Phi^{\sfb\bar\sfb}\Phi^{\sfc\bar\sfc}+\tfrac12\sfq^{\sfa}_{\sfb \sfc}\,\bar\sff_{\bar\sfb\bar\sfc}{}^{\bar\sfa}\Phi^{\sfb\bar\sfb}\Phi^{\sfc\bar\sfc}+\cdots~.
            \end{equation} 
            Note that $\Phi^{\sfa\bar\sfa}$ and all structure constants appearing here are completely independent of those in~\eqref{eq:left_theory}; we distinguish them only through their indices to keep the common structure manifest.
        \end{subequations}     
        We will refer to~\eqref{eq:left_theory} and~\eqref{eq:right_theory} as the \emph{left} and \emph{right} \emph{parent theories}, respectively. Note that the parent theories could indeed be identical.
        
        As discussed already in~\cite{Borsten:2020zgj,Borsten:2021hua}, we can produce offspring from our two \emph{parent theories} by combining either the $\alpha$ or $\bar\alpha$ components of all the fields and all the structure constants in the left parent theory with either the $\sfa$ or $\bar\sfa$ components of all the fields and the structure constants of the right parent theory. This process has clear analogies with the meiotic reproduction of diploid cells in biology, and we therefore term this process (as well as the resulting offspring theory) \emph{syngamy}. As familiar from biology, we now have the four syngamies:
        \begin{equation}\label{eq:gen_syngamies}
            \begin{aligned}
                \text{(i)}~~~&\Phi^{\alpha \sfa}~,~&&(\sfg_{\alpha\beta},\sfg_{\sfa\sfb},\sff_{\alpha\beta\gamma},\sff_{\sfa\sfb\sfc})~,~&&(\sfq^{\alpha}_{\beta},\sfq^{\sfa}_{\sfb},\ldots)~,
                \\
                \text{(ii)}~~~&\Phi^{\bar \alpha\sfa}~,~&&(\bar \sfg_{\bar\alpha\bar\beta},\sfg_{\sfa\sfb},\bar \sff_{\bar \alpha\bar \beta\bar \gamma},\sff_{\sfa\sfb\sfc})~,~&&(\bar\sfq^{\bar \alpha}_{\bar\beta},\sfq^{\sfa}_{\sfb},\ldots)~,
                \\
                \text{(iii)}~~~&\Phi^{\alpha \bar\sfa}~,~&&(\sfg_{\alpha\beta},\bar\sfg_{\bar\sfa\bar\sfb},\sff_{\alpha\beta\gamma},\bar\sff_{\bar\sfa\bar\sfb\bar\sfc})~,~&&(\sfq^{\alpha}_{\beta},\bar\sfq^{\bar\sfa}_{\bar\sfb},\ldots)~,
                \\
                \text{(iv)}~~~&\Phi^{\bar \alpha\bar\sfa}~,~&&(\bar \sfg_{\bar\alpha\bar\beta},\bar\sfg_{\bar\sfa\bar\sfb},\bar \sff_{\bar \alpha\bar \beta\bar \gamma},\bar\sff_{\bar\sfa\bar\sfb\bar\sfc})~,~&&(\bar\sfq^{\bar \alpha}_{\bar\beta},\bar\sfq^{\bar\sfa}_{\bar\sfb},\ldots)~.
            \end{aligned}
        \end{equation}
        Each syngamy evidently comes with its own action and BRST operator. For syngamy (i), for example, we have 
        \begin{subequations}
            \begin{equation}
                S=\tfrac12\sfg_{\alpha\beta}\,\sfg_{\sfa\sfb}\Phi^{\alpha \sfa}\wave\Phi^{\beta\sfb}+\tfrac1{3!}\sff_{\alpha\beta\gamma}\,\sff_{\sfa\sfb\sfc}\Phi^{\alpha\sfa}\Phi^{\beta \sfb}\Phi^{\gamma\sfc}
            \end{equation}
            as well as
            \begin{equation}\label{eq:DCQ}
                Q_\text{BRST}=Q^\text{L}_{\text{BRST}}+Q^\text{R}_{\text{BRST}}
            \end{equation}
            with 
            \begin{equation}
                \begin{aligned}
                    Q^\text{L}_{\text{BRST}}\Phi^{\alpha \sfa}&\coloneqq\sfq^\alpha_\beta\delta^\sfa_\sfb\Phi^{\beta\sfb}+\tfrac12\sfq^\alpha_{\beta\gamma}\,\sff_{\sfb\sfc}{}^\sfa\Phi^{\beta\sfb}\Phi^{\gamma\sfc}+\cdots~,
                    \\
                    Q^\text{R}_{\text{BRST}}\Phi^{\alpha\sfa}&\coloneqq\delta^\alpha_\beta\sfq^\sfa_\sfb\Phi^{\beta \sfb}+\tfrac12\sff_{\beta\gamma}{}^{\alpha}\,\sfq^\sfa_{\sfb\sfc}\Phi^{\beta \sfb}\Phi^{\gamma\sfc}+\cdots~.
                \end{aligned}
            \end{equation}
        \end{subequations}
        
        To illustrate the various syngamies in~\eqref{eq:gen_syngamies}, let us briefly consider the example in which both parent theories are CK--duality manifesting Yang--Mills theories that only differ in their color Lie algebras. As in \cref{ssec:CK_manifest_YM}, let us identify $\alpha$ and $\sfa$ with the kinematic DeWitt indices and $\bar\alpha$ and $\bar\sfa$ with the color DeWitt indices. We then have the following four cases:
        \begin{enumerate}[label=(\roman*)]\itemsep-2pt
            \item The left kinematic part is combined with the right kinematic part. This is the syngamy that is usually called double copy. The resulting theory has the same field content as $\caN=0$ supergravity, and, as we will show below, is quantum equivalent to this theory.
            \item The left color part is combined the right kinematic part. This syngamy has the same defining constants as, and is thus identical to, the left parent theory. 
            \item The left kinematic part is combined with the right color part. This syngamy is thus identical to the right parent theory. 
            \item The left color part is combined with the right color part. This syngamy is sometimes called the zeroth copy, and the resulting theory is (quantum) equivalent to a theory of biadjoint scalars, cf.~\cite{Borsten:2021hua}.
        \end{enumerate}
        
        Consider now a general syngamy with action $S$ and BRST operator $Q_\text{BRST}$. If $({Q}_\text{BRST})^2=0$ and ${Q}_\text{BRST}{S}=0$, we have a consistent theory, ready to be quantized.
        
        Let us assume that the terms ${Q}_\text{BRST}^2\phi$ for any field $\phi$ and ${Q}_\text{BRST}S$ are proportional to the \emph{generic} algebraic relations satisfied by the metric and structure constants $(\sfg_{\alpha\beta},\sff_{\alpha\beta\gamma}, \bar\sfg_{\bar\alpha\bar\beta}, \bar\sff_{\bar\alpha\bar\beta\bar\gamma})$ of general CK-dual theories, i.e.~anti-symmetry and the Jacobi identity of the Lie algebra structure constants as well as symmetry and invariance of the metric, cf.~\eqref{eq:alg_rel_NLSM} and~\eqref{eq:alg_rel_YM}. It is then clear that in all mixtures of two such CK-dual theories, the conditions $({Q}_\text{BRST})^2=0$ and ${Q}_\text{BRST}{S}=0$ are automatically satisfied because all defining constants are replaced by defining constants satisfying the same algebraic relations.
        
        Let us briefly mention that the above prescription for constructing the BRST generators in a syngamy actually manifestly applies to the generator of any continuous (super)symmetry, as long as the symmetry generators do not include explicit appearances of the Minkowski coordinate $x^\mu$.\footnote{Otherwise, $x^\mu$ fails to commute with differential operators, which causes complications.} Given this assumption, for every symmetry generator of the original theory one obtains left and right symmetry generators of the syngamy, which are guaranteed to (anti-)commute (but which are not guaranteed to be distinct nor non-anomalous). For example, the double copy of any theory on Minkowski space has \emph{left} and \emph{right} translation symmetry, which commute past each other --- in fact, they coincide. Similarly, supertranslations and R-symmetries can be double-copied. On the other hand, the generators of rotation or dilatation involve explicit appearance of the space--time coordinate $x^\mu$, so they lie outside the allowed ansatz.
        
        \subsection{Non-linear sigma model and the special galileon}\label{ssec:double_copy_NLSM}
        
        As a concrete application, we consider the relatively simple example of the syngamy of two copies of the non-linear sigma model that is usually called the double copy. We are interested in the syngamy that combines the left kinematic part with the right kinematic part. The resulting theory is the \emph{special galileon}. In this double copy construction, we can avoid the technical difficulties related to gauge symmetry appearing in Yang--Mills theory. 
        
        A galileon theory~\cite{Fairlie:1992nb,Fairlie:1992zn,Fairlie:1991qe,Dvali:2000hr,Nicolis:2008in} is a scalar field theory invariant under the Galilean-like symmetry transformation\footnote{not to be confused with Galilean symmetry on space--time: galileon theories are Lorentz-invariant, not Galilean-invariant}
        \begin{equation}
            \vartheta(x)\mapsto\vartheta(x)+c+b_{\mu}x^{\mu}
        \end{equation}
        with $c$ and $b_\mu$ constants. In $d$ space--time dimensions, there exist exactly $d+1$ independent invariants\footnote{up to total derivatives} $\scL^\text{Gal}_n$ of the Galilean-like transformations; the general galileon action in $d$ dimensions correspondingly reads as~\cite{Nicolis:2008in}
        \begin{equation}
            \begin{aligned}
                S^\text{Gal}&=\int\rmd^dx\,\sum_{n=1}^{d+1}\alpha_n\scL^\text{Gal}_n~,
                \\
                \scL^\text{Gal}_n&=\eps^{\mu_1\cdots\mu_d}\eps^{\nu_1\cdots\nu_d}\left(\prod_{j=n}^d\eta_{\mu_j\nu_j}\right)\, \left(\prod_{i=1}^{n-1}\partial_{\mu_i}\partial_{\nu_i}\vartheta\right) \vartheta~,
            \end{aligned}
        \end{equation}
        where $\alpha_n$ are real parameters. To avoid tadpole terms, we set $\alpha_1$ to 0, and we normalize $\alpha_2$ to have the standard kinetic term.
        
        Generically, theories with Lagrangians that feature derivatives of order greater than two are plagued by the Ostrogradsky instability, leading to a Hamiltonian unbounded from below. Nevertheless, this is not the case if the associated equations of motion are at most quadratic in the derivatives. The presence of the Galilean-like symmetry ensures this for galileon theories.
        
        The choice for the coupling constants $\alpha_n$ 
        \begin{equation}
            \alpha_{2n}=\frac1{2n}\binom d{2n-1}\mu^{2n-2}
            \eand
            \alpha_{2n+1}=0~,
        \end{equation}
        where $\mu$ has the mass dimension $-1-\frac d2$, corresponds to a particular instance of galileon theory that possesses several highly remarkable properties, known as the special galileon~\cite{Hinterbichler:2015pqa,Cheung:2014dqa,Cachazo:2014xea}. Importantly, the special galileon is equivalent to the double copy of the non-linear sigma model as was demonstrated at the tree level in~\cite{Cachazo:2014xea, Cheung:2016prv} and to all loop orders in~\cite{Borsten:2021hua}.
        
        The starting point of our construction is the local, cubic reformulation of the non-linear sigma model action that explicitly manifests loop-level CK duality, obtained from the algorithm presented in \cref{ssec:offshell_lift_NLSM}. We consider two copies of this theory, a left theory and a right theory, which may differ in their flavor Lie algebras. The double copy is now the syngamy where the kinematic indices, metric, and structure constants of the left theory are combined with the kinematic indices, metric, and structure constants from the right theory.
        
        Correspondingly, the double-copied field content is obtained by taking the tensor square of the flavor-stripped field content of the strictified, CK-dual non-linear sigma model, and the interaction vertices are the products of two of the interaction vertices of the non-linear sigma model. Except for the field $\vartheta\coloneqq\phi\otimes\phi$, all the double-copied fields are regarded as auxiliary. The action is then schematically of the following form:
        \begin{equation}
            {S}^\text{DC}=\tfrac12\Phi^{\sfi\bar\sfi}\sfg_{\sfi\sfj}\bar\sfg_{\bar\sfi\bar\sfj}\wave\Phi^{\sfj\bar\sfj}+\tfrac{1}{3!}{\sff}_{\sfi\sfj\sfk}{\bar\sff}_{\bar\sfi\bar\sfj\bar \sfk}\tilde\Phi^{\sfi\bar\sfi}\Phi^{\sfj\bar\sfj}\Phi^{\sfk\bar\sfk}~;
        \end{equation}
        since there is no gauge symmetry, there is only a trivial BRST operator.
        
        We now sketch an argument to show that the double copy theory is perturbatively quantum equivalent to the special galileon; for more details, see~\cite{Borsten:2021hua}. Since we are only interested in perturbative equivalence, it suffices to show that the loop-level scattering amplitudes of the double-copied theory and of the special galileon agree to any given order in the coupling constants and number of loops, up to counterterms. Thus, only a finite subset of the interaction vertices of the two theories has to be considered.
        
        After integrating out all auxiliary fields, the field contents of both theories are evidently the same. Moreover, it is known~\cite{Cachazo:2014xea,Cheung:2017ems,Cheung:2016prv,Du:2016tbc,Cheung:2017yef} that the tree-level scattering amplitudes of the special galileon and the double-copied theory agree. Equivalently, the double copy action has inherited its kinematic metric and structure constants from the CK duality manifesting parent action; they, therefore, obey the same identities as the flavor metric and structure constants. Correspondingly, the actions can only differ in interaction terms, and these must vanish on shell. Explicitly, the difference between the double-copied action ${S}^\text{DC}$ and the special galileon action must be of the form
        \begin{equation}\label{eq:DC_discrepancies}
            \int\rmd^dx\,\tilde F^i\wave\vartheta~,
        \end{equation}
        where the $\tilde F^i$, with $i$ running over some index set, are some expressions in the field $\vartheta$. Considering more closely the terms which are produced during the strictification procedure, these terms are local up to insertions of operators $\frac{1}{\wave}$. The potential discrepancies~\eqref{eq:DC_discrepancies} can all be absorbed iteratively, order by order in the number of external legs, by field redefinitions of the form 
        \begin{equation}
            \vartheta\mapsto\vartheta+\sum_i\tilde F^i~.
        \end{equation}
        Note again that we only need to correct finitely many vertices, and there are only finitely many required field redefinitions.
        
        \subsection{Yang--Mills BRST--Lagrangian double copy}\label{ssec:YM_double_copy}
        
        The BRST--Lagrangian double copy introduced in~\cite{Borsten:2020zgj,Borsten:2021hua} is the syngamy (i) in~\eqref{eq:gen_syngamies} with two Yang--Mills theories that may differ in the choice of gauge Lie algebra as parent theories. That is, we combine the kinematic DeWitt indices of the left Yang--Mills theory with the kinematic DeWitt indices of the right Yang--Mills theory, obtaining the field $\caH^{\sfi\sfj}$ and corresponding kinematic metric and structure constants. The field $\caH^{\sfi\sfj}$ is decomposed into off-shell (not necessarily irreducible) Lorentz representations given by the BRST fields of strictified $\caN=0$ supergravity as described in detail in~\cite{Borsten:2021hua},
        \begin{equation}
            (\caH^{\sfi\sfj})=\big(h_{\mu\nu}(x),B_{\mu\nu}(x),\varphi(x), X(x), Y^{m}(x)\big)~,
        \end{equation}
        where $h_{\mu\nu}$, $B_{\mu\nu}$, and $\varphi$ are the graviton, Kalb--Ramond 2-form, and dilaton, respectively. The full set of diffeomorphism and 2-form gauge symmetry ghosts and anti-ghosts of  $\caN=0$ supergravity is collectively denoted by $X$. Furthermore, $Y^{m}$ correspond to the bosonic and ghost strictification auxiliary fields~\cite{Borsten:2021hua}.
        
        The double-copied BRST action is
        \begin{equation}\label{eq:DCL}
            {S}^\text{DC}_\text{BRST}=\tfrac12 \sfg_{\sfi\sfk}\sfg_{\sfj\sfl}\caH^{\sfi\sfj}\wave\caH^{\sfk\sfl}+\tfrac1{3!} \sff_{\sfi\sfj\sfk}\sff_{\sfl\sfm\sfn}\caH^{\sfi\sfl}\caH^{\sfj\sfm}\caH^{\sfk\sfn}~,
        \end{equation}
        and we note that the metric and the kinematic structure constants are indeed double-copied. Similarly, the double-copied BRST operator reads as
        \begin{subequations}
            \begin{equation}\label{eq:DCQ_YM}
                {Q}^\text{DC}_\text{BRST}=Q^\text{L}_\text{BRST}+Q^\text{R}_\text{BRST}~,
            \end{equation}
            where 
            \begin{equation}
                \begin{aligned}
                    Q^\text{L}_\text{BRST}\caH^{\sfi\sfj}\ &\coloneqq\ \sfq^\sfi_\sfk\, \delta^{\sfj}_{\sfl}\,\caH^{\sfk\sfl}+\tfrac12\sfq_{\sfk\sfl}^\sfi \,{\bar\sff}_{\sfm\sfn}{}^{\sfj}\, \caH^{\sfk\sfm}\,\caH^{\sfl\sfn}~,
                    \\
                    Q^\text{R}_\text{BRST}\caH^{\sfi\sfj}\ &\coloneqq\ \delta_\sfk^\sfi\,\sfq^{\sfj}_{\sfl}\, \caH^{\sfk\sfl}+\tfrac12 \sff_{\sfk\sfl}{}^\sfi\, \sfq_{\sfm\sfn}^{\sfj}\,\caH^{\sfk\sfm}\,\caH^{\sfl\sfn}~.
                \end{aligned}
            \end{equation}
        \end{subequations}
        
        We claim that the double copy BRST action~\eqref{eq:DCL} and BRST operator~\eqref{eq:DCQ_YM} constitute a  well-defined perturbative BRST quantization,\footnote{Note that this does not mean that the double-copied ${Q}^\text{DC}_\text{BRST}$ is non-anomalous. As we note later, ${Q}^\text{DC}_\text{BRST}$ might differ from a canonically obtained BRST operator $Q^{\caN=0}_\text{BRST}$ by a trivial symmetry which may be anomalous.} quantum equivalent to the canonical BRST quantization of $\caN=0$ supergravity. The latter is described in detail in~\cite{Borsten:2020zgj,Borsten:2021hua}, see also~\cite{Borsten:2020xbt}. Concretely:
        \begin{enumerate}[label=(\roman*)]\itemsep-2pt
            \item We have ${Q}^\text{DC}_\text{BRST}{S}^\text{DC}_\text{BRST}=0$ up to a total derivative, as well as $({Q}^\text{DC}_\text{BRST})^2=0$ up to exact operator identities; we can trivially reintroduce Nakanishi--Lautrup fields to ensure $({Q}^\text{DC}_\text{BRST})^2=0$ if desired. This follows from the CK-dual Yang--Mills BRST action\footnote{The additional weak condition that ${Q}_\text{BRST}^2\phi$ for any field $\phi$ and ${Q}_\text{BRST}S$ are proportional to the generic algebraic relations are satisfied for Yang--Mills theory.}, as explained in \cref{ssec:BRST_double_copy}. Moreover, the linear parts of ${Q}^\text{DC}_\text{BRST}$ indeed agree with the expected form of the BRST operator of $\caN=0$ supergravity, as is evident from the explicit form given in~\cite{Borsten:2021hua}. 
            \item There is a gauge and a field redefinition such that, after putting the auxiliary fields $Y^{m}$ to zero, the kinetic terms agree~\cite{Borsten:2021hua},
            \begin{equation}
                {S}^\text{DC}_\text{BRST}\Big|_{\text{kin},\,Y^m=0}={S}^{\caN=0}_\text{BRST}\Big|_\text{kin}~.
            \end{equation}
            The linearized form of the double copy therefore evidently holds.
            \item The physical tree-level scattering amplitudes of~\eqref{eq:DCL} agree by construction with those of $\caN=0$ supergravity. The numerators of all physical tree-level scattering amplitudes derived from the Feynman diagrams of the action~\eqref{eq:DCL} are precisely those obtained by double-copying the physical tree-level kinematic numerators of the Yang--Mills theory. The action~\eqref{eq:DCL} is thus semi-classically equivalent to the $\caN=0$ supergravity action. This matching can be extended to all bosonic fields by a suitable choice of gauge (tuned such that non-transversely polarized fields cannot produce a discrepancy), cf.~\cref{ob:longitudinal_states}; then the (on-shell) BRST Ward identities of \cref{ob:onshell_Ward_identities} show that the matching extends to the full BRST-extended Hilbert space, including states containing ghosts and anti-ghosts.
            \item We now integrate out all auxiliary fields, which can lead to non-local interaction terms. The discrepancy between the interaction terms in both theories must be proportional to $\wave\Phi$ for some field $\Phi$ as they are invisible to the tree-level scattering amplitudes. As argued before in the context of CK duality, such differences can be produced by field redefinitions. These field redefinitions may be non-local, but they do not affect the consistency or quantum equivalence of the theory; they merely introduce well-defined additional counterterms in the renormalization procedure. In fact, these counterterms will be the additive inverses of those needed to ensure the unitarity of the loop amplitudes of the double copy action. Removing the local terms corresponds to removing the need for the corresponding counterterms.
        \end{enumerate}
        
        The field redefinitions and integrating out of the auxiliary fields lead to a transformed double-copied BRST operator ${Q}^\text{DC}_\text{BRST}$. Having matched the double-copied BRST action with the BRST action of $\caN=0$ supergravity, it is now natural to expect that also the corresponding BRST operators ${Q}^\text{DC}_\text{BRST}$ and $Q^{\caN=0}_\text{BRST}$ agree. This is almost the case. Because both $Q^{\caN=0}_\text{BRST}$ and ${Q}^\text{DC}_\text{BRST}$ describe symmetries of the action $S^{\caN=0}_\text{BRST}$ of $\caN=0$ supergravity, their difference is also a symmetry. This difference could, in principle, consist of any other symmetry in the theory, but we know that the linear parts of $Q^{\caN=0}_\text{BRST}$ and $\tilde Q^\text{DC}_\text{BRST}$ agree, and there are no candidates for perturbative, non-linear, non-trivial symmetries. The only remaining possibility is thus a \emph{trivial symmetry}, cf.~\cite[Section 2.5]{Henneaux:1990:47-105},~\cite[Theorem 3.1]{Henneaux:1992}, see also~\cite[Section 4.2]{Jurco:2018sby} for a large class of trivial symmetries that any theory possesses. These symmetries vanish on shell and do not lead to new Noether charges, so a potential difference by a trivial symmetry is irrelevant for most purposes.\footnote{Recall that trivial symmetries are general coordinate transformations on field space that leave the action invariant. Generically, however, their actions on the path integral measure will introduce non-trivial Jacobian determinants. Therefore, trivial symmetries, and hence ${Q}^\text{DC}_\text{BRST}$, may be anomalous.} See also~\cite{Campiglia:2021srh} for a double copy construction of symmetries in the  self-dual sectors of Yang--Mills theory and gravity,  including both perturbative and \emph{non-perturbative} symmetries, as well as the double copy of large gauge transformations that yields holomorphic supertranslations. 
        
        \subsection{Double copy of supersymmetric Yang--Mills theory}\label{ssec:DC_SUSY}
        
        Given CK duality for the BRST action of supersymmetric Yang--Mills theory, the BRST--Lagrangian double copy follows straightforwardly following the general principles laid out in the above sections. Nonetheless, there are several interesting new features that we will briefly expand upon here.    
        
        Although the arguments are generic, for definiteness we will focus on the double copy of supersymmetric Yang--Mills theory in $d=10$ space--time dimensions. This has the added benefit of implying the analogous results for all theories obtainable by toroidal dimensional reduction and field theory orbifolding, cf.~\cite{Chiodaroli:2013upa}. This includes, for example, the products of any pair of supersymmetric Yang--Mills theories in any $d\leq 10$, as summarized in~\cite{Anastasiou:2015vba}, as well as various supersymmetric Yang--Mills--matter theories, cf.~e.g.~\cite{Anastasiou:2016csv,Anastasiou:2017nsz,Bern:2019prr}.
        
        For supersymmetry, we add a Majorana--Weyl gluino $\psi^a_\alpha$ to the $d=10$ Yang--Mills BRST fields of~\eqref{eq:YM_field}
        \begin{equation}\label{eq:symfield}
            (\caA^{\sfi \bar\sfa})=\big(A^{a}_\mu(x),\psi^{a}_\alpha(x),\bar c^{a}(x),c^{a}(x),Y^{m a}(x)\big)~,
        \end{equation}
        where now $Y^{m a}$ also contains auxiliary superpartners. We adopt 32-component Majorana spinor conventions, so $\gamma_{11}\psi=\pm\psi$ according to the chirality chosen for the gluino. 
        
        The double copy is given by the same syngamy as before, cf.~\eqref{eq:DCL}. The result is IIA or IIB supergravity  if the chiralities of the gluinos are taken to be opposing or matching, respectively, as implied by the  tensor product of the  on-shell gluon and gluino states,
        \begin{equation}
            \begin{split}
                \text{IIA:}&\quad\bm{8}_v\otimes\bm{8}_s=\bm{56}_c\oplus\bm{8}_c\eand\bm{8}_c\otimes\bm{8}_v=\bm{56}_s\oplus\bm{8}_s~,
                \\
                \text{IIB:}&\quad\bm{8}_v\otimes\bm{8}_s=\bm{56}_c\oplus\bm{8}_c\eand\bm{8}_s\otimes\bm{8}_v=\bm{56}_c\oplus\bm{8}_c~.
            \end{split}
        \end{equation}
        
        Let us first consider the Neveu--Schwarz--Ramond (NS--R) and R--NS sectors. The additional physical field content in the NS--R sector is        
        \begin{equation}
            \caH_{\mu\beta}\ \sim\ A_\mu\otimes\psi_\beta\ \sim\ \Psi_{\mu\beta}\oplus\lambda_{\beta}~,
        \end{equation}
        where $\Psi_\mu$ and $\lambda$ are the NS--R gravitino and dilatino, respectively. The R--NS sector follows similarly. 
        
        In addition, there are ghosts and anti-ghosts for each of the two gravitini, which follow from the product of the Yang--Mills (anti-)ghosts with the gluini~\cite{Anastasiou:2014qba}. In the NS--R sector, we have 
        \begin{equation}
            \caH_{\tta\beta} \sim \bar c\otimes\psi \sim \bar\eta
            \eand
            \caH_{\ttg\beta} \sim c\otimes\psi \sim \eta~,
        \end{equation}
        where $\eta$ and $\bar\eta$ are the local supersymmetry ghost and anti-ghost, respectively. If we had not integrated out the Yang--Mills Nakanishi--Lautrup field $b$, its product with the right gluino would produce $\caH_{\ttn\beta}\sim b\otimes\psi\sim\chi$, the Nielsen--Kallosh auxiliary spinor. The R--NS ghosts and anti-ghosts are  constructed similarly. 
        
        To be able to apply our Lagrangian double copy prescription, one should first rewrite the gluino kinetic term so as to manifest the ansatz~\eqref{eq:leftYM}:
        \begin{equation}
            \bar\psi\slashed{\partial}\psi=\bar\psi\slashed{\partial}^{-1}\wave\psi~.
        \end{equation}
        This defines the extension of the kinematic metric~\eqref{eq:kin_metric} to the supersymmetric Yang--Mills BRST field space given by~\eqref{eq:symfield}. The double copy then proceeds as before. 
        
        The supercharges $\caQ^{\text L},\caQ^{\text R}$ of the left and right parent supersymmetric Yang--Mills theories generate the IIA/B supercharges~\cite{Bianchi:2008pu,Damgaard:2012fb,Anastasiou:2013hba,Anastasiou:2015vba}, in direct analogy to the BRST transformations~\eqref{eq:DCQ_YM}. Explicitly,
        \begin{equation}
            \caQ\caA^{\sfi\bar\sfa}=\caQ^\sfi_\sfj\ \delta^{\bar\sfa}_{\bar\sfb}\caA^{\sfj\bar\sfb}+\tfrac12\caQ^\sfi_{\sfj\sfk}\sff_{\bar\sfb\bar\sfc}{}^{\bar\sfa}\caA^{\sfj\bar\sfb}\caA^{\sfk\bar\sfc}
        \end{equation}
        gives rise to
        \begin{equation}\label{eq:leftsusy}
            \caQ^\text{L}\caH^{\sfi\sfj}=\caQ^\sfi_\sfk\delta^\sfj_\sfl\caH^{\sfk\sfl}+\tfrac12\caQ^\sfi_{\sfk\sfl}{\sff}_{\sfm\sfn}{}^{\sfj}\caH^{\sfk\sfm}\caH^{\sfl\sfn}
        \end{equation}
        with analogous formulas for $\caQ^\text{R}$. The invariance of the double copy Lagrangian under supersymmetry then follows from a fully analogous argument to the discussion of BRST invariance. Note that the local supersymmetry parameter, not included in~\eqref{eq:leftsusy}, is identified with the local supersymmetry ghost $\eta$. 
        
        As an explicit example, consider 
        \begin{equation}
            \caQ_\alpha A_\mu^a=\delta^a_b\gamma_\mu{}_\alpha{}^\beta\psi_\beta^b+\cdots~,
        \end{equation}
        where the higher order terms represented by the ellipsis are induced by the field redefinitions required for manifest CK duality. Setting $(\sfi\sfj)=(\mu\nu)$ for instance, this double-copies to
        \begin{equation}
            \caQ^{\text{L}}_\alpha\caH_{\mu\nu}=\gamma_\mu{}_\alpha{}^\beta\caH_{\beta\nu}+\cdots~.
        \end{equation}
        Applying $\caQ^\text{L}$ to the graviton $h_{\mu\nu}$ and $B_{\mu\nu}$, which are given by  $\caH^{\sfi\sfj}$ as in~\cite{Borsten:2021hua}, then determines the NS--R gravitino in terms of $\caH_{\alpha\nu}$. The same applies for the R--NS gravitino, using $\caQ^\text{R}$. Similarly, the NS--R and R--NS dilatini are determined by $\caQ^\text{L}$ and $\caQ^\text{R}$ acting on $B_{\mu\nu}$ and the dilaton, given in terms of $\caH^{\sfi\sfj}$ as in~\cite{Borsten:2021hua}.
        
        The logic of the preceding discussion did not differ from the case of Yang--Mills theory, save for the graviton, the diffeomorphism gauge potential, replaced by the gravitino, the local supersymmetry gauge potential. However, the Ramond--Ramond (R--R) sector is rather more subtle and requires some new conceptual ingredients.  
        
        The key difference is that the product of two spinors in the context of a field theoretic double copy gives \emph{field strengths}, not potentials~\cite{Nagy:2014jza}\footnote{We are also grateful to A.~Anastasiou, M.~J.~Duff, S.~Nagy, and M.~Zoccali for this observation, which was originally made in (as yet) unpublished joint work.}. Roughly speaking, an R--R double copy field $\caH_{\alpha\beta}\sim\psi_\alpha\otimes \psi_\beta$  is a bispinor that decomposes into a sum of $p$-form field strengths $F_p$, 
        \begin{subequations}\label{eq:bispinor_decomp}
            \begin{equation}
                \caH_{\alpha\beta}=\sum_{p=0}^d\frac{1}{p!}(\gamma^{\mu_1\cdots\mu_p}C)_{\alpha\beta}F_{\mu_1\cdots\mu_p}
            \end{equation}
            with
            \begin{equation}
                \gamma^{\mu_1\cdots\mu_p}{}_\alpha{}^\beta=\gamma^{[\mu_1}{}_\alpha{}^{\alpha_1}\gamma^{\mu_2}{}_{\alpha_1}{}^{\alpha_2}\cdots\gamma^{\mu_p]}{}_{\alpha_p}{}^\beta~,
            \end{equation}
        \end{subequations}
        where $C$ is the charge conjugation matrix and $\star F_p\equiv F_{d-p}$. The particular set of field strengths generated depends on the space--time dimension and the class of spinors considered, but can be read off from the relevant $\frso(d)$ tensor product decomposition. For example, in our cases of interest, we have
        \begin{equation}
            \begin{split}
                \text{IIA:}&\quad\bm{\overline{16}}\otimes\bm{16}=\bm{1}\oplus\bm{45}\oplus\bm{210}~,
                \\
                \text{IIB:}&\quad\bm{16}\otimes\bm{16}=\bm{10}\oplus\bm{120}\oplus\bm{126}~.
            \end{split}
        \end{equation}
        For type IIA, the $\bm{45}$ and $\bm{210}$ correspond to the usual R--R 2-form and 4-form field strengths, respectively. The 0-form field strength corresponding to the singlet lacks degrees of freedom and can be integrated out (although it is tempting to regard it as the Romans mass). For type IIB, the $\bm{10}$, $\bm{120}$, and $\bm{126}$ correspond to the usual R--R 1-form, 3-form, and self-dual 5-form field strengths, respectively.   
        
        The necessity of identifying $\psi\otimes\psi$ with field strengths follows from three related observations. First, as is clear from the above example of type IIA/B supergravity, the tensor product of the off-shell gluino spinor representations yields tensor representations corresponding to those carried by the field strengths, not the potentials. Second, the BRST transformation of the gluino has no linear contribution, $Q_\text{BRST}\psi=[c,\psi], $ so $\psi\otimes\psi$ cannot transform as a gauge potential under the double copy of the BRST transformations. Finally, the background R--R fields couple to the type II superstring worldsheet  action through their field strengths only --- the R--R superstring vertex operators, constructed from the products of open superstring vertex operators, correspond to field strengths. This is reflected in the scattering amplitudes involving R--R external states, which are always proportional to a power of their momenta and so have vanishing soft limits --- superstrings are uncharged with respect to the R--R gauge potentials.  
        
        The key implication of the latter comment for the present discussion is that the IIA/B supergravity actions can be written exclusively in terms of the R--R field strengths, once they are suitably defined. For example, the IIA supergravity R--R sector Lagrangian is
        \begin{subequations}
            \begin{equation}\label{eq:IIARR}
                F_2\wedge{\star F_2}+\tilde F_4\wedge{\star F_4}+B_2\wedge F_4\wedge F_4~,
            \end{equation}
            where
            \begin{equation}
                F_2=\rmd C_1~,
                \quad
                F_4=\rmd C_3~,
                \quad
                \tilde F_4=\rmd C_3+H_3\wedge C_1~,
                \quad             
                H_3=\rmd B_2~,
            \end{equation}
            with the Bianchi identity
            \begin{equation}
                \rmd\tilde F_4=-H_3\wedge F_2~.
            \end{equation}
        \end{subequations}
        Up to total derivatives,~\eqref{eq:IIARR} can be rewritten as 
        \begin{equation}
            F_2\wedge{\star F_2}+\tilde F_4\wedge{\star F_4}+B_2\wedge\tilde F_4\wedge\tilde F_4+B_2\wedge B_2\wedge F_2\wedge\tilde F_4-\tfrac13B_2\wedge B_2\wedge B_2\wedge F_2\wedge F_2~,
        \end{equation}
        so the action may be formulated purely in terms of $F_2$ and $\tilde F_4$, with no bare R--R potentials appearing. Similarly, the type IIB action admits a formulation in terms of only R--R field strengths~\cite{Bergshoeff:2001pv}. 
        
        This is implied by the double copy construction of the R--R sector: the identification of $\psi\otimes\psi$ with field strengths implies that there must exist a formulation of the type II supergravity action with no bare R--R potentials. In fact, more is required by the double copy. Since the double copy action is written purely in terms of the R--R field strengths, they must be treated as \emph{elementary} fields and still correctly reproduce  scattering amplitudes involving the R--R sector through their Feynman diagrams. Interestingly, as we will explain below, the double copy Lagrangian achieves this automatically through a generalization of Sen's mechanism~\cite{Sen:2015uaa,Sen:2015nph} for imposing self-duality on $\frac d2$-form field strengths in $d\equiv2\pmod 4$ space--time dimensions. 
        
        The double copy R--R sector is, in terms of the bispinor $\caH_{\alpha\beta}$ and its Majorana conjugate $\overline{\caH}^{\alpha\beta}$, given by 
        \begin{equation}
            {S}_\text{R--R}^\text{DC}=\underbrace{\int\rmd^dx\,\overline{\caH}^{\alpha\beta}\wave{}^{-1}\slashed{\partial}_\alpha{}^{\alpha'}\slashed{\partial}_\beta{}^{\beta'}\caH_{\alpha'\beta'}}_{\eqqcolon\,{S}_\text{R--R,\,kin}^\text{DC}}+{S}^\text{DC}_\text{R--R,\,int}~.
        \end{equation}
        Next, upon decomposing the bispinor in terms of the field strengths as in~\eqref{eq:bispinor_decomp}, the kinetic terms for each elementary $p$-form field strength are of the form
        \begin{equation}\label{eq:nonlocal_sen_gen}
            {S}_\text{R--R,\,kin}^\text{DC}=-\tfrac12\int\left\{F\wedge{\star F}-\rmd F\wedge{\star\wave}{}^{-1}\rmd F\right\}+\cdots~,
        \end{equation}
        where we only displayed the terms relevant for our discussion below. In the free theory, the equation of motion for $F$ implies the Maxwell equation $\rmd{\star F}=0$ and the Bianchi identity $\rmd F=0$. In terms of the Hodge-dual field strength ${\star F}$, the kinetic term can be written as
        \begin{subequations}
            \begin{equation}\label{eq:dual_nonlocal_sen_gen}
                \int\rmd^dx\,({\star F})_{\mu_1\cdots\mu_{q+1}}P^{\mu_1\cdots\mu_{q+1}}{}_{\nu_1\cdots\nu_{q+1}}({\star F})^{\nu_1\cdots\nu_{q+1}}~,
            \end{equation}
            where the appearance of the operator 
            \begin{equation}
                P^{\mu_1\cdots \mu_{q+1}}{}_{\nu_1\cdots \nu_{q+1}}=\left(\wave\delta^{\mu_1}_{\nu_1}\cdots\delta^{\mu_{q+1}}_{\nu_{q+1}}+\overset{\leftarrow}{\partial}{}^{[\mu_1}\overset{\rightarrow}{\partial}{}_{[\nu_1}\delta^{\mu_2}_{\nu_2}\cdots\delta^{\mu_{q+1}]}_{\nu_{q+1}]}\right)\wave{}^{-1}
            \end{equation}
        \end{subequations}
        can intuitively be interpreted as inserting a projector on each internal line of the Feynman diagrams that removes the unphysical modes, ensuring that the tree-level scattering amplitudes match those of the conventional Abelian $p$-form theory. The double copy yields a formulation of the R--R sector in terms of elementary gauge invariant field strengths and, accordingly, with none of the usual gauge-fixing machinery. There simply are no double copy fields that could be identified with R--R ghosts or anti-ghosts. Since the double copy R--R sector tree-level scattering amplitudes match, by construction, those of IIA/B supergravity, they are semi-classically equivalent. The complete absence (and irrelevance) of R--R gauge-fixing and ghost sectors implies that the matching of the physical tree-level scattering amplitudes alone suffices for full quantum equivalence. 
        
        This formulation of the R--R sector in terms of elementary $(p+1)$-form field strengths $F_{p+1}$, is perhaps less familiar, and it is not immediately obvious that it yields  physics equivalent to that of the usual theory of Abelian $p$-form  gauge potentials $A_p$. A first natural step towards establishing this equivalence would be to make the action~\eqref{eq:nonlocal_sen_gen} local. This can be done by introducing $q$-form auxiliary fields $B_{q}$, where $q=d-p-2$, putatively Hodge-dual to the $A_p$. In doing so, one discovers that the double copy automatically reproduces a generalization of Sen's mechanism. 
        
        \subsection{Sen's mechanism from the double copy}\label{ssec:sen_mech}
        
        Sen's mechanism is a method to construct action principles for differential form fields obeying a self-duality constraint, motivated by IIB string field theory~\cite{Sen:2015uaa,Sen:2015nph}, where the R--R sector is naturally given in terms of bispinors. Given the close kinship  between the double copy and $\text{closed}=\text{open}\otimes\text{open}$ string relations, one may also expect this mechanism to appear in the current context. Indeed, the use of bispinors for the R--R sector has essentially the same origin in each case. 
        
        To illustrate this, let us start from Sen's mechanism generalized to arbitrary (as opposed to self-dual) field strengths. Consider a $(p+1)$-form field strength $F$ that is treated as elementary, but required to be physically equivalent to an Abelian $p$-form potential $A$ coupled to the other fields of the theory, collectively denoted by $\Phi$, only through its field strength. The generalization of Sen's mechanism implementing this demand is given by the action
        \begin{equation}\label{eq:sen_gen2}
            S^\text{genSen}=\int\left\{-\tfrac12F\wedge{\star F}-\xi B\wedge\rmd F+\tfrac12\rmd B\wedge{\star\rmd B}\right\}+S^\text{genSen,\,int}[F,\Phi]~,
        \end{equation}
        where we have introduced a second $q=(d-p-2)$-form $B$ and a real parameter $\xi$. Note that $\Phi$ does not include $B$. In the absence of the $\rmd B\wedge{\star\rmd B}$ term, $B$ is just the familiar Lagrange multiplier enforcing $\rmd F=0$, and for $S^\text{genSen}_\text{int}$ linear in $F$ we can  integrate out $F$ to obtain the dual $(d-p-2)$-form Lagrangian in terms of $B$.\footnote{Since we are interested here in perturbative scattering amplitudes on a Minkowski background, we can ignore all subtleties regarding quantum equivalence due to topology and non-perturbative effects, cf.~\cite{Borsten:2021pte} and references therein.} When $d\equiv2\pmod 4$ and $F$ is self-dual, then $F{\wedge\star F}=0$, and we recover Sen's original ansatz, which is applicable to self-dual 5-form of IIB supergravity~\cite{Sen:2015nph} and approaches to the 6-dimensional $\caN=(2,0)$ theory~\cite{Lambert:2019diy,Rist:2020uaa}. Generically, the equation of motion for $B$ implies that the combination $\xi F+(-1)^{q}{\star\rmd B}$ is closed, which determines $B$ partly in terms of $F$ with the remaining part decoupling from the system. The equation of motion for $F$ then implies the Bianchi identity $\rmd\hat F=0$ and the equation of motion $\rmd{\star \hat F}=\frac{\xi^2}{1+\xi^2}\rmd R$, where $\hat F\coloneqq F+\frac{(-1)^{(d+1)(p+1)}}{1+\xi^2}{\star R}$ and $R$ is defined by the variation $S^\text{genSen,\,int}[F+\delta F,\Phi]\eqqcolon\int\delta F\wedge R+\caO((\delta F)^2)$ up to total derivatives.
        
        To make contact with the double-copied action~\eqref{eq:nonlocal_sen_gen}, note that~\eqref{eq:sen_gen2} remains invariant under the usual $q$-form gauge symmetry $\delta B_q\coloneqq\rmd\Lambda_{q-1}$, $\delta\Lambda_{q-1}\coloneqq\rmd\Lambda_{q-2}$, \ldots, and $\delta\Lambda_1\coloneqq\rmd\Lambda_0$~\cite{Rist:2020uaa}. Let us apply the standard Abelian $q$-form BRST quantization, choosing Feynman gauge and integrating out the Nakanishi--Lautrup fields, to leave 
        \begin{equation}\label{eq:sen_gen}
            S^\text{genSen}_\text{BRST}=\int\big\{-\tfrac12F\wedge{\star F}-\xi B\wedge\rmd F-\tfrac12B\wedge{\star\wave}B+\cdots\big\}~,
        \end{equation}
        where we have omitted the tower of ghost fields which decouple in Feynman gauge. Upon integrating out $B$ by its equation of motion 
        \begin{equation}
            B=(-1)^{q(d-q)}\xi\wave{}^{-1}{\star\rmd F}~,
        \end{equation}
        we recover the double-copied action~\eqref{eq:nonlocal_sen_gen} for an appropriate choice of $\xi$. 
        
        Even though we did not ask for it, the double copy produces (a generalization of) Sen's mechanism automatically, reinforcing its naturalness as an approach to describing the physics of gauge potentials in terms of elementary field strengths.
        
        \section{Conclusions and outlook}\label{sec:conclusion}
        
        In this paper, we have constructed a toolkit, not evident from a purely on-shell perspective, for boosting on-shell tree-level CK duality to off-shell loop-level CK duality up to counterterms. This toolkit uses the action of the underlying quantum field theory as an ordering principle, and full CK duality  becomes a manifest feature of the action. This new perspective on CK duality implies a very direct proof of the double copy at the loop level.
        
        Much as focusing on physical on-shell scattering amplitudes has dramatically advanced our understanding, stepping back off shell once again sheds new light on the scattering amplitudes themselves. One might say that \emph{off-shell is the new on-shell}.\footnote{We also remark that the homotopy algebraic perspective on field theories unifies scattering amplitudes and action principles.} 
        
        CK duality links the internal color symmetry to a kinematic symmetry algebra acting on space--time, and the Coleman--Mandula theorem seems to imply a fundamental dichotomy. In this paper, we have shown that both symmetries are indeed on an equal footing such that the Yang--Mills field $A_\mu^a(x)$ is biadjoint in terms of the color index $a$ and the kinematic index $\mu$. We have explained in \cref{ssec:coleman-mandula} how this perspective does not violate the Coleman--Mandula theorem.
        
        Although our paper is couched in the language of actions, we stress that what we do is purely \emph{perturbative}, that is, it concerns scattering amplitudes and correlators, in particular the usual scattering amplitude-theoretic CK duality and double copy conjectures. We do not claim here any direct implications for (among others) classical double copy of exact classical solutions~\cite{Monteiro:2014cda, Luna:2015paa,Luna:2016due,White:2016jzc,Berman:2018hwd, White:2020sfn,Chacon:2021wbr} and other non-perturbative double-copy-type relations~\cite{Alawadhi:2019urr, Banerjee:2019saj,Alawadhi:2021uie}.
        
        An important point we stressed throughout the paper but may require further highlighting is the following. Step~\ref{item:counterterms} in our algorithm in \cref{sec:outline} implies that counterterms arising from the Jacobian of the non-local field redefinitions may be required to ensure unitarity on the physical Hilbert space. Correspondingly the CK-dual loop integrands  generated by the Feynman diagrams of~\eqref{eq:YM_prepared_action_intro} manifestly  satisfy all desirable properties, as described in e.g.~\cite{Bern:2015ooa}, but may violate unitarity, which must be restored by counterterms. In particular, we do not contradict the result of~\cite{Bern:2015ooa} that there is no CK-dual 4-point, 2-loop integrand satisfying \emph{all} properties described therein. Due to the lack of unitarity before adding counterterms in our case, the standard proof~\cite{Bern:2010yg} of the loop-level validity of the double copy is clearly not applicable directly, but the argument  of~\cite{Borsten:2020zgj,Borsten:2021hua}, streamlined here in \cref{ssec:YM_double_copy}, is fully independent\footnote{Note that our argument in~\cite{Borsten:2020zgj,Borsten:2021hua} does assume the existence of a proof of the usual tree-level double copy for physical states. One such proof is certainly the standard proof of~\cite{Bern:2010yg}.} and valid nonetheless. The off-shell CK duality of~\eqref{eq:YM_prepared_action_intro} is sufficient for the double copy. Again, we stress that we do not double-copy counterterms, which nonetheless can be systematically accounted for. The latter observation is made clear by starting from $S^{\caN=0}_{\text{BRST}}$ and considering the field redefinitions required to produce the non-local terms that appear in ${S}^\text{DC}_\text{BRST}$ once the auxiliary fields have been integrated out. The counterterms required for unitary are just those induced by the Jacobian determinants of these field redefinitions. Similarly,  the `labeling problem'\footnote{See, for example, the discussion of~\cite{Casali:2020knc} and the references therein.} is naturally avoided by working at the level of the action.
        
        To summarize: we have lifted the notion of CK duality from on-shell gluon scattering amplitudes to the underlying BRST gauge-fixed action itself, up to the Jacobian counterterms. From this point of view, CK duality is rendered a manifest symmetry, in the usual sense, of an action that is perturbatively quantum equivalent to standard Yang--Mills theory.  
        
        The Feynman diagrams of this classical BRST action generate amplitude integrands that satisfy CK duality to all loop orders. However, the potential Jacobian counterterms, required to ensure unitarity beyond one loop, will generically break this CK duality. We have seen no reason to think that the Jacobian counterterm contributions to the integrands can be made to manifest CK duality, at least not while preserving the properties inherited from a direct Feynman diagram construction. CK duality can thus be understood as a classical symmetry that is generically anomalous. It is only in the case of very special field theories, such as $\caN=4$ super Yang--Mills theory, that there is evidence~\cite{Borsten:2022vtg} that our procedure can be implemented in a way that avoid problematic counterterms.
        
        Let us reiterate that the CK-anomaly is harmless, at least for most applications and purposes, as explained in \cref{ssec:absence_unitarity}; quantum consistency is ensured from the outset. More precisely, the required counterterms exist, are unique in the appropriate sense, and preserve BRST invariance.  
            
        We close by mentioning a few points that we plan to address in future work. The results of this paper beg to be formulated in the language of homotopy algebras.\footnote{See~\cite{Macrelli:2019afx,Jurco:2019yfd,Jurco:2020yyu,Saemann:2020oyz} for a discussion of scattering amplitudes and Berends--Giele recursion relations in terms of homotopy algebras for both tree- and loop-level scattering amplitudes. See also~\cite{Arvanitakis:2019ald} for related discussions of the S-matrix in the language of homotopy algebras,~\cite{Lopez-Arcos:2019hvg,Gomez:2020vat} for the tree-level perturbiner expansion, and~\cite{Nutzi:2018vkl} for a homotopy algebra interpretation of tree-level on-shell recursion relations.} The main result of the present contribution  translates into the assertion that, with the requisite gauge fixing, field redefinitions, and auxiliary fields, Yang--Mills theory --- and any other theory for which CK duality holds at tree level --- is described by a homotopy algebra that contains a so-called BV$^{\wave}$-algebra, which is the strict version of the concept of a BV$^{\wave}_\infty$-algebra as defined in~\cite{Reiterer:2019dys}; this algebra captures precisely the strong form of off-shell CK duality described in this paper. Furthermore, the double copy can also be naturally described in this formalism. We defer a detailed discussion of this formalism to an upcoming paper~\cite{Borsten:2022aa}.
        
        As mentioned several times, the results of this paper ignore CK duality and the double copy of counterterms. We argue in \cref{ssec:absence_unitarity} that this is largely irrelevant, and in particular it does not matter for determining the UV behavior of supergravity theories. It is clear that we have to validate this claim with detailed computations, which we intend to perform in future work once the simplifying homotopy algebraic framework~\cite{Borsten:2022aa} is fully set up. We note that the precise form of the counterterms depends on the regularization and renormalization schemes, such that any statements one makes would depend on the scheme, which is clearly undesirable. For certain quantum field theories, however, nature does give us a regulator in the form of string theory: the infinite tower of stringy modes come in to render the Yang--Mills integrands finite. In this regard, a natural way to prove CK duality for Yang--Mills theory with counterterms should be to prove CK duality at the loop level for open string amplitudes\footnote{See~\cite{Geyer:2021oox} for an approach to go in the other direction: to employ the loop-level scattering equations~\cite{Adamo:2013tsa,Geyer:2015bja}, the ambitwistor string~\cite{Mason:2013sva}, and modular invariance to uplift the double copy construction of supergravity loop amplitudes into superstring amplitudes.}.
        
        
    \end{body}
    
\end{document}